\documentclass[twocolumn,pra,aps,showpacs,showkeys,superscriptaddress,
 floatfix,reprint]{revtex4-1}

\usepackage{graphicx}
\usepackage{bm}
\usepackage[latin1]{inputenc}
\usepackage{amsmath}
\usepackage{float}
\allowdisplaybreaks
\usepackage{amssymb}
\usepackage{color}
\usepackage{caption}
\usepackage[english]{babel}
\usepackage{hyperref}
\usepackage{balance}

\DeclareMathOperator{\Tr}{Tr} 
\newcommand{\nn}{\nonumber}
\def\transt#1{{\cal T}^{({\rm #1})}}
\def\trans#1{{\cal T}^{(#1)}}
\def\scatt#1{{\cal S}^{({\rm #1})}}

\def\E#1{E^{({\rm #1})}}
\def\BS{{({\rm BS})}}
\allowdisplaybreaks
\newcommand{\lk}{\left(}
\newcommand{\rk}{\right)}
\newcommand{\lsz}{\left[}
\newcommand{\rsz}{\right]}
\newcommand{\lka}{\left\{}
\newcommand{\rka}{\right\}}
\usepackage[dvipsnames]{xcolor} 

\newcommand{\Hop}{\hat{H}}
\newcommand{\rhop}{\hat{\rho}}
\newcommand{\aop}{\hat{a}}
\newcommand{\adop}{\hat{a}^\dagger}

\newcommand{\bop}{\hat{b}}
\newcommand{\bdop}{\hat{b}^\dagger}
\newcommand{\sigop}{\hat{\sigma}}
\newcommand{\D}[1]{\mathcal{D}[#1]}
\newcommand{\Lb}[3]{\mathcal{L}^{#1}_{#2}[#3]}
\newcommand{\Ed}{\mathcal{E}}
\newcommand{\ev}[1]{\left\langle #1 \right\rangle}

\begin{document}
\title{Transfer matrix approach to determining the linear response of all-fiber networks of cavity-QED systems}
\author{Nikolett N\'emet}
\email[]{nnem614@aucklanduni.ac.nz}
\affiliation{The Dodd-Walls Centre for Photonic and Quantum Technologies, New
Zealand}
\affiliation{Department of Physics, 
             University of Auckland, Auckland, New Zealand}
\author{Donald White}
\email[]{donald@aoni.waseda.jp}
\affiliation{Department of Applied Physics, Waseda University, Tokyo, Japan}
\author{Shinya Kato}
\affiliation{Department of Applied Physics, Waseda University, Tokyo, Japan}
\affiliation{JST PRESTO, 4-1-8 Honcho, Kawaguchi, Saitama 332-0012, Japan}
\author{Scott Parkins}
\email[]{s.parkins@auckland.ac.nz}
\affiliation{The Dodd-Walls Centre for Photonic and Quantum Technologies, New
Zealand}
\affiliation{Department of Physics, 
             University of Auckland, Auckland, New Zealand}
\author{Takao Aoki}
\email[]{takao@waseda.jp}
\affiliation{Department of Applied Physics, Waseda University, Tokyo, Japan}

\date{\today}

\begin{abstract}
A semiclassical model is presented for characterizing the linear response of elementary quantum optical systems involving cavities, optical fibers, and atoms. Formulating the transmission and reflection spectra using a scattering-wave (transfer matrix) approach, the calculations become easily scalable. To demonstrate how useful this method is, we consider the example of a simple quantum network, i.e., two cavity-QED systems connected via an optical fiber. Differences between our quasi-exact transfer matrix approach and a single-mode, linearized quantum-optical model are demonstrated for parameters relevant to recent experiments with coupled nanofiber-cavity-QED systems.
\end{abstract}

\maketitle
\section{Introduction}

Fiber-optic systems are excellent candidates for the implementation of large-scale quantum networks owing to low propagation losses and to the variety of optical components that can be readily incorporated in a given setup \cite{Kimble2008}.
These include, as a result of recent developments in the field, short lengths of tapered nanofiber with waists on the order of $400$~nm, which are carefully tailored using a heat-and-pull method \cite{Nayak2007,Solano2017,Nayak2018}. These in turn enable efficient trapping of atoms in the evanescent field of the nanofiber using laser fields at red- and blue-detuned magic wavelengths \cite{Grimm2000,LeKien2004}. Efficient coupling of atomic excitations into guided fiber modes, and vice versa, have introduced new ways of implementing key elements of quantum communication or computation setups. For example, a quantum memory for single-photon states has been demonstrated using a cloud of cold atoms around a nanofiber \cite{Gouraud2015,Sayrin2015}, while chiral coupling of atoms to waveguide modes has been proposed~\cite{Lodahl2017}, and signatures of super- and subradiant behaviour of atoms coupled to a nanofiber field have been observed~\cite{Kien2005,Kien2012,SolanoNatCom2017}.

In order to increase the coupling between the trapped atoms and the light field, fiber Bragg grating (FBG) mirrors confine the optical modes longitudinally \cite{Nayak2011,Wuttke2012,Kato2015}. An important characteristic of fiber quantum-electrodynamic (fiber-QED) systems is that the cooperativity does not depend on the length of the fiber because of its continuous mode structure. Thus, as the effective mode area is greatly reduced in a tapered nanofiber, even a cavity with relatively low finesse and long length can reach the strong coupling regime~\cite{Kato2015,Ruddell2017}.

Recent experiments have measured the transmission spectra of a network of cavity quantum electrodynamic (cavity-QED) systems, where two ensembles containing some tens of (cesium) atoms trapped in the evanescent fields of nanofiber cavities were connected by a standard single-mode optical fiber. The dressed states of the atoms with the optical normal modes (cavity-fiber-cavity) were identified in the output spectra~\cite{Kato2019}. In the same experiment a fiber-dark mode was observed, which could serve as a robust and coherent coupling channel between the two cavity-QED systems. Reflection and transmission spectra from a similar setup showed signatures of normal modes consisting of atomic and optical excitations. Of particular interest was a cavity-dark mode, which couples the atoms via the connecting fiber mode, but without any photons in the cavities~\cite{White2019}.

Theoretical output spectra of fiber-optic systems can be obtained from the coherent dynamics of single optical modes and atoms, with incoherent losses accounted for in a master equation approach. However, in the case of the above-mentioned experiments, strong coupling was achieved with cavity lengths on the order of a meter and mirror reflectances as low as $60\%$. Thus, even though one can obtain good qualitative agreement between theory and experiment, some quantitative discrepancies between single-mode models and the experimental results point to an issue with the usual assumption of high mirror reflectances (i.e., low bandwidth) \cite{Kato2019,White2019}.

In this work, we present an alternative description of the output spectra in the weak driving limit using the well-known concept of transfer matrices. In this regime the atoms can be approximated as linearly polarizable systems \cite{Deutsch1995, Xuereb2009, Xuereb2010, Dombi2013}. This description is analogous to the one obtained using the input-output relations for weak driving \cite{Goban2015}. The transfer matrices characterize the propagation and scattering of traveling waves; thus, they are well-suited to the analysis of large-scale networks. While this method is restricted to the linear regime, where the atoms are not excited substantially, it offers an important check of the validity of single-mode quantum-optical models, which can of course also be applied in the stronger driving, non-linear regime. The present work also offers significant computational advantages over the single-mode quantum models when considering systems with multiple cavities, coupling fibers, and atoms.

The paper is structured as follows. First we introduce the formalism that we apply to describe the various network components using a transfer matrix approach. Then, we proceed to consider the cases of (i) an empty cavity, (ii) a single cavity-QED system, (iii) two fiber-coupled cavities, and (iv) two fiber-coupled cavity-QED systems, using both the transfer matrix model and the simplest quantum optical model, in which the cavities and coupling fiber are each modeled by single modes. We also show that in certain cases where differences emerge between output spectra obtained from the two models, better agreement can be recovered by simultaneously driving counter-propagating fields in the transfer matrix model in order to better ``mimick'' excitation of a (single) standing-wave mode in the quantum optical model.

\section{Scattering and transfer matrices for a linear four-port device}
\begin{figure}[h!]
\includegraphics[scale=0.43, viewport=0 0 570 123]{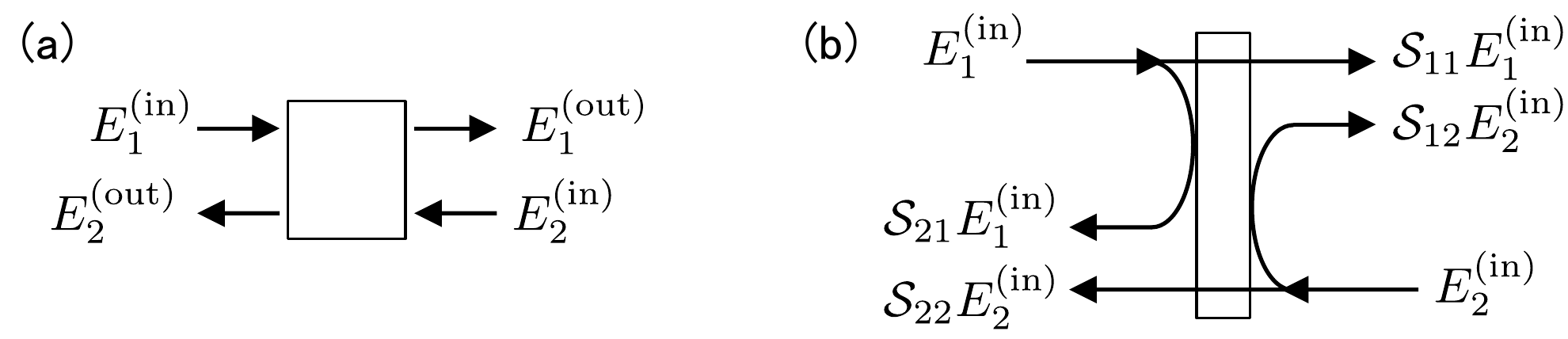}%
\caption{(a) A linear four-port device. (b) Amplitude transmission/reflection coefficients.
\label{fig:four-port}}
\end{figure}

First we review the scattering and transfer matrices for a linear four-port device (2 inputs and 2 outputs, e.g., a beam splitter), as shown schematically in Fig.~\ref{fig:four-port}(a).
The scattering matrix ${\cal S}$ describes the relationship between the two output fields $E_{1,2}^{({\rm out})}$ and the two input fields $E_{1,2}^{({\rm in})}$, 
\begin{align}
\begin{pmatrix}
E_1^{({\rm out})} \\
E_2^{({\rm out})}
\end{pmatrix}
=
{\cal S}
\begin{pmatrix}
E_1^{({\rm in})} \\
E_2^{({\rm in})}
\end{pmatrix}. 
\end{align}
The scattering matrix elements ${\cal S}_{ij}$ are the amplitude transmission/reflection coefficients defined in Fig.~\ref{fig:four-port}(b),
\begin{align}
{\cal S}
=
\begin{pmatrix}
{\cal S}_{11} & {\cal S}_{12} \\
{\cal S}_{21} & {\cal S}_{22}
\end{pmatrix}
.
\end{align}

On the other hand, the transfer matrix $\cal T$ relates the fields on the left-hand side, $E_1^{({\rm in})}$ and $E_2^{({\rm out})}$, to the fields on the right-hand side, $E_1^{({\rm out})}$ and $E_2^{({\rm in})}$,
\begin{align}
\begin{pmatrix}
E_1^{({\rm in})} \\
E_2^{({\rm out})}
\end{pmatrix}
=
{\cal T}
\begin{pmatrix}
E_1^{({\rm out})} \\
E_2^{({\rm in})}
\end{pmatrix}
. 
\end{align}
The transfer matrix elements ${\cal T}_{ij}$ can be expressed in terms of ${\cal S}_{ij}$ via
\begin{align}
{\cal T}
=
\begin{pmatrix}
{\cal T}_{11} & {\cal T}_{12} \\
{\cal T}_{21} & {\cal T}_{22}
\end{pmatrix}
=
\frac{1}{{\cal S}_{11}}
\begin{pmatrix}
1 & -{\cal S}_{12} \\
{\cal S}_{21} & {\cal S}_{11}{\cal S}_{22} - {\cal S}_{12}{\cal S}_{21}
\end{pmatrix}
,
\end{align}
while the scattering matrix elements  ${\cal S}_{ij}$ are expressed in terms of  ${\cal T}_{ij}$ as
\begin{align}
{\cal S}
=
\frac{1}{{\cal T}_{11}}
\begin{pmatrix}
1 & -{\cal T}_{12} \\
{\cal T}_{21} & {\cal T}_{11}{\cal T}_{22} - {\cal T}_{12}{\cal T}_{21}
\end{pmatrix}
.\label{eq:TtoS}
\end{align}

\begin{figure}[h!]
\includegraphics[scale=0.5, viewport=0 0 373 116]{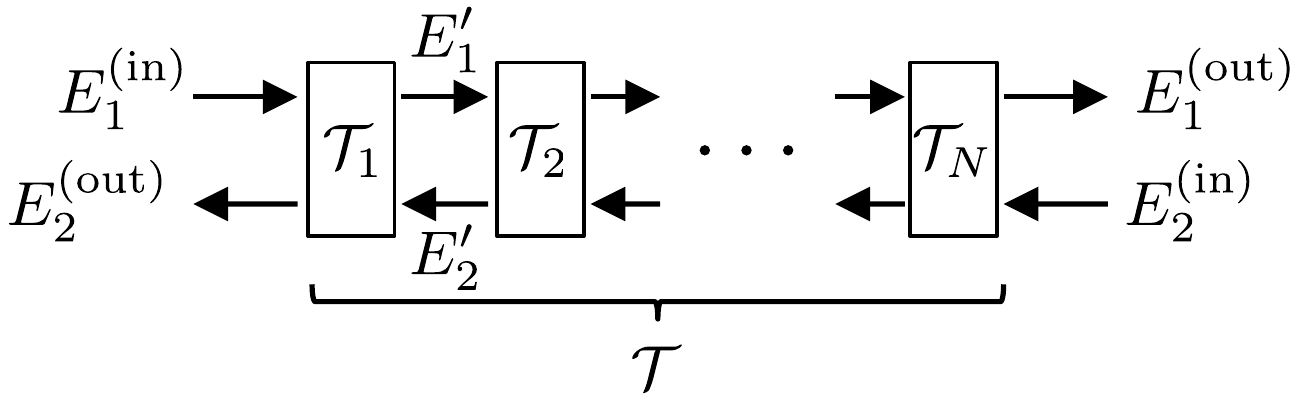}%
\caption{A series of $N$ linear four-port devices.
\label{fig:four-port-series}}
\end{figure}

In general, considering a series of linear four-port devices (Fig.~\ref{fig:four-port-series}), the fields $E_1^{({\rm in})}$ and $E_2^{({\rm out})}$ on the left-hand side are related to those on the right-hand side, $E_1^{({\rm out})}$ and $E_2^{({\rm in})}$, by
\begin{align}
\begin{pmatrix}
E_1^{({\rm in})} \\
E_2^{({\rm out})}
\end{pmatrix}
=
{\cal T}_1
\begin{pmatrix}
E_1^\prime \\
E_2^\prime
\end{pmatrix}
= \cdots 
= {\cal T}_1 {\cal T}_2 \cdots {\cal T}_N
\begin{pmatrix}
E_1^{({\rm out})} \\
E_2^{({\rm in})}
\end{pmatrix}
.
\end{align}
Therefore, the transfer matrix for this series of devices is given by the product of the transfer matrices of the individual elements,
\begin{align}
{\cal T} = {\cal T}_1 {\cal T}_2 \cdots {\cal T}_N.
\end{align}
In the following, we introduce the scattering and transfer matrices for specific fiber-network components. 

\subsection{Beam splitter/Mirror}
The first example is a lossless beam splitter. The scattering matrix of such a device also applies for a partially reflecting mirror and can be expressed by a unitary matrix \cite{Campos1989},
\begin{align}
{\cal S}^{(\rm M)}
=
e^{i\phi_0}
\begin{pmatrix}
e^{i\phi_{\rm t}} \sqrt{T} & e^{i\phi_{\rm r}} \sqrt{R} \\
-e^{-i\phi_{\rm r}} \sqrt{R} & e^{-i\phi_{\rm t}} \sqrt{T}
\end{pmatrix}
,
\end{align}
where $R$ and $T=1-R$ are the reflectance and transmittance of the beam splitter, respectively. For the case of a lossy beam splitter, $T+R<1$ and ${\cal S}^{(\rm M)}$ is no longer unitary.

In the following, for the sake of simplicity, we take $\phi_0 = -\pi/2$, $\phi_{\rm t} = \pi$, $\phi_{\rm r} = \pi/2$, so that the amplitude transmission/reflection coefficients for both input ports are the same and the reflection coefficients are real, ${\cal S}_{11} = {\cal S}_{22}$, ${\cal S}_{12}={\cal S}_{21} \in \Re$, i.e.,
\begin{align}
\label{eq:S_M}
{\cal S}^{(\rm M)}
=
\begin{pmatrix}
i\sqrt{T} & \sqrt{R} \\
\sqrt{R} & i\sqrt{T}
\end{pmatrix}
.
\end{align}
The corresponding transfer matrix is given by
\begin{align}
{\cal T}^{(\rm M)}
=
\frac{i}{\sqrt{T}}
\begin{pmatrix}
-1 & \sqrt{R} \\
-\sqrt{R} & T+R
\end{pmatrix}
.
\end{align}

\subsection{Free propagation}
\begin{figure}[h!]
\includegraphics[scale=0.5, viewport=0 0 397 121]{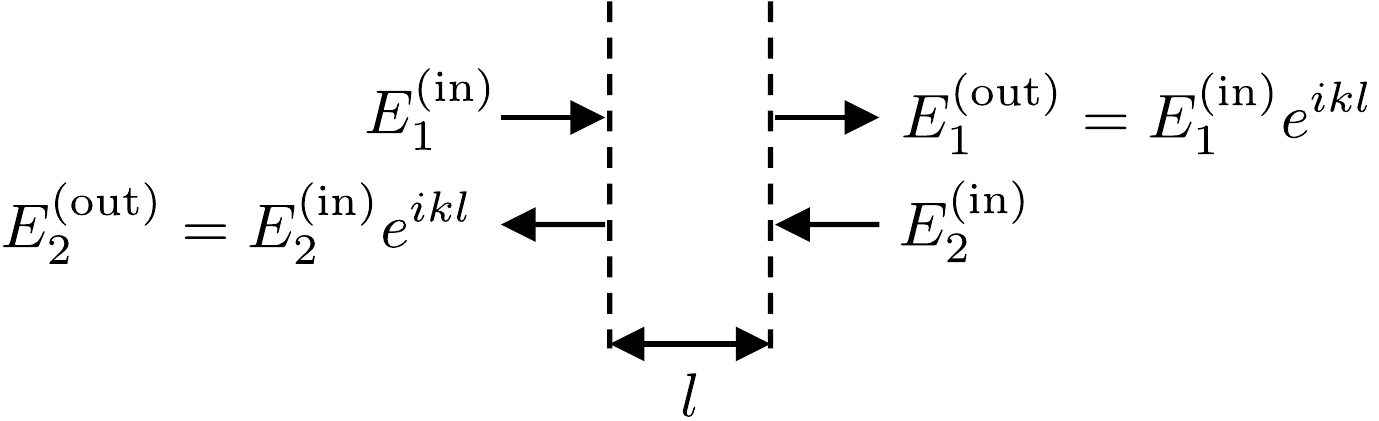}%
\caption{Propagation over a distance $l$.
\label{fig_propagation}}
\end{figure}
The transfer matrix for free propagation in a lossless medium over a distance $l$ (Fig.~\ref{fig_propagation}) is given by
\begin{align}
{\cal T}^{(l)}
=
\begin{pmatrix}
e^{-ikl} &0 \\
0 & e^{ikl}
\end{pmatrix}
=
\begin{pmatrix}
e^{-i\omega l/c} &0 \\
0 & e^{i\omega l/c}
\end{pmatrix}
\label{propagation}
,
\end{align}
where $k, \omega, c$ are the wave number, the angular frequency, and the velocity of the propagating wave, respectively. 

\begin{figure}[h!]
\includegraphics[scale=0.5, viewport=0 0 429 131]{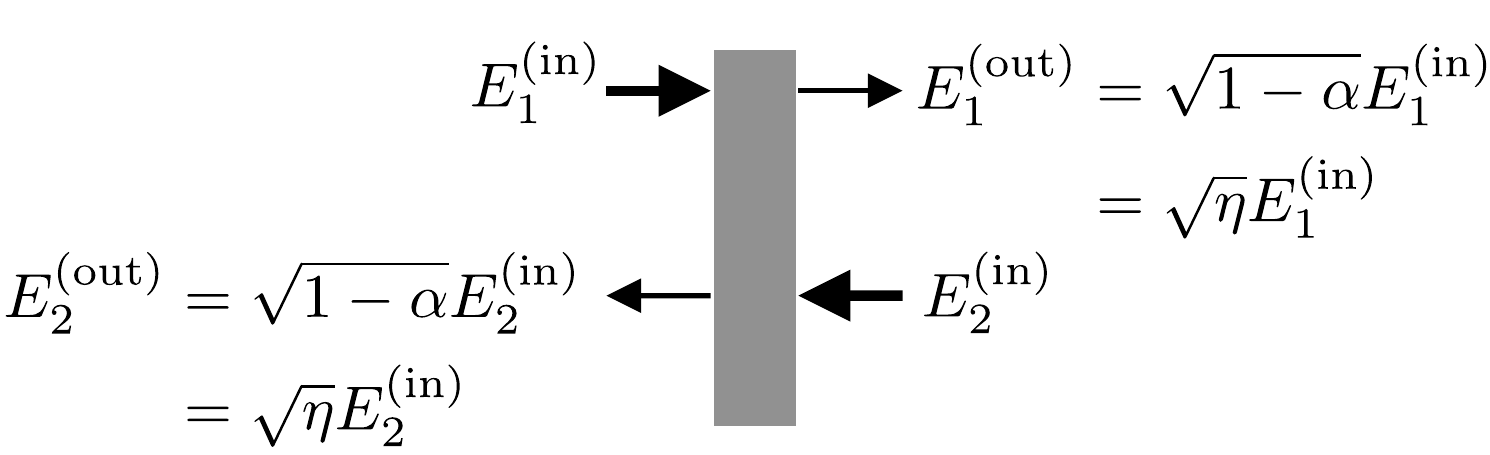}%
\caption{Linear loss.
\label{fig:loss}}
\end{figure}
The transfer matrix for a linear loss $\alpha$ (Fig.~\ref{fig:loss}) is given by
\begin{align}
{\cal T}^{(\alpha)}
=
\begin{pmatrix}
1/\sqrt{1-\alpha} &0 \\
0 & \sqrt{1-\alpha}
\end{pmatrix}
=
\begin{pmatrix}
1/\sqrt{\eta} &0 \\
0 & \sqrt{\eta}
\end{pmatrix}
,
\end{align}
where $\eta = 1-\alpha$ is the transmission efficiency. One can also incorporate a linear loss by replacing the real wave number $k$ in Eq.~(\ref{propagation}) with the complex wave number $\tilde{k}_j$, 
\begin{align}
\tilde{k}_j = k + ik^\prime_j, \quad\quad k^\prime_j =  {\rm Im} \tilde{k}_j = -\frac{\ln \eta_j}{2l}\nn \\ \left( \left| e^{i\tilde{k}_jl} \right|^2 = e^{-2k^\prime_j l}= \eta_j \right).
\end{align}
Here the indices $j$ refer to the distinct spatial regions of the setup (e.g., cavities) from left to right.

\subsection{Two-level atom in a waveguide}
The coupling rate between an atom and an optical mode describes the strength of the interaction between the dipole moment $d$ of the considered atomic transition and the local electric field produced by the optical mode. Therefore, after mode quantization, it takes the following form (see, e.g., \cite{Walls2008}),
\begin{align}
g = d\sqrt{\frac{\omega_{\rm C}}{2\varepsilon_0 \hbar V}},
\end{align}
where $V=\frac{Al}{2}$~\footnote{The factor $1/2$ comes from the standing-wave spatial mode of the Fabry-P\'erot cavity.} for a Fabry-P\'erot cavity of length $l$ and cross-sectional area $A$.
Coupling an atom to a waveguide can be described similarly, but a continuous spectrum of optical modes has to be considered. The transition from one regime to the other can be described via \cite{Blow1990,Domokos2002}
\begin{align}
g_{\rm W} = \frac{1}{\sqrt{\Delta k}}g = \sqrt{\frac{l^\prime}{2\pi}}g = d\sqrt{\frac{\omega_{\rm W}}{4\pi \varepsilon_0 \hbar A}},
\end{align}
where the mode volume for a waveguide of length $l^\prime$ is $V=Al^\prime$. The above expression also highlights the fact that the coupling strength between the atoms and the waveguide is independent of the length. Thus, we have the following relationship between the two coupling strengths:
\begin{align}
g = g_{\rm W} \sqrt{\frac{4\pi}{l}}.
\end{align}
\begin{figure}[h!]
\vglue -.4cm
\includegraphics[scale=0.5, viewport=0 0 457 114]{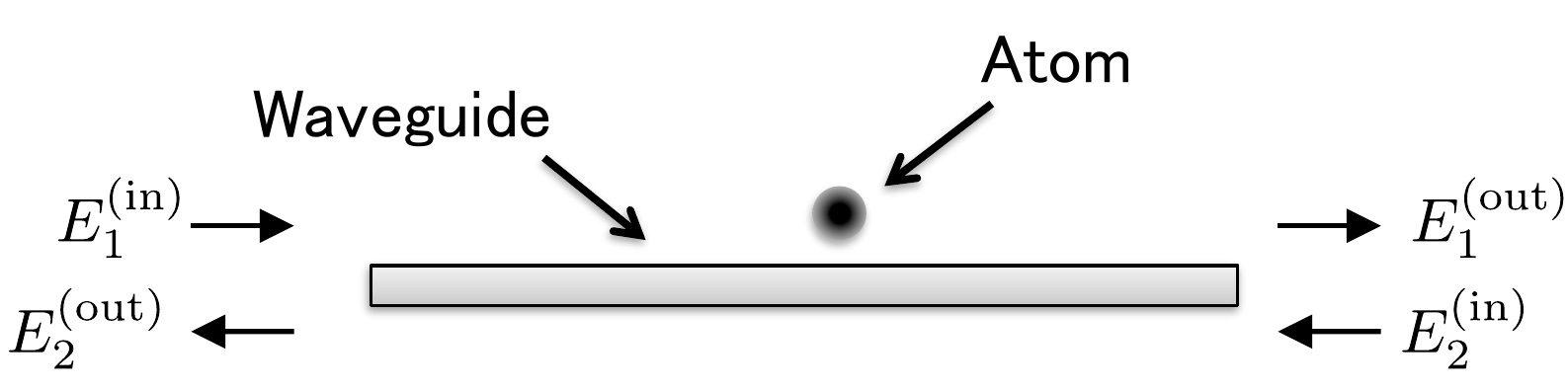}%
\caption{Two-level atom coupled to a one-dimensional waveguide.
\label{fig:atom}}
\end{figure}

Let us assume that an atom couples to forward and back-propagating one-dimensional guided modes (Fig.~\ref{fig:atom}). Then, considering weak coherent driving fields, the atomic dynamics becomes linear with respect to these fields. Using a real-space version of the input-output relations leads to the following transfer matrix \cite{Goban2015},
\begin{align}
\label{eq:T_atom}
{\cal T}^{(\rm A)}
=
\begin{pmatrix}
1-i\xi & -i\xi \\
i\xi & 1+i\xi
\end{pmatrix}
,
\end{align}
where 
\begin{align}
\xi = -\frac{\Gamma_{\rm 1D}/\Gamma^\prime}{i + 2\Delta_{\rm A}/\Gamma^\prime}.
\label{atom_xi}
\end{align}
In Eq.~(\ref{atom_xi}), $\Gamma_{\rm 1D}$ is the radiative intensity decay rate into the guided mode, given by
\begin{align}
\Gamma_{\rm 1D} &= \frac{4\pi g_{\rm W}^2}{v_g} ,
\end{align}
where $g_{\rm W}$ and $v_g$ are the above defined atom-waveguide coupling rate and the group velocity, respectively.

$\Gamma^\prime$ is the energy decay rate into all the other modes, and we can usually assume that this decay rate is the same as that in free space, $\Gamma^\prime \approx \Gamma_0 (= 2\gamma)$. Finally, 
$\Delta_{\rm A} = \omega - \omega_{\rm A}$ is the probe-atom detuning, where $\omega_{\rm A}$ is the atomic transition frequency.

The narrow widths of nanofibers support a special instance of atomic decay into guided modes that is commonly cited as chiral coupling \cite{Lodahl2017}. In this case the atom is only coupled to the field propagating in one direction and not the other. This is included in the above formalism by setting $\xi$ to 0 in the first row of Eq.~\ref{eq:T_atom} when the atom is coupled to the left-propagating mode or in the second row when it is coupled to the right-propagating one.

It is worth comparing the Purcell effect for a waveguide with that for a cavity. As mentioned above, a waveguide enhances the spontaneous emission of an atom coupled to the guided mode, and its decay rate is given by $\Gamma_{\rm 1D}$. On the other hand, a cavity with a large field decay rate $\kappa\gg (g,\gamma )$ gives rise to a cavity-enhanced spontaneous emission rate given by 
\begin{align}
\Gamma_{\rm cav} = 2C\Gamma_0 = 4C\gamma,
\end{align}
where $C=g^2/(\kappa\gamma)$
is the cooperativity parameter. We can see that
\begin{align}
\Gamma_{\rm cav} &= \frac{4g^2}{\kappa} = \frac{8\pi g_{\rm W}^2}{\kappa l} = \frac{8\pi g_{\rm W}^2}{v_g (T_1+T_2+2\alpha)/4} ,
\end{align}
where $T_{1,2}$ are the transmittances of the two cavity mirrors and $\alpha$ the intrinsic cavity loss. 
For $T_1=T_2=1$ and $\alpha=0$ we find $\Gamma_{\rm cav}=4\Gamma_{1D}$. The factor of 4 is related to the difference between the travelling-wave and standing-wave description of the interaction. More specifically, a factor of 2 comes from the difference between the mode volume for a cavity and a waveguide, as discussed above. The other factor of 2 comes from the fact that in a waveguide two counter-propagating modes are considered, rather than a single cavity mode \cite{Goban2015}. Therefore the same amount of intensity decays into two modes instead of one.

In the following we apply this formalism to simple examples of all-fiber networks. The similarities and differences between the results obtained by the transfer matrix approach and the quantum optical model are highlighted. Note that the specific parameter values that we use for the presented numerical results are largely inspired by the experimental platforms of \cite{Kato2019,White2019}.

\section{Empty Fabry-P\'erot cavity}

\begin{figure}[h!]
\includegraphics[scale=0.5, viewport=0 0 363 171]{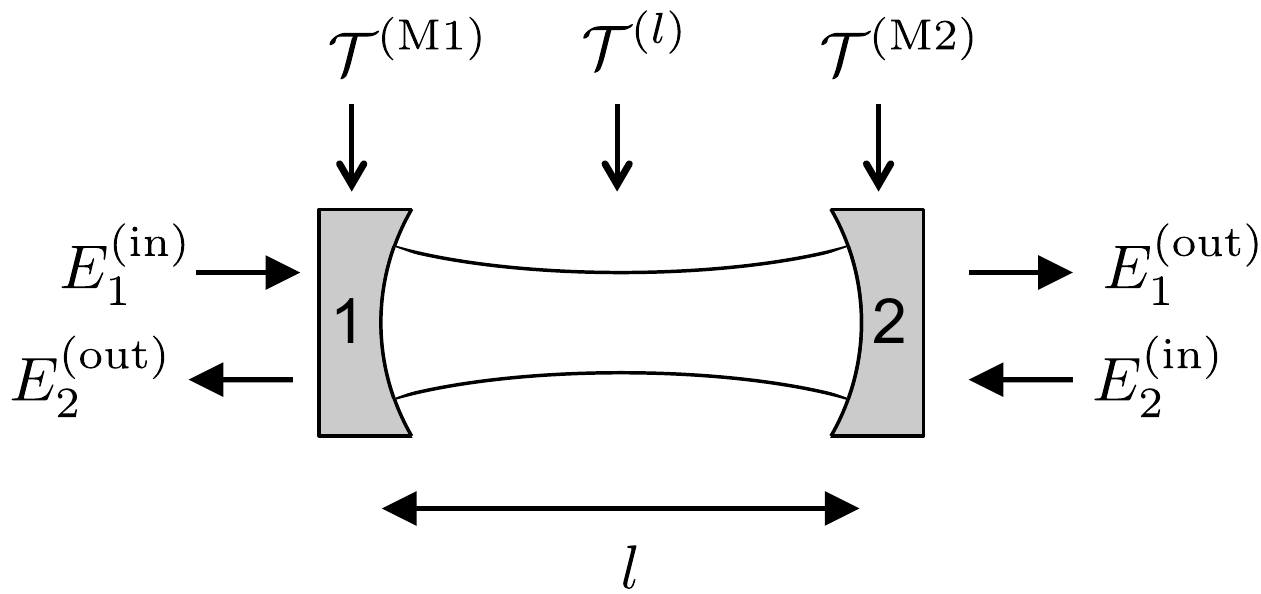}%
\caption{Empty Fabry-P\'erot cavity.
\label{fig:empty-fabry-perot}}
\end{figure}

Here we consider a standing-wave, (fiber) Fabry-P\'erot cavity bounded by two FBG mirrors at a distance $l$, as shown in Fig.~\ref{fig:empty-fabry-perot}. 

\subsection{Transfer matrix approach}
The transfer matrix for the cavity is given by
\begin{align}
{\cal T}^{({\rm FP})}
&=
{\cal T}^{(\rm M1)}
{\cal T}^{(l)}
{\cal T}^{(\rm M2)}
\nn\\
&=
\frac{i}{\sqrt{T_1}}
\begin{pmatrix}
-1 & \sqrt{R_1} \\
-\sqrt{R_1} & T_1+R_1
\end{pmatrix}
\begin{pmatrix}
e^{-i\tilde{k} l} &0 \\
0 & e^{i\tilde{k} l}
\end{pmatrix}\nn\\
&\qquad\times 
\frac{i}{\sqrt{T_2}}
\begin{pmatrix}
-1 & \sqrt{R_2} \\
-\sqrt{R_2} & T_2+R_2
\end{pmatrix} .
\end{align}
The elements of this matrix are evaluated as
\begin{align}
{\cal T}^{({\rm FP})}_{11} &= \frac{{\cal C}_{FP}(\omega)}{\sqrt{\eta T_1T_2}}\left[1 - \sqrt{R_1 R_2} \eta e^{i{\Phi}(\Delta_{\rm C})}\right] ,\\
{\cal T}^{({\rm FP})}_{12} &= \frac{{\cal C}_{FP}(\omega)}{\sqrt{\eta T_1T_2}}\left[- \sqrt{R_2} \right.\nn\\
&\left.\hspace{1.7cm}+\sqrt{R_1}(T_2+R_2)\eta e^{i{\Phi}(\Delta_{\rm C})}\right],\\
{\cal T}^{({\rm FP})}_{21} &= \frac{{\cal C}_{FP}(\omega)}{\sqrt{\eta T_1T_2}}\left[\sqrt{R_1}  \nn\right.\\
&\left.\hspace{1.7cm}-(T_1+R_1) \sqrt{R_2} \eta e^{i{\Phi}(\Delta_{\rm C})}\right],\\
{\cal T}^{({\rm FP})}_{22} &= \frac{{\cal C}_{FP}(\omega)}{\sqrt{\eta T_1T_2}}\left[-\sqrt{R_1 R_2} \right.\nn\\
&\left.\hspace{1.7cm}+(T_1+R_1)(T_2+R_2)\eta e^{i{\Phi}(\Delta_{\rm C})} \right]
, 
\end{align}
with
\begin{align}\label{C_FP}
{\cal C}_{FP}(\omega) &= -e^{-i\pi \omega / \omega_{\rm FSR}}, & {\Phi}(\Delta_{\rm C}) &= 2\pi \frac{\Delta_{\rm C} }{\omega_{\rm FSR}},
\end{align}
where $R_{1,2}$ and $T_{1,2}$ are the reflectance and transmittance of mirrors $1$ and $2$, 
and $l$ and $\eta$ are the cavity length and effective transmission of single-pass propagation inside the cavity. 
In Eq.~(\ref{C_FP}), we have introduced the free-spectral range (FSR) of the cavity,
\begin{align}
\omega_{\rm FSR} = 2\pi \frac{c}{2l} = \frac{\pi c}{l} ,
\label{FSR}
\end{align}
and the probe-cavity detuning
\begin{align}
\label{eq:cav_det}
\Delta_{\rm C} = \omega - \omega_{\rm C},
\end{align}
where $\omega_{\rm C}$ is the cavity resonance frequency that satisfies $e^{i2\pi \omega_{\rm C} / \omega_{\rm FSR}}=1$, and therefore 
\begin{align}
    e^{i2kl} = e^{i2\pi \omega/\omega_{\rm FSR}} = e^{i2\pi \Delta_{\rm C}/\omega_{\rm FSR}}.
\end{align}

\begin{figure*}[tb!]
\centering
\includegraphics[width=0.75\textwidth]{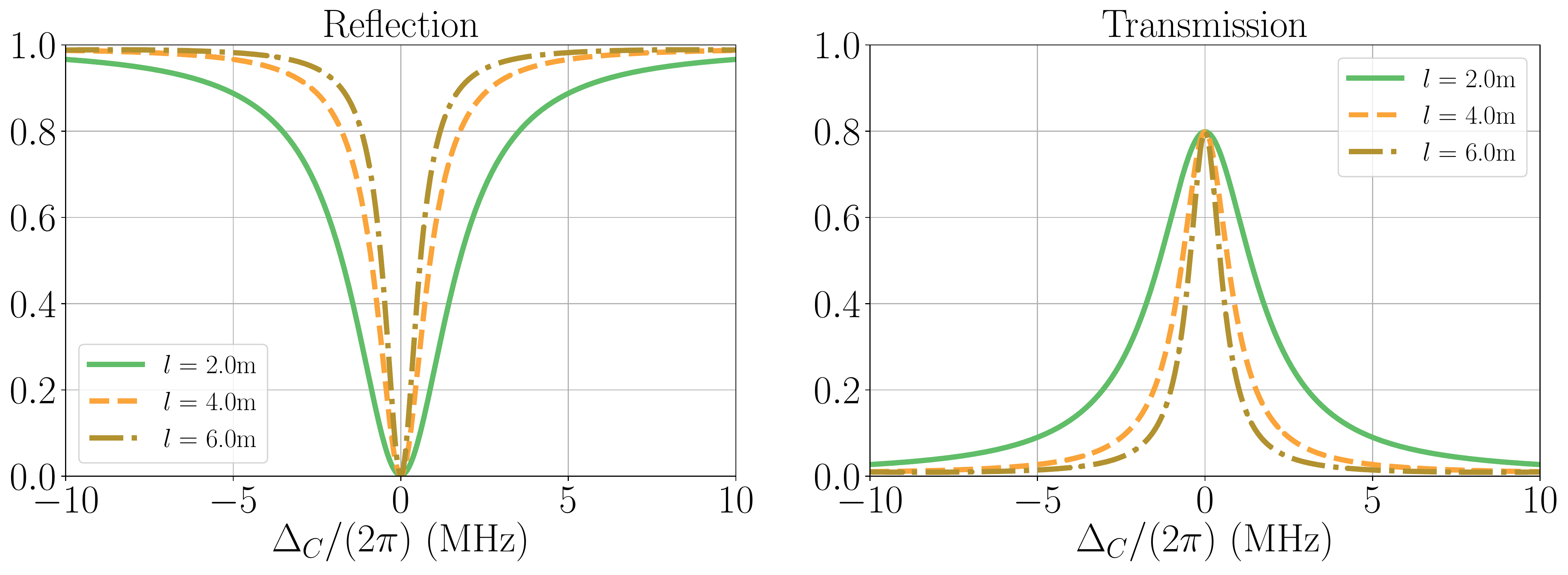}

\includegraphics[width=0.8\textwidth]{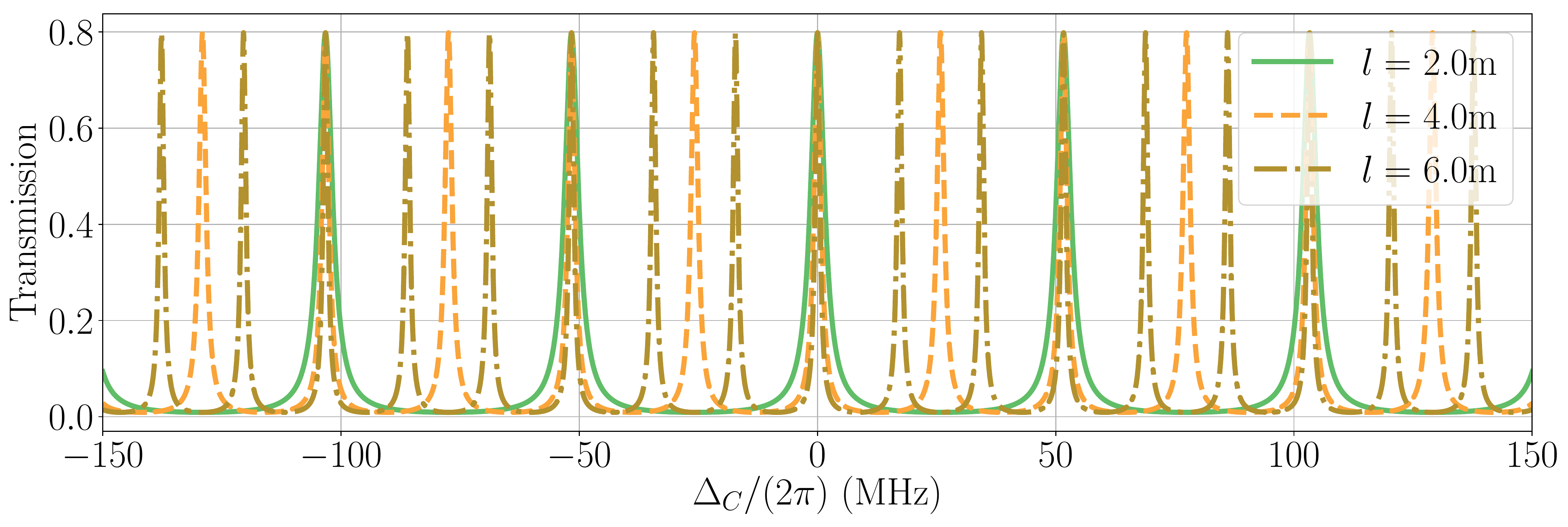}%
\vspace{-.3cm}
\caption{Transmission and reflection spectra of an empty Fabry-P\'erot cavity using the TM model. The setup is weakly driven from the left. The different lines correspond to different cavity lengths, and thus to different free spectral ranges. Parameters: $R_1 =0.8, R_2=0.85$, $\eta=0.98$ (i.e., $2\%$ single-pass transmission loss in the cavity).
\label{fig:FP_TM_l}}
\vspace{3mm}
\includegraphics[width=0.8\textwidth]{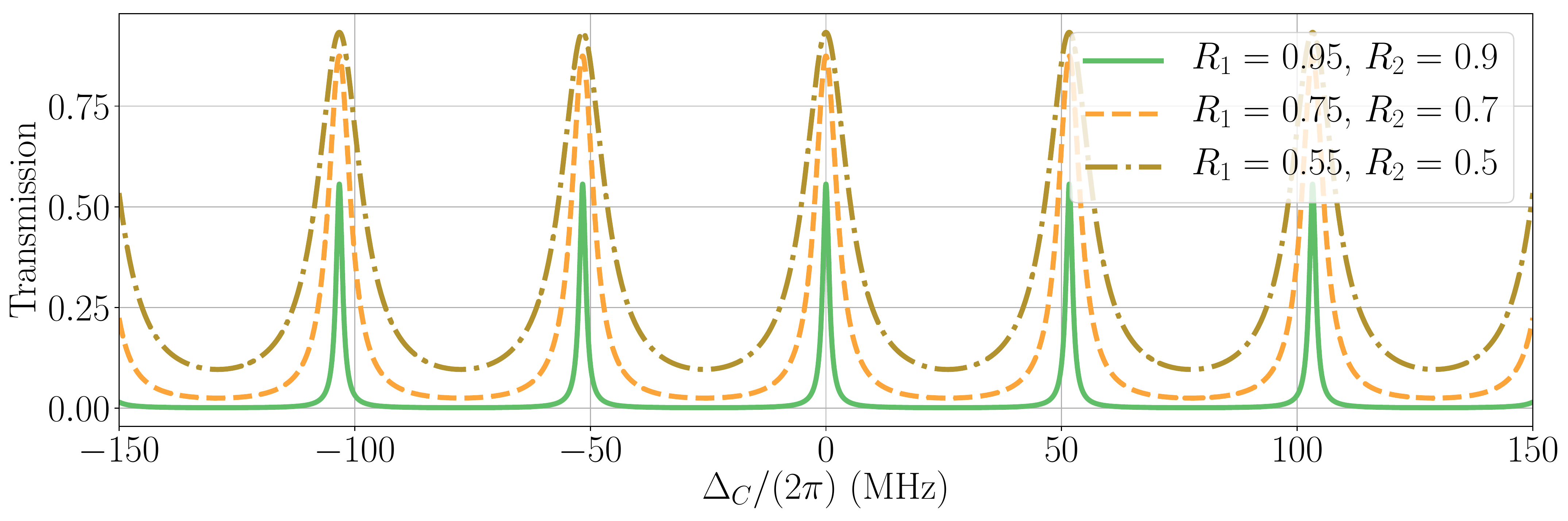}%
\vspace{-.3cm}
\caption{Same as in Fig.~\ref{fig:FP_TM_l}, but the different lines correspond to different reflectances for the cavity mirrors. Parameters: $l=2$ m, $\eta=0.98$.
\label{fig:FP_TM_R_far}}
\vspace{-3mm}
\end{figure*}

We obtain the transmission and reflection spectra of the cavity as
\begin{align} 
T^{(\rm FP)}(\omega) &= \left| {\cal S}^{({\rm FP})}_{11} \right|^2 = \left| \frac{1}{{\cal T}^{({\rm FP})}_{11}} \right|^2\nn\\ %
&= \left| \frac{\sqrt{\eta T_1T_2}}{1 -\eta \sqrt{R_1R_2}e^{i\Phi(\Delta_C)} } \right|^2,
\label{T_FP} \\
R^{(\rm FP)}(\omega) &= \left| {\cal S}^{({\rm FP})}_{21} \right|^2 = \left| \frac{{\cal T}^{({\rm FP})}_{21}}{{\cal T}^{({\rm FP})}_{11}} \right|^2\nn\\ %
&= \left| \frac{\sqrt{R_1} -\eta \left(T_1+R_1\right) \sqrt{R_2} e^{i\Phi(\Delta_C)}}{1 -\eta \sqrt{R_1R_2}e^{i\Phi(\Delta_C)} } \right|^2
.
\label{R_FP}
\end{align}

Let us examine these spectra in more detail. The upper panel of Fig.~\ref{fig:FP_TM_l} shows that decreasing the length of the cavity increases the effective linewidth and the free spectral range becomes larger (see also the lower panel in Fig.~\ref{fig:FP_TM_l} where higher-order resonances are visible). Lowering the reflectances also increases the linewidth and leads to overlapping resonances (Fig.~\ref{fig:FP_TM_R_far}).

\begin{figure*}[tb]
\centering
\includegraphics[width=0.75\textwidth]{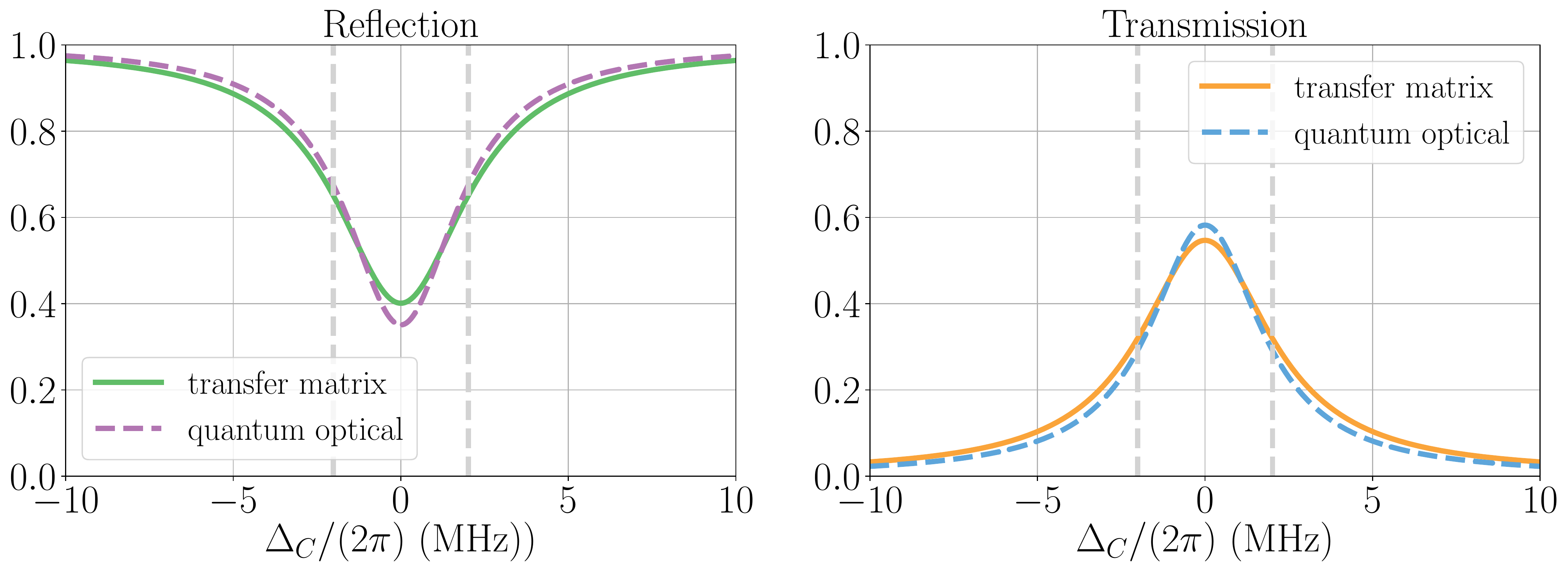}%
\vspace{-.3cm}
\caption{Comparison between the TM and single-mode quantum-optical (QO) models. The setup is weakly driven from the left. Transmission and reflection spectra of an empty Fabry-P\'erot cavity are shown within 1 FSR, where, for one of the mirrors, the condition $T\ll 1$ does not hold. Grey, vertical dashed lines are shown at $\pm\kappa_{\rm C}/(2\pi)$. Parameters: (TM) $R_1 =0.9$, $R_2=0.65$, $l=2$ m ($\omega_{\rm FSR}/(2\pi)=51.635$ MHz), $\eta=0.98$. (QO) $\kappa_1/(2\pi) = 0.411$ MHz, $\kappa_2/(2\pi) = 1.438$ MHz, $\kappa_{\rm C,i}/(2\pi) = 0.166$ MHz.\label{fig:FP_QM_close}}
\end{figure*}
\begin{figure*}[tb]
\includegraphics[width=0.8\textwidth]{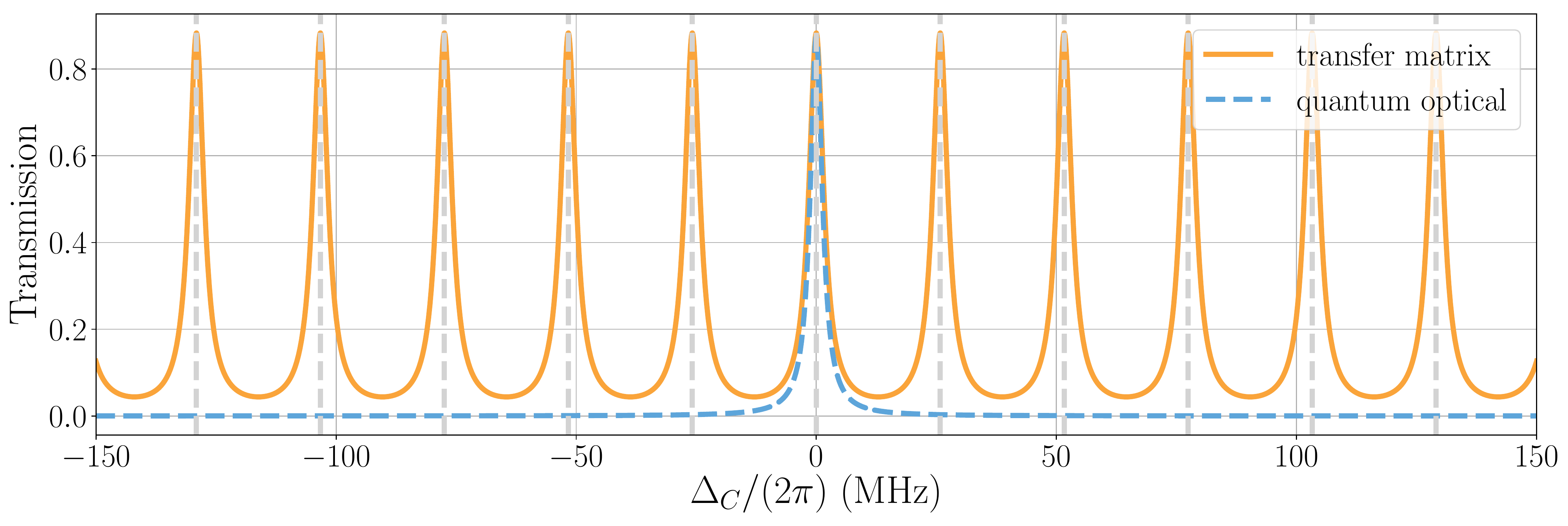}%
\vspace{-.3cm}
\caption{Comparison between TM and QO models, shown over a wider range of detunings. Parameters: (TM) $R_1 =0.7$, $R_2=0.6$, $l=4$ m ($\omega_{\rm FSR}/(2\pi)=25.818$ MHz), $\eta=0.98$. (QO) $\kappa_1/(2\pi) = 0.616$ MHz, $\kappa_2/(2\pi) = 0.821$ MHz, $\kappa_{\rm C,i}/(2\pi) = 0.083$ MHz.\label{fig:FP_QM_far}}
\vspace{-.3cm}
\end{figure*}

If we take the limit of high finesse, $\{ \alpha , \, 1-R_1, \,1-R_2\} \ll 1$, and small detunings ($|\Delta_{\rm C}| \ll \omega_{\rm FSR}$, i.e., $\Phi(\Delta_C)\ll1$), Eqs.~(\ref{T_FP}) and (\ref{R_FP}) reproduce the results obtained from the single-mode, input-output formalism (as we will see in the following subsection),
\begin{align}
\label{eq:R_cav}
R^{(\rm FP)}(\omega) &= \left| 1 - \frac{2\kappa_1 }{\kappa_{\rm C} - i \Delta_{\rm C} } \right|^2,
\\\label{eq:T_cav}
T^{(\rm FP)}(\omega) &= \left| \frac{2\sqrt{\kappa_1 \kappa_2}}{\kappa_{\rm C} - i \Delta_{\rm C} } \right|^2
,
\end{align}
where $\kappa_{j}$ are the cavity field decay rates through the mirrors $j=\lka1,2\rka$, given by
\begin{align}
\label{eq:kappa_def}
\kappa_{j} = \frac{1}{2}\frac{c}{2l}T_{j} = \frac{1}{2}\frac{\omega_{\rm FSR}}{2\pi}T_{j}
,
\end{align}
and $\kappa_{\rm C}$ is the total cavity field decay given by
\begin{align}
\kappa_{\rm C} &= \frac{1}{2}\frac{c}{2l}\big[ (1-R_1) + (1-R_2) + 2\alpha \big] \nn\\
&= \frac{1}{2}\frac{\omega_{\rm FSR}}{2\pi}\big[ (1-R_1) + (1-R_2) + 2\alpha \big]
,
\end{align}
which is related to the linewidth of the resonance, i.e., the full width at half maximum is given by $2\kappa_{\rm C}$ (see Fig.~\ref{fig:FP_QM_close} and below).

\vspace{-.5cm}

\subsection{Quantum-optical model}
The quantum-optical model is built on the assumption that wherever a Fabry-P\'erot cavity is considered, standing wave modes are able to build up. Therefore, most frequently a mode operator $\aop$ is assigned to the dominant (or most relevant) single standing-wave mode.

The standard way of describing the dynamics of an open quantum optical system is to use the master equation for the reduced density operator, $\rhop$, of the system. For a cavity driven by a probe laser of frequency $\omega$ and strength $\Ed$, this is given by (in a frame rotating at frequency $\omega$, and setting $\hbar=1$)
\begin{align}
\frac{d}{dt}\rhop &= -i\lsz\Hop^{({\rm cav})},\rhop\rsz +\Lb{({\rm cav})}{}{\rhop} ,
\end{align}
where
\begin{align}
\label{eq:Hcav}
\Hop^{({\rm cav})} &= -\Delta_{\rm C}\adop\aop+\Ed\lk\aop+\adop\rk ,\\
\label{eq:Lcav}
\Lb{\rm (cav)}{}{\rhop} &= (\kappa_1+\kappa_2+\kappa_{\rm C,i})\D{\aop}\rhop=\kappa_{\rm C}\D{\aop}\rhop ,
\end{align}
and $\D{\hat{O}} = 2\hat{O}\rhop\hat{O}^\dagger-\hat{O}^\dagger\hat{O}\rhop-\rhop\hat{O}^\dagger\hat{O}$.

Here, $\kappa_{\rm C,i}$ denotes internal (or intrinsic) loss within the cavity, which, if associated with propagation loss, is related to $\alpha$ by
\begin{align}
\kappa_{\rm C,i} = -\frac{1}{2}\frac{c}{l}\ln (1-\alpha ) ,    
\end{align}
which reduces to $\kappa_{\rm C,i}=c\alpha/(2l)$ for $\alpha\ll 1$.

The time evolution of the expectation value of the cavity field amplitude can be determined from the master equation as
\begin{align}
\frac{d}{dt}\ev{\aop} &= \frac{d}{dt}\Tr\lsz\aop\rhop(t)\rsz =\Tr\lsz\aop\frac{d}{dt}\rhop(t)\rsz \nn\\
&=(i\Delta_{\rm C}-\kappa_{\rm C})\ev{\aop}-i\Ed.
\end{align}

The reflection and transmission spectra can be obtained by expressing the steady state of these equations as a function of the driving frequency $\omega$ \cite{Parkins2013},
\begin{align}
\ev{\aop}_{\rm ss}&=-i\frac{\Ed}{\kappa_{\rm C}-i\Delta_{\rm C}} ,
\end{align}
which gives a Lorentzian lineshape for the intracavity intensity as a function of $\Delta_{\rm C}$.

The reflection and transmission spectra can be calculated when driving from the left by using the steady-state intracavity field combined with the coherent input field via the input-output formalism as
\begin{align}
\label{eq:R}
R &= \frac{|\ev{\aop_{in,1}}+\sqrt{2\kappa_1}\ev{\aop}_{\rm ss}|^2}{|\ev{\aop_{in,1} }|^2+|\ev{\aop_{in,2} }|^2} ,\\
\label{eq:T}
T &= \frac{|\ev{\aop_{in,2}}+\sqrt{2\kappa_2}\ev{\aop}_{\rm ss}|^2}{|\ev{\aop_{in,1} }|^2+|\ev{\aop_{in,2} }|^2},
\end{align}
where
\begin{align}
\ev{\aop_{in,1}} &= i\frac{\Ed}{\sqrt{2\kappa_1}} , \qquad
\ev{\aop_{in,2}} = 0,
\end{align}
which give back the formulae (\ref{eq:R_cav},\ref{eq:T_cav}). 

Whenever the assumptions of high finesse and small detuning become marginal, results from the quantum-optical model can be expected to start deviating from the more accurate transfer matrix approach. This is illustrated in Figs.~\ref{fig:FP_QM_close} and \ref{fig:FP_QM_far}, where the two models are compared for relatively low-finesse cavities. We note again that the transfer matrix model describes a scattering problem and therefore incorporates multiple frequency modes for each optical element (Fig.~\ref{fig:FP_QM_far}). This causes broadening of the central Lorentzian in Fig.~\ref{fig:FP_QM_close}. In contrast, the quantum-optical model considers only a single mode for the cavity, and therefore, captures the spectrum only in the close vicinity of the cavity resonance frequency. Fig. \ref{fig:FP_QM_far} also highlights that as the mirror reflectances and the FSR of the cavity decrease, overlaps between neighbouring resonances develop, and this is of course only accounted for in the TM description.

\vspace{1cm}
\section{Single cavity-QED system}

\begin{figure}[h!]
\includegraphics[scale=0.5, viewport=0 0 363 190]{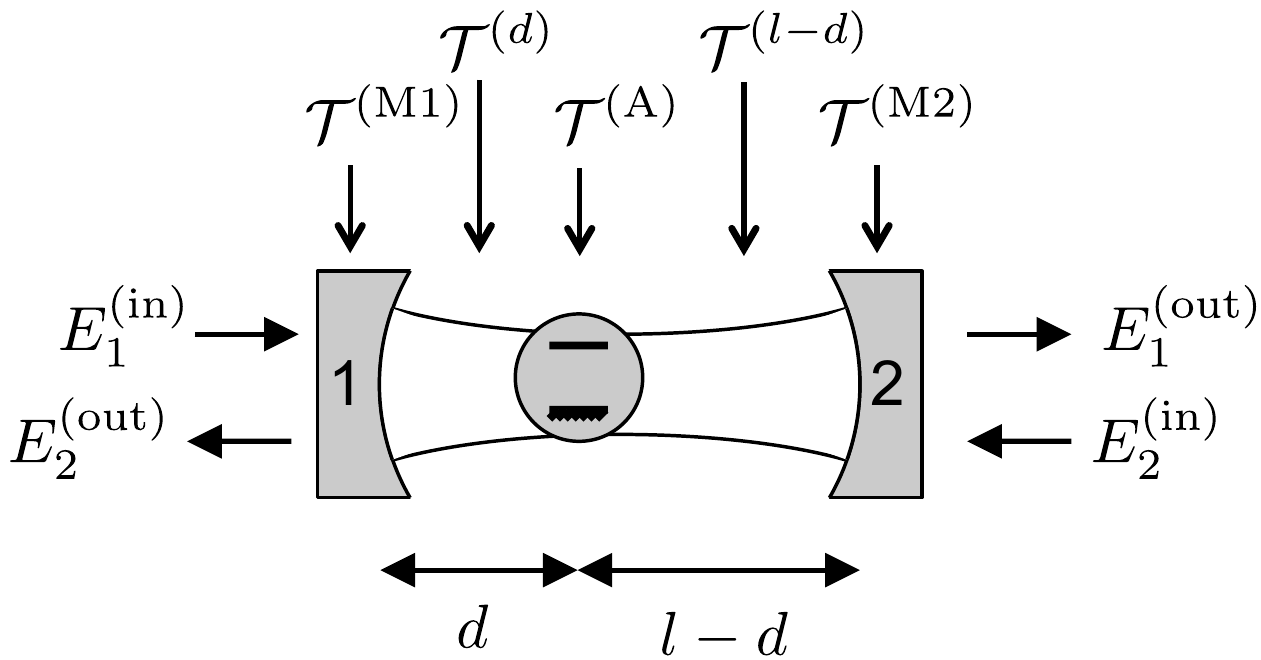}%
\caption{Single-atom cavity QED setup.
\label{cQED}}
\vspace{-1cm}
\end{figure}

In this section we consider a simple cavity-QED system, where a single atom is trapped in a cavity field, as depicted in Fig.~\ref{cQED}.

\subsection{Transfer matrix approach}
In this case, the transmission and reflection spectra are given by
\begin{align} 
\label{T_cQED}
T^{(\rm cQED)}(\omega) &= \left| \frac{1}{{\cal T}^{({\rm cQED})}_{11}} \right|^2 , \\ %
\label{R_cQED}
R^{(\rm cQED)}(\omega) &= \left| \frac{{\cal T}^{({\rm cQED})}_{21}}{{\cal T}^{({\rm cQED})}_{11}} \right|^2 ,%
\end{align}
where the transfer matrix corresponding to the whole system is

\begin{figure*}[tb]
\includegraphics[width=0.75\textwidth]{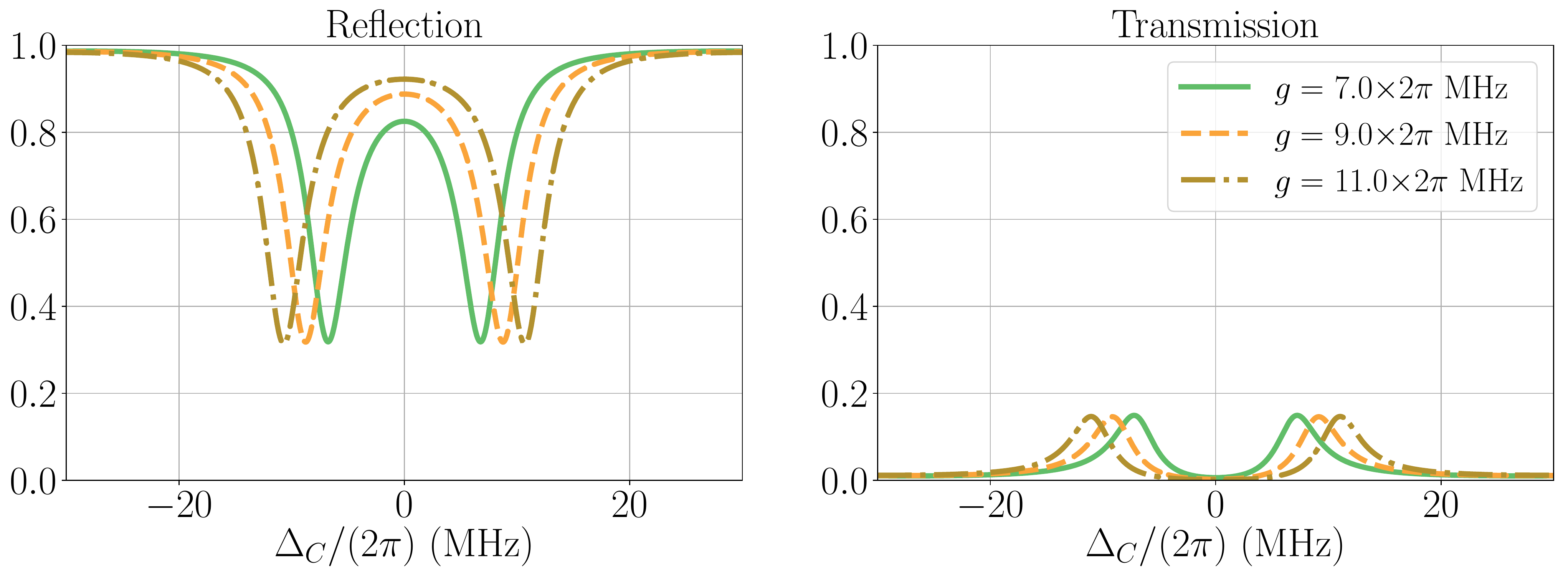}%
\vspace{-.3cm}
\caption{Transmission and reflection spectra of a Fabry-P\'erot cavity with an atom inside using the TM model near the atomic resonance frequency $\omega_{\rm A}=\omega_{\rm C}$ ($\Delta_{\rm A}=\Delta_{\rm C}\equiv\Delta$). The setup is weakly driven from the left. Parameters: $R_1 =0.8$, $R_2=0.85$, $l=2$ m ($\omega_{\rm FSR}/(2\pi)=51.64$~MHz), $\eta=0.98$, $\gamma /(2\pi)=2.65$~MHz.
\label{fig:cQED_TM_g_close}}
\vspace{3mm}

\includegraphics[width=0.8\textwidth]{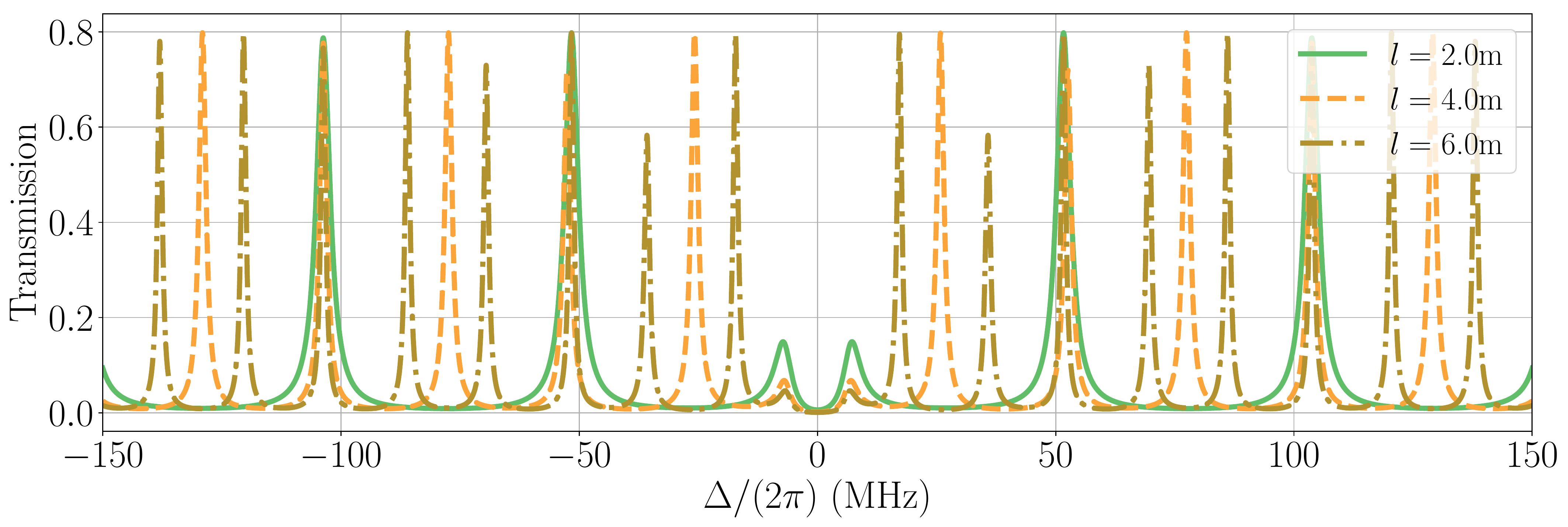}%
\vspace{-.3cm}
\caption{Same as in Fig.~\ref{fig:cQED_TM_g_close} but over a wider range of detunings. Parameters: $R_1 =0.8$, $R_2=0.85$, $\eta=0.98$, $g=7\cdot 2\pi$~MHz, $\gamma /(2\pi)=2.65$~MHz.
\label{fig:cQED_TM_l_far}}
\vspace{-.3cm}
\end{figure*}

\begin{widetext}
\begin{align}
{\cal T}^{({\rm cQED})}
&=
{\cal T}^{(\rm M1)}
{\cal T}^{(d)}
{\cal T}^{({\rm A})}
{\cal T}^{(l-d)}
{\cal T}^{(\alpha)}
{\cal T}^{(\rm M2)}
\nn\\
&=
\frac{i}{\sqrt{T_1}}
\begin{pmatrix}
-1 & \sqrt{R_1} \\
-\sqrt{R_1} & T_1+R_1
\end{pmatrix}
\begin{pmatrix}
e^{-i\tilde{k} d} &0 \\
0 & e^{i\tilde{k} d}
\end{pmatrix}
\begin{pmatrix}
1-i\xi & -i\xi \\
i\xi & 1+i\xi
\end{pmatrix}
\begin{pmatrix}
e^{-i\tilde{k} (l-d)} &0 \\
0 & e^{i\tilde{k} (l-d)}
\end{pmatrix}
\frac{i}{\sqrt{T_2}}
\begin{pmatrix}
-1 & \sqrt{R_2} \\
-\sqrt{R_2} & T_2+R_2
\end{pmatrix}
,
\end{align}
with elements
\begin{align}
&{\cal T}^{({\rm cQED})}_{11}
=
-\frac{e^{-i\tilde{k} l}}{\sqrt{T_1T_2}} 
\left\{
\left[ (1-i\xi) - i\xi  \sqrt{R_1}e^{2i\tilde{k} d} \right]  -\left[ i\xi+ (1+i\xi) \sqrt{R_1}e^{2i\tilde{k} d} \right] \sqrt{R_2} e^{2i\tilde{k} (l-d)}
\right\}
, \label{TcQED11} \\
&{\cal T}^{({\rm cQED})}_{12}
=-\frac{e^{-i\tilde{k} l}}{\sqrt{T_1T_2}} 
\left\{
\left[ -(1-i\xi) + i\xi  \sqrt{R_1}e^{2i\tilde{k} d} \right] \sqrt{R_2}  +\left[ i\xi+ (1+i\xi) \sqrt{R_1}e^{2i\tilde{k} d} \right](T_2+R_2) e^{2i\tilde{k} (l-d)} 
 \right\}
, \label{TcQED12} \\
&{\cal T}^{({\rm cQED})}_{21}
=-\frac{e^{-i\tilde{k} l}}{\sqrt{T_1T_2}} 
\left\{
\left[ (1-i\xi)\sqrt{R_1} - i\xi (T_1+R_1)e^{2i\tilde{k} d} \right]   -\left[ i\xi \sqrt{R_1} + (1+i\xi )(T_1+R_1)e^{2i\tilde{k} d} \right] \sqrt{R_2} e^{2i\tilde{k} (l-d)}
\right\}
, \label{TcQED21} \\
&{\cal T}^{({\rm cQED})}_{22}
=
%
-\frac{e^{-i\tilde{k} l}}{\sqrt{T_1T_2}} 
\left\{
 \left[-(1-i\xi)\sqrt{R_1} + i\xi (T_1+R_1)e^{2i\tilde{k} d} \right] \sqrt{R_2} \right.\nn\\
&\hspace{3.5cm}\left. + \left[ i\xi \sqrt{R_1} + (1+i\xi )(T_1+R_1)e^{2i\tilde{k} d} \right] (T_2+R_2) e^{2i\tilde{k} (l-d)}
 \right\}
. \label{TcQED22} 
 \end{align}
${\cal T}_{ij}^{({\rm cQED})}$ depend on position $d$ of the atom, which reflects the spatial variation of the atom-field coupling rate associated with the standing-wave nature of the cavity mode [$g(d) \sim g \cos (kd)$]
\footnote{Another origin is the propagation efficiency. This can be understood by, for example,  considering the reflection by an atom with no mirrors ($T_1 =T_2 = 1$). The incoming field experiences propagation through a distance $d$ (with an efficiency of $\eta = e^{-2k^\prime d}$) before reaching the atom and after reflecting off the atom and returning to the initial position. The amplitude of the reflected field would obviously depend on $d$. }.
For simplicity, we can take $d=0$ to obtain simpler expressions:
\begin{align}
{\cal T}^{({\rm cQED})}_{11}
&=
\frac{{\cal C}_{FP}(\omega)}{\sqrt{\eta T_1T_2}}
\left\{
\left[ (1-i\xi) - i\xi  \sqrt{R_1}\right]  -\left[ i\xi+ (1+i\xi) \sqrt{R_1} \right] \sqrt{R_2} \eta e^{i {\Phi}(\Delta_{\rm C})}
\right\} ,
\\
{\cal T}^{({\rm cQED})}_{12}
&=
\frac{{\cal C}_{FP}(\omega)}{\sqrt{\eta T_1T_2}}
\left\{
\left[ -(1-i\xi) + i\xi  \sqrt{R_1} \right] \sqrt{R_2} + \left[ i\xi+ (1+i\xi) \sqrt{R_1} \right](T_2+R_2) \eta e^{i {\Phi}(\Delta_{\rm C})} 
 \right\} ,\\
{\cal T}^{({\rm cQED})}_{21}
&=
\frac{{\cal C}_{FP}(\omega)}{\sqrt{\eta T_1T_2}}
\left\{
\left[ (1-i\xi)\sqrt{R_1} - i\xi (T_1+R_1) \right] -\left[ i\xi \sqrt{R_1} + (1+i\xi )(T_1+R_1) \right] \sqrt{R_2} \eta e^{i {\Phi}(\Delta_{\rm C})}
\right\} ,
\\
{\cal T}^{({\rm cQED})}_{22}
%
&=
%
\frac{{\cal C}_{FP}(\omega)}{\sqrt{\eta T_1T_2}}
\left\{
 \left[-(1-i\xi)\sqrt{R_1} + i\xi (T_1+R_1) \right] \sqrt{R_2} + \left[ i\xi \sqrt{R_1} + (1+i\xi )(T_1+R_1) \right] (T_2+R_2) \eta e^{i{\Phi}(\Delta_{\rm C})}
 \right\} .
\end{align}
\end{widetext}
\begin{figure*}[tb]
\includegraphics[width=0.7\textwidth]{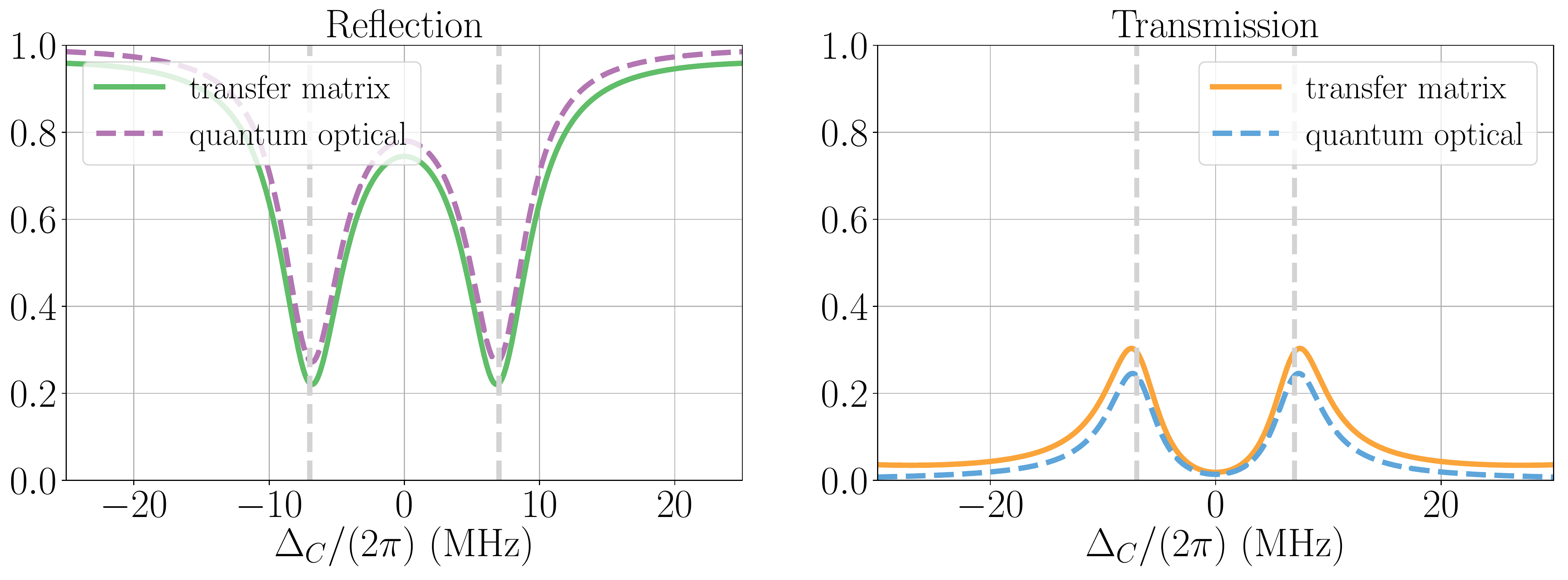}%
\vspace{-.3cm}
\caption{Comparison of the transmission and reflection spectra using the TM and single-mode QO models for a Fabry-P\'erot cavity with an atom inside, considering frequencies near the atomic resonance frequency $\omega_{\rm A}=\omega_{\rm C}$ ($\Delta_{\rm A}=\Delta_{\rm C}\equiv\Delta$). The setup is weakly driven from the left. Grey vertical dashed lines are drawn at $\pm g/(2\pi)$. Parameters: (TM) $R_1 =0.7$, $R_2=0.7$, $l=2$ m ($\omega_{\rm FSR}/(2\pi)=51.635$~MHz), $\eta =0.98$. (QM) $\kappa_1/(2\pi) = \kappa_2/(2\pi) = 1.233$~MHz, $\kappa_{\rm C,i}/(2\pi) = 0.166$~MHz. For both models $g/(2\pi)=7$~MHz and $\gamma /(2\pi)=2.65$~MHz.
\label{fig:cQED_QM_close}}
\vspace{3mm}
\includegraphics[width=0.8\textwidth]{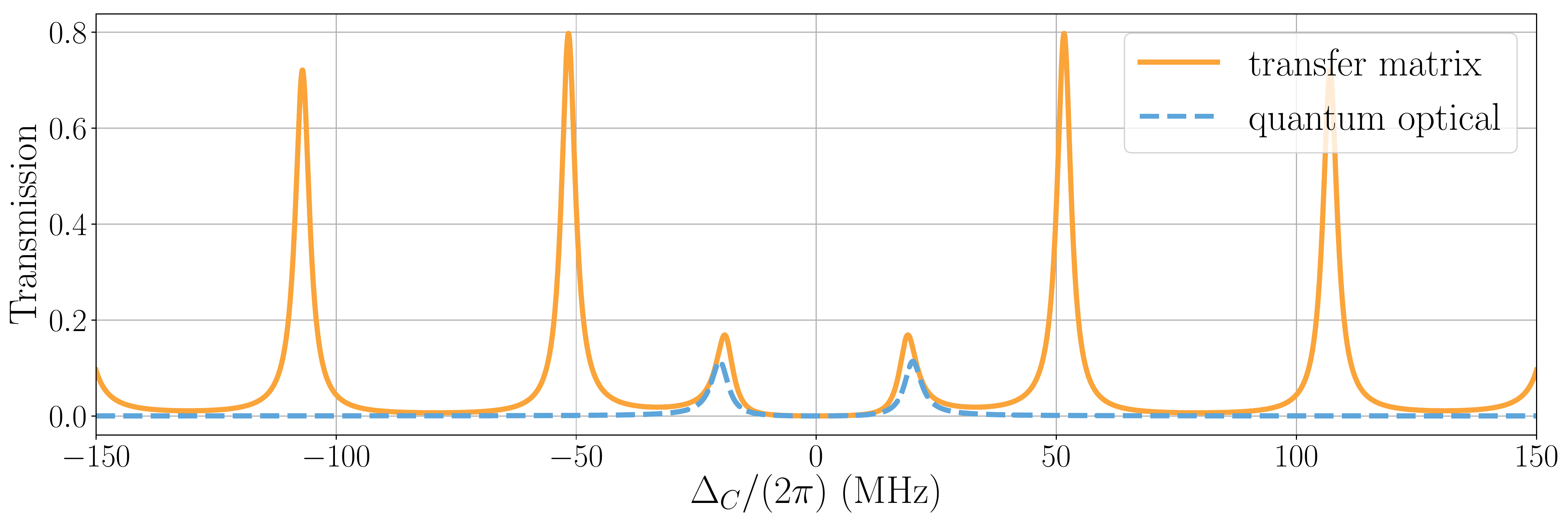}%
\vspace{-.3cm}
\caption{Comparison of the transmission and reflection spectra of a cavity-QED system using the TM and QO models for a wider range of frequencies ($\Delta_{\rm A}=\Delta_{\rm C}\equiv\Delta$). The setup is weakly driven from the left. The QO model overestimates the splitting of the central resonance as it does not account for higher-order resonances. Parameters: (TM) $R_1 =0.8$, $R_2=0.85$, $l=2$ m ($\omega_{\rm FSR}/(2\pi)=51.635$~MHz), $\eta=0.98$. (QM) $\kappa_1/(2\pi) = 0.822$~MHz, $\kappa_2/(2\pi) = 0.616$~MHz, $\kappa_{\rm C,i}/(2\pi) = 0.166$~MHz. For both models $g/(2\pi)=20$~MHz and $\gamma /(2\pi)=2.65$~MHz.
\label{fig:cQED_QM_far}}
\vspace{-.3cm}
\end{figure*}

Examples of spectra are shown in Figs.~\ref{fig:cQED_TM_g_close} and \ref{fig:cQED_TM_l_far} for the case in which $\omega_{\rm C}=\omega_{\rm A}$ (so $\Delta_{\rm C}=\Delta_{\rm A}\equiv \Delta$). Increasing the effective coupling between the atom and the cavity gives enhanced Rabi splitting of the central resonance (Fig.~\ref{fig:cQED_TM_g_close}), while changing the length of the cavity alters the free spectral range and the intensities of the peaks. We note that, even though only the peak closest to atomic resonance is subject to Rabi splitting, it is clear in Fig.~\ref{fig:cQED_TM_l_far} that the coupled atom also has a non-trivial effect on higher order resonances, especially when the FSR is decreased (i.e., cavity length increased).

If we assume 
$\{\alpha , \, 1-R_1, \,1-R_2\} \ll 1$, $|\Delta_{\rm C}| \ll \omega_{\rm FSR}$, and $\Gamma^\prime = 2\gamma$, then using the relation
\begin{align}
\Gamma_{\rm 1D} = \frac{l}{v_g}g^2 = \frac{\pi}{\omega_{\rm FSR}}g^2,
\end{align}
Eqs.~(\ref{T_cQED}) and (\ref{R_cQED}) give
\begin{align}
T^{(\rm cQED)}(\omega) &= \left| 
\frac{2\sqrt{\kappa_1 \kappa_2} (\gamma - i \Delta_{\rm A} )}{(\kappa_{\rm C} - i \Delta_{\rm C})(\gamma - i \Delta_{\rm A} ) + g^2} 
\right|^2
, \\
R^{(\rm cQED)}(\omega) &= \left| 
1 - \frac{2\kappa_1 (\gamma - i \Delta_{\rm A} )}{(\kappa_{\rm C} - i \Delta_{\rm C})(\gamma -  i \Delta_{\rm A} ) + g^2} 
 \right|^2
,
\end{align}
which reproduce the single-mode, quantum-optical results in the linear regime (\cite{Parkins2013}, and next subsection).

\begin{figure*}[t!]
\includegraphics[width=0.7\textwidth]{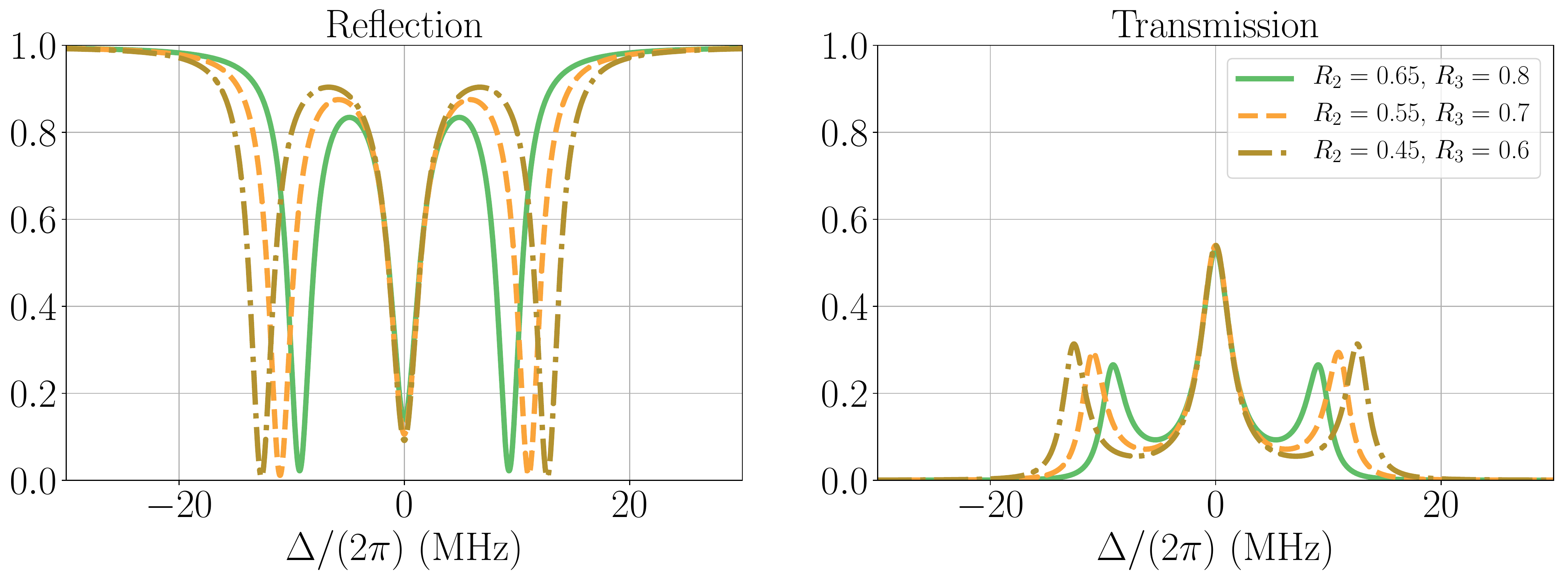}%
\vspace{-.3cm}
\caption{Transmission and reflection spectra calculated using the TM approach for two Fabry-P\'erot cavities coupled by an optical fiber. Weak driving is considered from the left. The various curves represent the spectra with different mid-mirror reflectances. The parameters can be found in Table~\ref{tab:expTM_params} of Appendix \ref{sec:params}, unless specified otherwise.
\label{fig:cav-cav_R+}}
\includegraphics[width=0.7\textwidth]{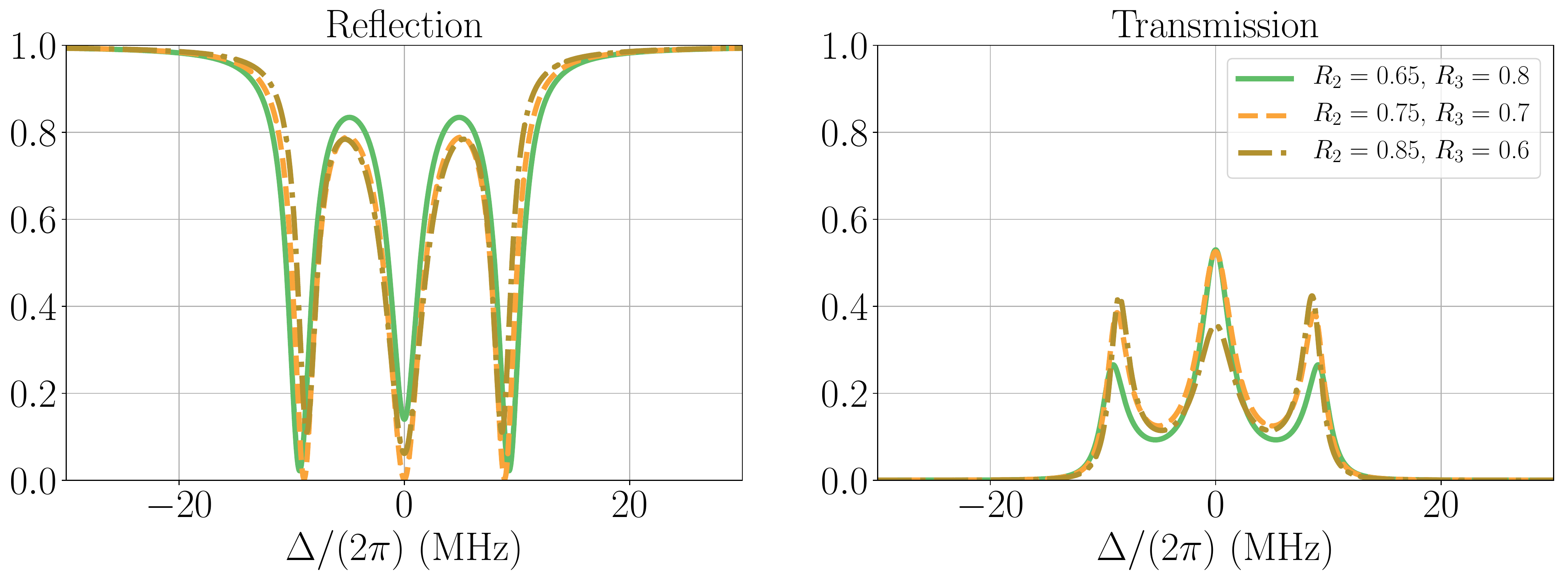}%
\vspace{-.3cm}
\caption{The same transmission and reflection spectra as in Fig.~\ref{fig:cav-cav_R+}, but with increasing (decreasing) reflectance of mirror 2 (mirror 3).
\label{fig:cav-cav_R-}}
\vspace{-.3cm}
\end{figure*}

\subsection{Quantum optical model}

The quantum optical description combines the free time-evolution of a single cavity mode and the field of an atom with the evolution due to their interaction. As only two levels of the atom's energy spectrum are considered, the spin-1/2 or Pauli matrices $\lk\hat{\sigma}_z, \hat{\sigma}^+,\hat{\sigma}^-\rk$ are used to represent the atomic excitations \cite{Walls2008}. Extending the master equation of the previous section with this contribution, we obtain
\begin{align}
\frac{d}{dt}\rhop &= -\frac{i}{\hbar}\lsz\Hop^{(\rm cQED)},\rhop\rsz +\Lb{(\rm cQED)}{}{\rhop} ,
\end{align}
with
\begin{align}
\Hop^{(\rm cQED)} &= -\Delta_{\rm C}\adop\aop-\Delta_{\rm A}\sigop^+\sigop^- \nn\\
&\hspace{.3cm}+g\lk\adop\sigop^-+\sigop^+\aop\rk+\Ed\lk\aop+\adop\rk ,\\
\Lb{(\rm cQED)}{}{\rhop} &= \kappa_{\rm C}\D{\aop}\rhop +\gamma\D{\sigop^-}\rhop .
\end{align}
From this, the equations of motion for the cavity and atomic amplitudes can be derived as
\begin{align}
\frac{d}{dt}\ev{\aop} 
&=(i\Delta_{\rm C}-\kappa_{\rm C})\ev{\aop}-ig\sigop^--i\Ed,
\\
\frac{d}{dt}\ev{\sigop^-} 
&=\lk i\Delta_{\rm A}-\gamma\rk\ev{\sigop^-}+ig\ev{\sigop_z\aop}\nn
\\
&\simeq \lk i\Delta_{\rm A}-\gamma\rk\ev{\sigop^-}-ig\ev{\aop},
\end{align}
where $\kappa_{\rm C}=\kappa_1+\kappa_2+\kappa_{\rm C,i}$ and $\gamma$ is the spontaneous emission rate of the atom into free space. The last equation is obtained by considering the weak driving limit, where we assume that the excited state of the atom is not significantly populated and, therefore, $\left\langle\hat{\sigma}_z\hat{a}\right\rangle\simeq \left\langle\hat{\sigma}_z\right\rangle\left\langle\hat{a}\right\rangle\simeq -\left\langle\hat{a}\right\rangle$.

The transmission and reflection spectra can be calculated similarly to the empty cavity case. The  steady-state cavity field amplitude is given by
\begin{align}
\ev{\aop}_{\rm ss} = -i\frac{\Ed\lk\gamma-i\Delta_{\rm A}\rk}{g^2+\lk\gamma-i\Delta_{\rm A}\rk\lk\kappa_{\rm C}-i\Delta_{\rm C}\rk} ,
\end{align}
which can be used in Eqs.~(\ref{eq:R}, \ref{eq:T}) to give $R$ and $T$.
The main characteristic of the obtained transmission and reflection spectra in the strong coupling regime ($g\gg\kappa_{\rm C},\gamma$) is the splitting of the central peak.

In contrast with the quantum-optical model, the transfer matrix approach enables the atom to couple to multiple modes, which results in a higher rate of transmission away from resonance (Fig.~\ref{fig:cQED_QM_close}). Note that only the central resonance shows Rabi splitting, which is mostly captured, but slightly overestimated, by the quantum model. This is due to the longer length of the cavity, which results in a decreased FSR and means that multiple resonances interact with the atom. As the QO model only captures the contribution of the central resonance, it overestimates the effect of the atom. The intensity variations and shifts of the higher-order resonances, missing from the QO model, can also only be described by the TM approach (Fig.~\ref{fig:cQED_QM_far}). 



\section{Connected empty cavities}
\begin{figure}[h!]
\includegraphics[width=0.48\textwidth, viewport=0 0 638 172]{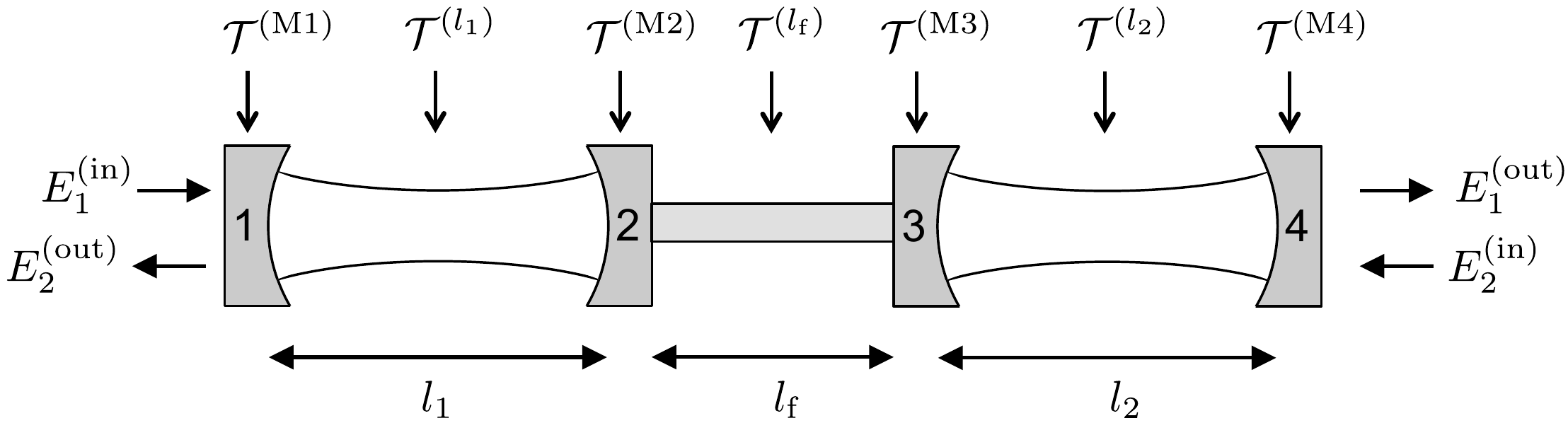}%
\caption{Connected empty cavities.
\label{fig:connected-empty-fabry-perot}}
\end{figure}

\begin{figure*}[tb!]
\includegraphics[width=0.7\textwidth]{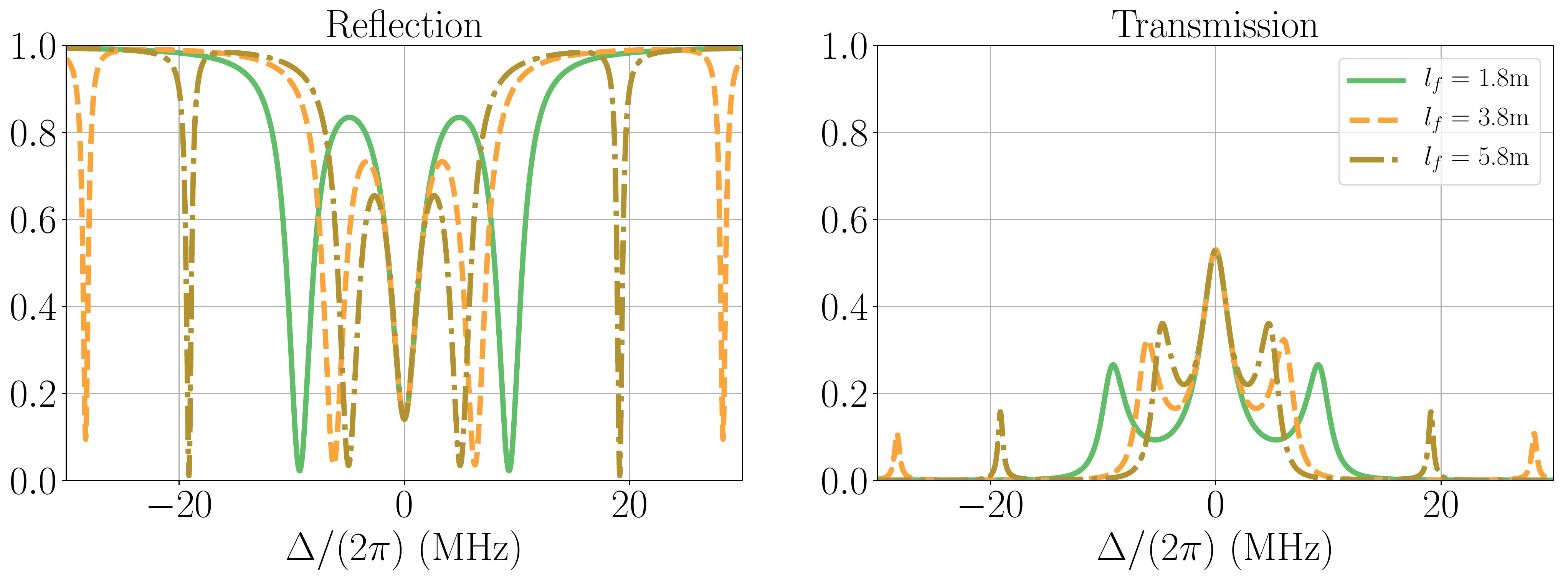}%
\vspace{-.3cm}
\caption{Coupled-cavity transmission and reflection spectra for increasing connecting-fiber length, i.e., increasing distance between the cavities. The parameters can be found in Table~\ref{tab:expTM_params} of Appendix \ref{sec:params}, unless specified otherwise.
\label{fig:cav-cav_lf}}
\vspace{3mm}
\includegraphics[width=0.75\textwidth]{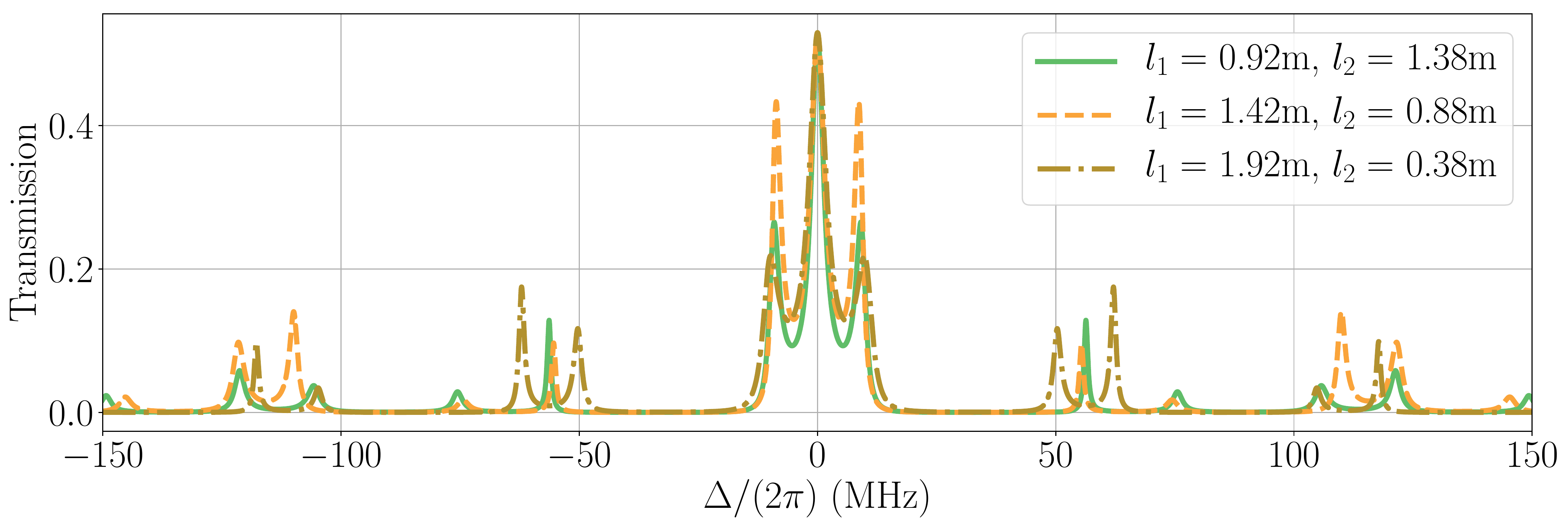}%
\vspace{-.3cm}
\caption{Transmission spectra of the coupled-cavity system for increasing (decreasing) length of cavity 1 (cavity 2). The parameters can be found in Table~\ref{tab:expTM_params} of Appendix \ref{sec:params}, unless specified otherwise.
\label{fig:cav-cav_lc}}
\vspace{-.3cm}
\end{figure*}

Having examined the main building blocks in detail, let us consider an elementary fiber network. From now on we focus mostly on setups that involve the same (or similar) parameters as recent, coupled fiber-cavity-QED experiments \cite{Kato2019,White2019} (see Tables~\ref{tab:expTM_params} and \ref{tab:expQO_params} in Appendix \ref{sec:params}). As a first step, we consider two empty Fabry-P\'erot cavities connected by an optical fiber, as shown in Fig.~\ref{fig:connected-empty-fabry-perot}.

\subsection{Transfer matrix approach}
The transfer matrix of the coupled-cavity system is calculated from
\begin{align}
{\cal T}^{({\rm 2FP})}
&=
{\cal T}^{(\rm M1)}
{\cal T}^{(l_1)}
{\cal T}^{(\rm M2)}
{\cal T}^{(l_{\rm f})}
{\cal T}^{(\rm M3)}
{\cal T}^{(l_2)}
{\cal T}^{(\rm M4)}
\nn\\
%
%
&=
\frac{{\cal C}_{\rm 2FP}(\omega)}{\sqrt{\eta_1 \eta_{\rm f} \eta_2 T_1 T_2 T_3 T_4}}
\begin{pmatrix}
-1 & \sqrt{R_1} \\
-\sqrt{R_1} & T_1+R_1
\end{pmatrix}\nn\\
&\hspace{.5cm}\times
\begin{pmatrix}
1 &0 \\
0 & \eta_1 e^{i \frac{2\pi \Delta_{\rm C1}}{\omega_{\rm FSR1}}}
\end{pmatrix}\begin{pmatrix}
-1 & \sqrt{R_2} \\
-\sqrt{R_2} & T_2+R_2
\end{pmatrix}
\nn\\
&\hspace{.5cm}\times
\begin{pmatrix}
1 &0 \\
0 & \eta_{\rm f} e^{i \frac{2\pi \Delta_{\rm Cf}}{\omega_{\rm FSRf}}}
\end{pmatrix}
\begin{pmatrix}
-1 & \sqrt{R_3} \\
-\sqrt{R_3} & T_3+R_3
\end{pmatrix}
\nn\\
&\hspace{.5cm}\times
\begin{pmatrix}
1 &0 \\
0 & \eta_2 e^{i \frac{2\pi \Delta_{\rm C2}}{\omega_{\rm FSR2}}}
\end{pmatrix}\begin{pmatrix}
-1 & \sqrt{R_4} \\
-\sqrt{R_4} & T_4+R_4
\end{pmatrix}
,
\end{align}
where
\begin{align*}
    {\cal C}_{\rm 2FP}(\omega) = e^{-i\pi \omega (1/ \omega_{{\rm FSR}1}+1/ \omega_{\rm FSRf}+1/ \omega_{{\rm FSR}2})},
\end{align*}
and 
\begin{align*}
\Delta_{{\rm C}1} &= \omega - \omega_{{\rm C}1}, & \Delta_{\rm Cf} &= \omega - \omega_{\rm Cf}, & \Delta_{{\rm C}2} &= \omega - \omega_{{\rm C}2}
\end{align*}
are the detunings of the probe for cavity 1, the connecting fiber (which also constitutes a multimode cavity), and cavity 2, respectively.


Transmission and reflection spectra of the coupled-cavities system are shown in Figs.~\ref{fig:cav-cav_R+}-\ref{fig:cav-cav_lc} for a variety of parameters, with all optical modes on resonance ($\omega_{{\rm C}1}=\omega_{{\rm C}2}=\omega_{\rm Cf}$). A defining feature of the spectra, for the chosen parameters, is the triplet structure centered around zero detuning. This can be interpreted in terms of three normal modes -- two symmetric ``bright'' modes and one resonant, anti-symmetric ``fiber-dark'' mode -- that result from coupling the resonant cavities via the connecting-fiber ``cavity'' \cite{Serafini2006}. 

\begin{figure*}[tb!]
\includegraphics[width=1\textwidth]{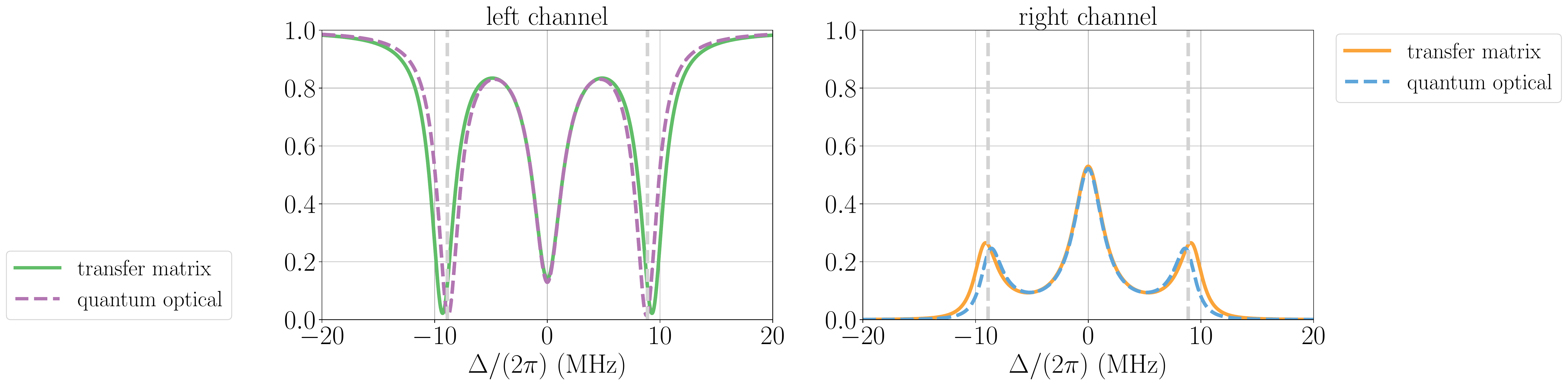}%
\vspace{-.3cm}
\caption{Comparison of spectra for the coupled-cavity system using the TM and QO models, when driving from the left ($\Ed_1\neq 0$, $\Ed_2=0$). The vertical, grey dashed lines indicate the frequencies $\pm\sqrt{v_1^2+v_2^2}/(2\pi)$. The parameters can be found in Tables~\ref{tab:expTM_params} and \ref{tab:expQO_params} of Appendix \ref{sec:params} ($g_1=g_2=0$).
\label{fig:cav-cav_QM_left}}
\vspace{3mm}
\includegraphics[width=1\textwidth]{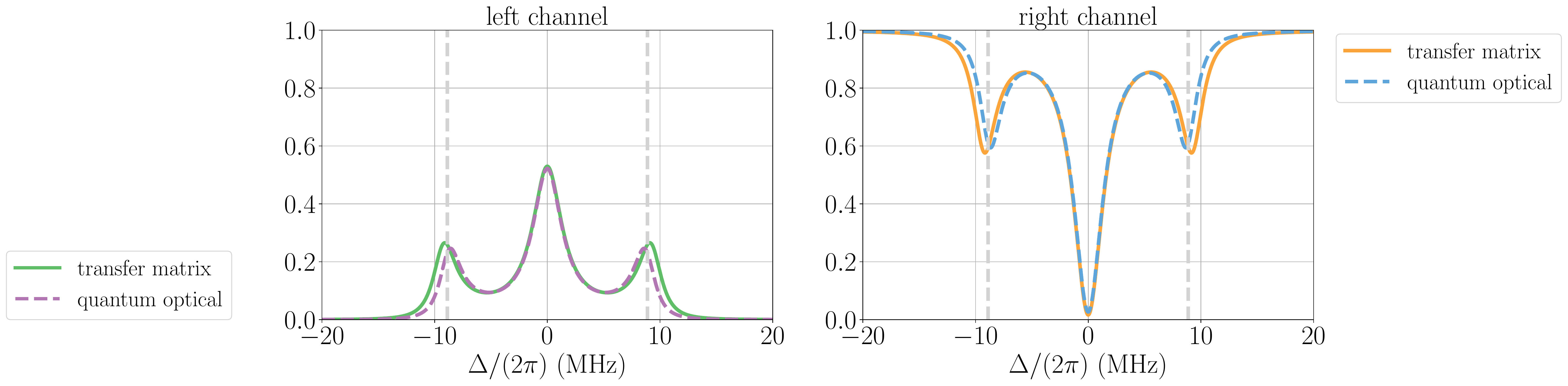}%
\vspace{-.3cm}
\caption{Same as in  Fig.~\ref{fig:cav-cav_QM_left}, but with driving from the right ($\Ed_1=0$, $\Ed_2\neq0$).
\label{fig:cav-cav_QM_right}}
\vspace{-.3cm}
\end{figure*}

Focusing only on the central triplet, a number of additional features illustrated by Figs.~\ref{fig:cav-cav_R+}-\ref{fig:cav-cav_lc} should also be noted. In particular, decreasing the reflectivities of mirrors 2 and 3 shifts the side peaks further away from resonance while having no effect on the central peak (see Fig.~\ref{fig:cav-cav_R+}). The same effect can be observed when the length of the connecting fiber is decreased, i.e., when the free spectral range of that section is increased (see Fig.~\ref{fig:cav-cav_lf}). However, when the reflectivity of one mirror is increased and the other is decreased in appropriate proportions, the side peaks stay at approximately the same position (see Fig.~\ref{fig:cav-cav_R-}). A similarly suppressed effect can be seen if the length of one cavity is increased while the length of the other cavity is decreased (see Fig.~\ref{fig:cav-cav_lc}). In summary, the positions of the side peaks in the central triplet depend essentially on the overall transmission rate through the connecting fiber, which is determined by a number of factors, including the lengths of the cavities.

\subsection{Quantum-optical model}

Assuming short enough lengths and, thus, sufficiently large FSR's for both the cavities and the connecting fiber, it is reasonable to model the fields of the cavities and the coupling fiber using single modes. The master equation for this system can then be written as
\begin{align}
\frac{d}{dt}\rhop&=-\frac{i}{\hbar}\lsz\Hop^{({\rm cav})}_1+\Hop^{({\rm cav})}_2+\Hop^{({\rm fibre})},\rhop\rsz \nn\\
&\hspace{.5cm}+\Lb{({\rm cav})}{1}{\rhop}+\Lb{({\rm cav})}{2}{\rhop}+\Lb{({\rm fibre})}{}{\rhop},
\end{align}
where $\Hop^{({\rm cav})}_i$ and $\Lb{({\rm cav})}{i}{\rhop}$ are defined similarly to the single empty-cavity case using cavity-mode operators $\aop_i$. The additional fiber terms incorporate the independent dynamics of the fiber as well as coherent excitation exchange at rates $v_1$ and $v_2$ between this mode (operator $b$) and the cavity modes,
\begin{align}
\Hop^{({\rm fiber})}&=-\Delta_{\rm Cf}\bdop\bop+v_1\lk\bdop\aop_1+\adop_1\bop\rk \nn\\
&\hspace{.3cm}+v_2\lk\bdop\aop_2+\adop_2\bop\rk,\\
\Lb{({\rm fiber})}{}{\rhop}&=\kappa_{{\rm Cf,i}}\D{\bop}\rhop,
\end{align}
where $\kappa_{{\rm Cf,i}}$ is the intrinsic loss rate of the coupling fiber mode.

This master equation gives the following equations of motion,
\begin{align}
\frac{d}{dt}\ev{\aop_1} &= (i\Delta_{{\rm C}1}-\kappa_{{\rm C}1})\ev{\aop_1}-iv_1\ev{\bop}-i\Ed_1,\\
\frac{d}{dt}\ev{\bop} &= (i\Delta_{\rm Cf}-\kappa_{\rm Cf,i})\ev{\bop}-iv_1\ev{\aop_1}-iv_2\ev{\aop_2},\\
\frac{d}{dt}\ev{\aop_2} &= (i\Delta_{\rm C2}-\kappa_{\rm C2})\ev{\aop_2}-iv_2\ev{\bop}-i\Ed_2 ,
\end{align}
where $\Ed_1$ and $\Ed_2$ are the coherent driving strengths of cavities 1 and 2, respectively (i.e., this allows for probe driving through both $E_1^{\rm (in)}$ and $E_2^{\rm (in)}$). The overall decay rate for cavity 1 is $\kappa_{\rm C1}=\kappa_1+\kappa_{\rm C1,i}$, where $\kappa_1$ characterizes mirror 1 and $\kappa_{\rm C1,i}$ represents the intrinsic fiber loss due to absorption and scattering in cavity 1 \cite{Kato2019,White2019}. Similarly, $\kappa_{\rm C2}=\kappa_4+\kappa_{\rm C2,i}$. Note that $\kappa_1$ and $\kappa_4$ are defined as before in (\ref{eq:kappa_def}), but have different cavity lengths, $l_1$ and $l_2$ respectively, in their expressions.

The obtained reflection and transmission spectra in Figs.~\ref{fig:cav-cav_QM_left} and \ref{fig:cav-cav_QM_right} show an interesting systematic shift of the side peaks compared to the transfer matrix approach. This can be understood by the model provided in \cite{Serafini2006}. In particular, the coherent coupling strengths are proportional to the decay rates into free space via mirrors 2 and 3, respectively. These effective decay rates are broadened more than can be captured by the single-mode model for the given parameters. This was also shown for a single cavity in Fig.~\ref{fig:FP_QM_close}. Thus, the emerging multimode nature of the field in the connecting fiber means more modes coupling to the cavities, increasing the effective cavity-fiber coupling strengths. As the position of the side peaks is determined by these coupling strengths, they shift further away from resonance \cite{Serafini2006}. 

By extending the model in \cite{Serafini2006} to the case of unequal $\kappa_2$ and $\kappa_3$, the splitting between the center peak and the side peaks is given by 
\footnote{In \cite{Serafini2006}, the free spectral range is expressed as $\omega_{\rm FSR} = 2\pi c/l$, which must be corrected as $\omega_{\rm FSR} = \pi c/l$ as in Eq.~(\ref{FSR}). Furthermore, $\bar{\nu}$ seems to be defined as frequency, whereas $\bar{\nu}$ and FSR are angular frequency.}

\begin{align}
\sqrt{v_1^2 + v_2^2}
&= \sqrt{2\left(\frac{\kappa_2}{2\pi}\right)\omega_{\rm FSRf}+ 2\left(\frac{\kappa_3}{2\pi}\right)\omega_{\rm FSRf}}
\nn\\ 
&= \sqrt{\frac{\kappa_2 + \kappa_3}{\pi}\omega_{\rm FSRf}}
\nn\\
&= \sqrt{\frac{1}{\pi}\left(\frac{1}{2} \frac{\omega_{\rm FSR1}}{2\pi}T_2 + \frac{1}{2} \frac{\omega_{\rm FSR2}}{2\pi}T_3  \right)\omega_{\rm FSRf}}
\nn\\
&= \frac{1}{2\pi}\sqrt{\left( \omega_{\rm FSR1} T_2 + \omega_{\rm FSR2} T_3 \right)\omega_{\rm FSRf}} .
\end{align}
This expression explains the difference in the behavior of the side peaks as a result of changing the central mirror reflectances in Figs.~\ref{fig:cav-cav_R+} and \ref{fig:cav-cav_R-}. When both of the mirror reflectances are decreased, the result is a net increase of the side peak's distance from resonance. On the other hand, when only one of the reflectances is decreased, and the other is increased, the net change in the side peak's position is negligible. Both figures show that the expected side-peak position of $(v_1^2 + v_2^2)^{1/2}$ approximately matches the numerically-obtained curves. Note that as mirror 2 has a lower reflectance than mirror 3, the light field effectively couples less strongly to the fiber field when incident from the right. This results in less intense signatures from the bright modes (which involve the fiber mode) in the reflection spectrum (Fig.~\ref{fig:cav-cav_QM_right}).


\newpage
\section{Connected cavity QED systems}

\begin{figure}[H]
\includegraphics[scale=0.38, viewport=0 0 638 233]{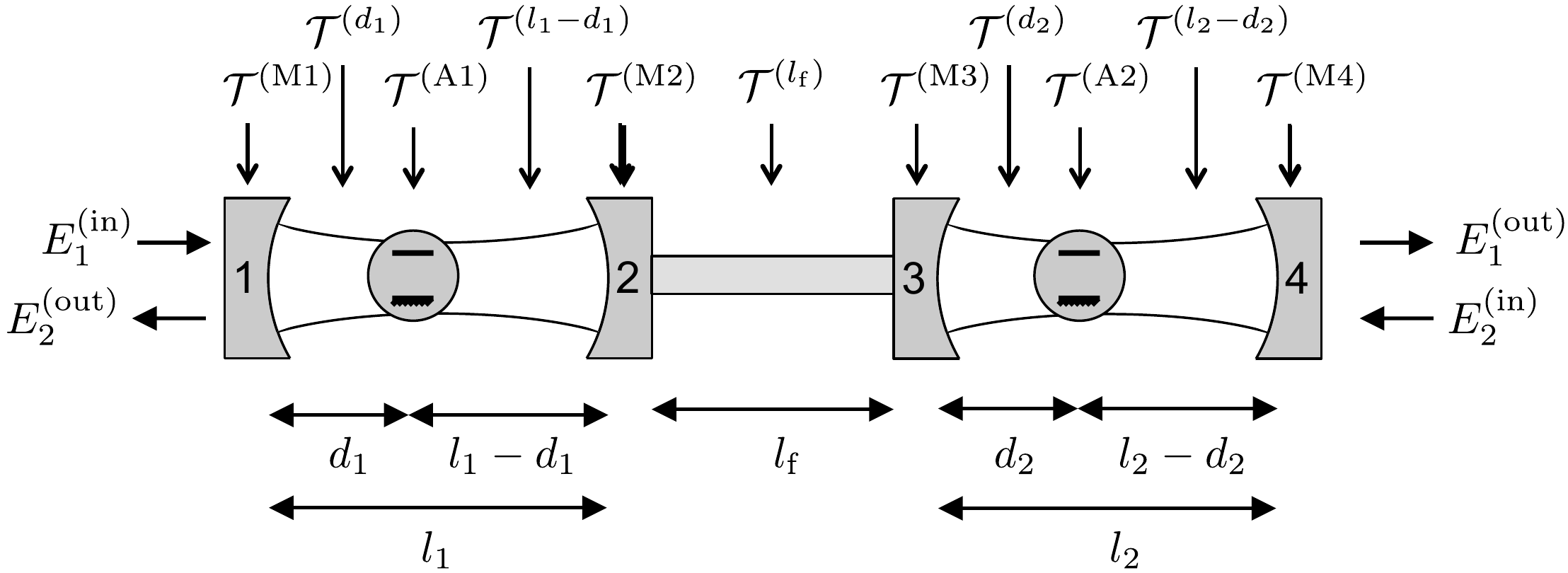}%
\caption{Two connected cavity QED systems.
\label{fig:connected-cQED}}
\end{figure}

In the context of quantum networks, an elementary, but important and topical example is a pair of cavity-QED systems connected by an optical fiber, as shown in Fig.~\ref{fig:connected-cQED}. Recent experiments have indeed explored transmission and reflection spectra of this system via various ports \cite{Kato2019,White2019}. In this section we demonstrate how the TM description can differ to various extents from the standard QO approach for the parameters considered in these experiments.

\subsection{Transfer matrix approach}

\begin{figure*}[bt]
\includegraphics[width=0.7\textwidth]{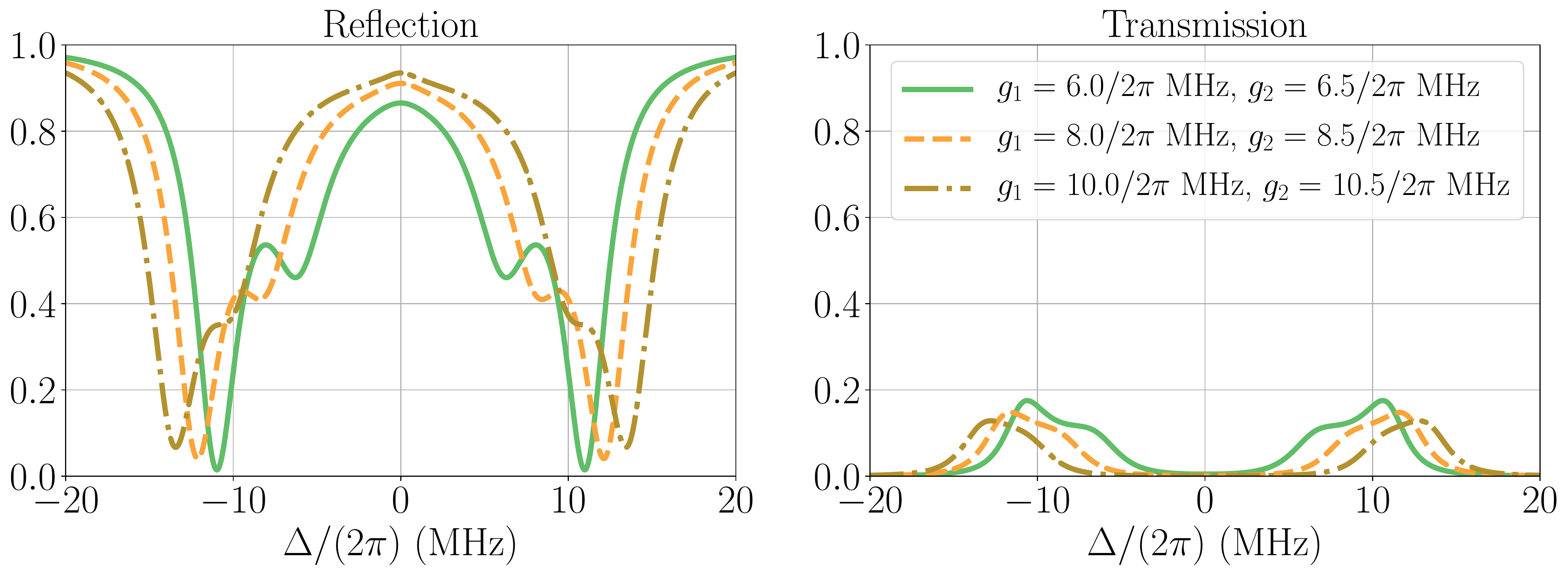}%
\vspace{-.3cm}
\caption{Transmission and reflection spectra of two cavity-QED systems coupled to each other by an optical fiber, computed using the TM model. Weak driving is considered from the left. The various curves are for varying atom-cavity coupling strengths. The parameters can be found in Table~\ref{tab:expTM_params} of Appendix \ref{sec:params}, except $\eta_1=0.99$, $\eta_2=0.99$, $\eta_{\rm f}=0.99$.
\label{fig:cQED-cQED_TM_g+}}
\includegraphics[width=0.7\textwidth]{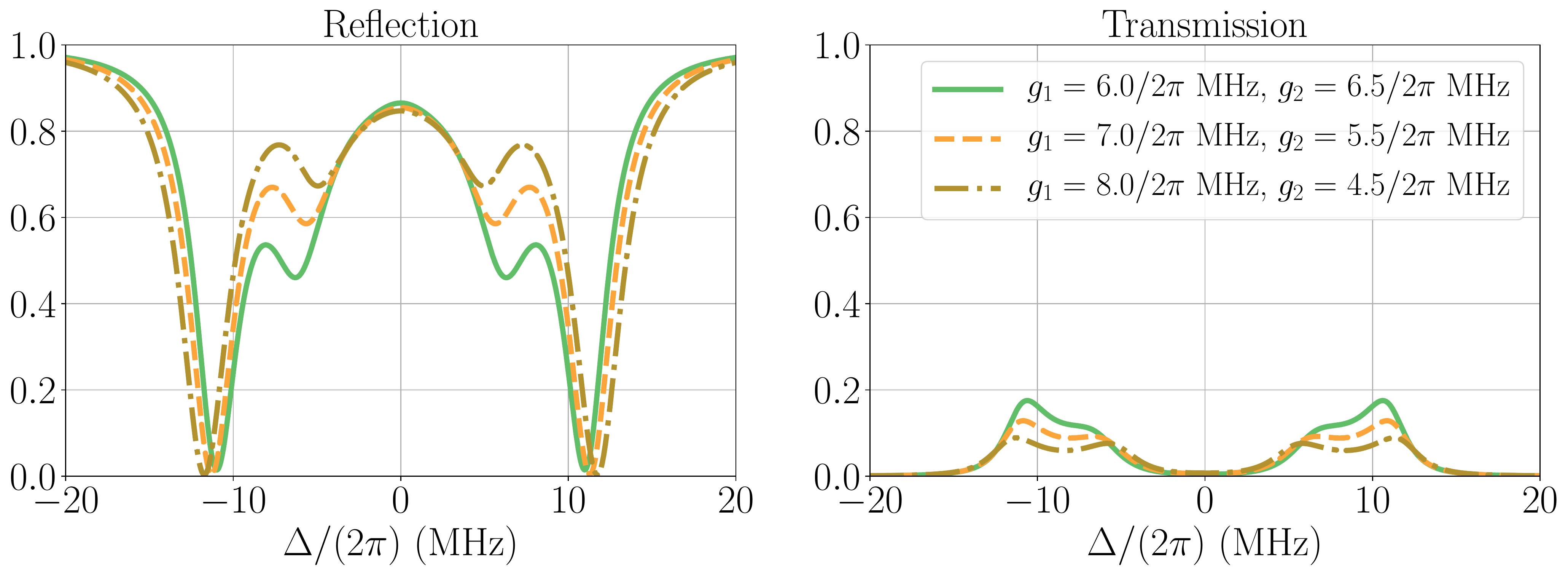}%
\vspace{-.3cm}
\caption{Transmission and reflection spectra for the same setup as in Fig.~\ref{fig:cQED-cQED_TM_g+}, but increasing the atom-cavity coupling strength only in cavity 1, while decreasing it in the other cavity.
\label{fig:cQED-cQED_TM_g-}}
\vspace{-.3cm}
\end{figure*}

The transfer matrix for the whole system can be calculated as follows:
\begin{widetext}
\begin{align}
{\cal T}^{({\rm 2cQED})}
&=
{\cal T}^{(\rm M1)}
{\cal T}^{(d_1)}
{\cal T}^{(\rm A1)}
{\cal T}^{(l_1-d_1)}
{\cal T}^{(\rm M2)}
{\cal T}^{(l_{\rm f})}{\cal T}^{(\rm M3)}
{\cal T}^{(d_2)}
{\cal T}^{(\rm A2)}
{\cal T}^{(l_2-d_2)}
{\cal T}^{(\rm M4)}
\nn\\
&=
\frac{i}{\sqrt{T_1}}
\begin{pmatrix}
-1 & \sqrt{R_1} \\
-\sqrt{R_1} & T_1+R_1
\end{pmatrix}
\begin{pmatrix}
e^{-i\tilde{k}_1 d_1} &0 \\
0 & e^{i\tilde{k}_1 d_1}
\end{pmatrix}
\begin{pmatrix}
1-i\xi_1 & -i\xi_1 \\
i\xi_1 & 1+i\xi_1
\end{pmatrix}
\begin{pmatrix}
e^{-i\tilde{k}_1 (l_1-d_1)} &0 \\
0 & e^{i\tilde{k}_1 (l_1-d_1)}
\end{pmatrix}\nn\\
&\quad\times
\frac{i}{\sqrt{T_2}}
\begin{pmatrix}
-1 & \sqrt{R_2} \\
-\sqrt{R_2} & T_2+R_2
\end{pmatrix}
\begin{pmatrix}
e^{-i\tilde{k}_{\rm f} l_{\rm f}} &0 \\
0 & e^{i\tilde{k}_{\rm f} l_{\rm f}}
\end{pmatrix}
\frac{i}{\sqrt{T_3}}
\begin{pmatrix}
-1 & \sqrt{R_3} \\
-\sqrt{R_3} & T_3+R_3
\end{pmatrix}
\begin{pmatrix}
e^{-i\tilde{k}_2 d_2} &0 \\
0 & e^{i\tilde{k}_2 d_2}
\end{pmatrix}\nn\\
&\quad\times
\begin{pmatrix}
1-i\xi_2 & -i\xi_2 \\
i\xi_2 & 1+i\xi_2
\end{pmatrix}
\begin{pmatrix}
e^{-i\tilde{k}_2 (l_2-d_2)} &0 \\
0 & e^{i\tilde{k}_2 (l_2-d_2)}
\end{pmatrix}
\frac{i}{\sqrt{T_4}}
\begin{pmatrix}
-1 & \sqrt{R_4} \\
-\sqrt{R_4} & T_4+R_4
\end{pmatrix}
.
\end{align}

\noindent Taking $d_1=d_2=0$ for simplicity gives
\begin{align}
{\cal T}^{({\rm 2cQED})}
&=
{\cal T}^{(\rm M1)}
{\cal T}^{(\rm A1)}
{\cal T}^{(l_1)}
{\cal T}^{(\rm M2)}
{\cal T}^{(l_{\rm f})}
{\cal T}^{(\rm M3)}
{\cal T}^{(\rm A2)}
{\cal T}^{(l_2)}
{\cal T}^{(\rm M4)}
\nn\\
&=
\frac{{\cal C}_{2FP}(\omega)}{\sqrt{\eta_1 \eta_{\rm f} \eta_2 T_1 T_2 T_3 T_4}}
\begin{pmatrix}
-1 & \sqrt{R_1} \\
-\sqrt{R_1} & T_1+R_1
\end{pmatrix}
\begin{pmatrix}
1-i\xi_1 & -i\xi_1 \\
i\xi_1 & 1+i\xi_1
\end{pmatrix}
\begin{pmatrix}
1 &0 \\
0 & \eta_1 e^{i \frac{2\pi \Delta_{\rm C1}}{\omega_{\rm FSR1}}}
\end{pmatrix}
\begin{pmatrix}
-1 & \sqrt{R_2} \\
-\sqrt{R_2} & T_2+R_2
\end{pmatrix}\nn\\
&\quad\times
\begin{pmatrix}
1 &0 \\
0 & \eta_{\rm f} e^{i \frac{2\pi \Delta_{\rm Cf}}{\omega_{\rm FSRf}}}
\end{pmatrix}
\begin{pmatrix}
-1 & \sqrt{R_3} \\
-\sqrt{R_3} & T_3+R_3
\end{pmatrix}
\begin{pmatrix}
1-i\xi_2 & -i\xi_2 \\
i\xi_2 & 1+i\xi_2
\end{pmatrix}
\begin{pmatrix}
1 &0 \\
0 & \eta_2 e^{i \frac{2\pi \Delta_{\rm C2}}{\omega_{\rm FSR2}}}
\end{pmatrix}
\begin{pmatrix}
-1 & \sqrt{R_4} \\
-\sqrt{R_4} & T_4+R_4
\end{pmatrix}
.
\end{align}
\end{widetext}

\begin{figure*}[tb]
\includegraphics[width=1\textwidth]{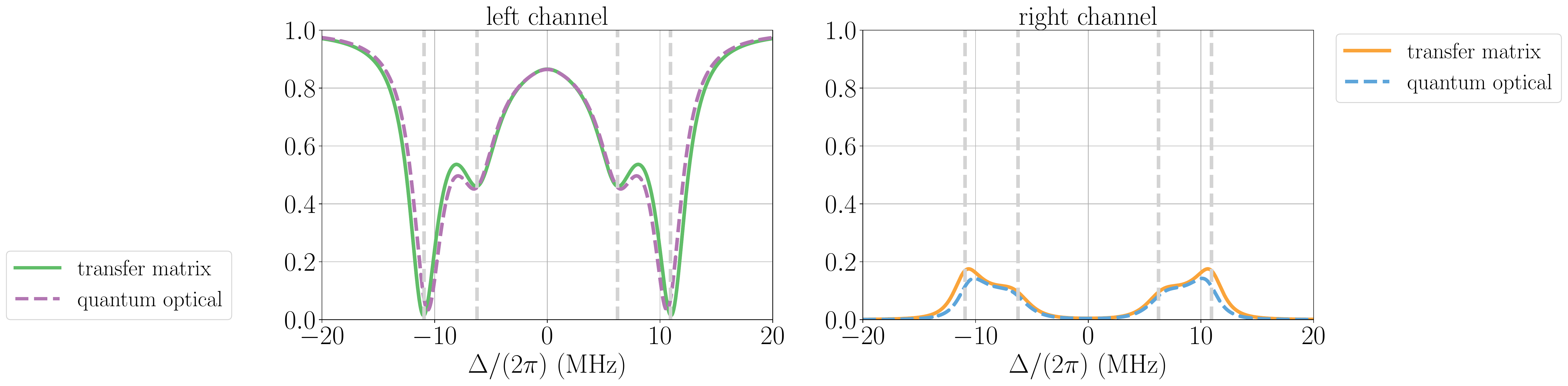}%
\vspace{-.3cm}
\caption{Comparing the spectra of the coupled cavity-QED system in the TM and QM approaches. The setup is weakly driven from the left ($\Ed_1\neq0,\Ed_2=0$). The grey dashed lines are frequencies obtained from a normal mode analysis of the QM model (see the Supplementary Material of \cite{White2019}). The parameters can be found in Tables~\ref{tab:expTM_params} and \ref{tab:expQO_params} of Appendix \ref{sec:params}, except for $g_1/(2\pi)=6$ MHz, $g_2/(2\pi)=6.5$ MHz.
\label{fig:cQED-cQED_QM_left}}

\includegraphics[width=1\textwidth]{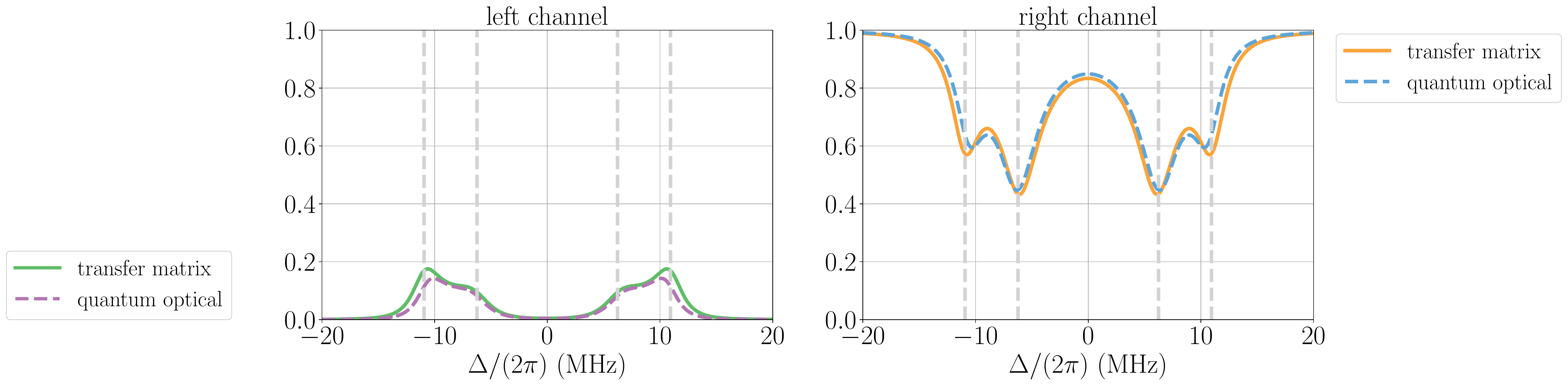}%
\vspace{-.3cm}
\caption{Same as Fig.~\ref{fig:cQED-cQED_QM_left} but considering driving from the right instead of the left ($\Ed_1=0,\Ed_2\neq0$).
\label{fig:cQED-cQED_QM_right}}
\vspace{-.3cm}
\end{figure*}

We see that the central peak observed in the previous section is split into two peaks, similar to the vacuum Rabi splitting for the one-cavity case. These peaks represent the anti-symmetric fiber-dark modes \cite{Kato2019,White2019}. Meanwhile, the side peaks corresponding to the symmetric bright modes are shifted further away from resonance. As pointed out in \cite{Serafini2006}, the central peak is an anti-symmetric superposition of the modes of cavities 1 and 2 and has no contribution from the connecting fiber. Therefore we can interpret this splitting as the vacuum Rabi splitting for the case of placing atoms in one effective cavity with length $l_0 = l_1 + l_2$. 
Since the atom-cavity coupling rate is inversely proportional to the square root of the cavity length, $g \propto 1/\sqrt{l}$, the coupling rate between one atom and the effective cavity should be given by 
\begin{align}
    g^\prime_{} = \sqrt{g_{1}^2 g_{2}^2/(g_{1}^2 + g_{2}^2)}.
\end{align}

In order to demonstrate the similarity with the Rabi splitting in a single cavity-QED system, we show in Fig.~\ref{fig:cQED-cQED_TM_g+} how the spectra change if $g_1$ and $g_2$ are increased simultaneously. In this case, the splitting of the central peak increases together with the distance of the other two side peaks. The overall transmission decreases. If one of the coupling strengths is increased, while the other is decreased in a suitable proportion, on the other hand, the splittings stay around the same value, while the overall transmission still decreases, as shown in Fig.~\ref{fig:cQED-cQED_TM_g-}. This further emphasizes the delocalized nature of the normal modes \cite{Kato2019}.

\subsection{Quantum-optical model}

The master equation for this system can be written as
\begin{align}
\frac{d}{dt}\rhop&=-i\lsz\Hop^{({\rm cQED})}_1+\Hop^{({\rm cQED})}_2+\Hop^{({\rm fiber})},\rhop\rsz \nn\\
&\hspace{.4cm}+\Lb{({\rm cQED})}{1}{\rhop}+\Lb{({\rm cQED})}{2}{\rhop}+\Lb{({\rm fibre})}{}{\rhop},
\end{align}
where $\Hop^{({\rm cQED})}_j$ and $\Lb{({\rm cQED})}{j}{\rhop}$ are defined as in the single cavity case using cavity operators $\aop_j$, and where an atom is included with spin operators $\sigop^-_j$. The additional fiber term incorporates the independent dynamics of the fiber as well as the excitation exchange between this mode and the cavities, similarly to the previous section. 

This master equation results in a similar set of dynamical equations as in the coupled-cavity case. In the weak driving limit (linearized regime), these are
\begin{align}
\frac{d}{dt}\ev{\sigop^-_1} &= \lk i\Delta_{\rm A}-\gamma\rk\ev{\sigop^-_1}-ig_1\ev{\aop_1} ,
\\
\frac{d}{dt}\ev{\aop_1} &= (i\Delta_{\rm C1}-\kappa_{\rm C1})\ev{\aop_1}-ig_1\ev{\sigop^-_1}\nn
\\
&\hspace{.5cm}-iv_1\ev{\bop}-i\Ed_1 ,
\\
\frac{d}{dt}\ev{\bop} &= (i\Delta_{\rm Cf}-\kappa_{\rm Cf,i})\ev{\bop}-iv_1\ev{\aop_1}\nn\\
&\hspace{.5cm}-iv_2\ev{\aop_2} ,
\\
\frac{d}{dt}\ev{\aop_2} &= (i\Delta_{\rm C2}-\kappa_{\rm C2})\ev{\aop_2}-ig_2\ev{\sigop^-_2}\nn
\\
&\hspace{.5cm}-iv_2\ev{\bop}-i\Ed_2 ,
\\
\frac{d}{dt}\ev{\sigop^-_2} &= \lk i\Delta_{\rm A}-\gamma\rk\ev{\sigop^-_2}-ig_2\ev{\aop_2} .
\end{align}

Similarly to the previous subsection, using the steady-state expectation values of the cavity fields, we can determine the reflection and transmission spectra. Comparing the two approaches in Figs.~\ref{fig:cQED-cQED_QM_left} and \ref{fig:cQED-cQED_QM_right}, we see a similar shift in the position of the outermost peaks to the coupled-cavity case. This is due to the fact that these peaks originate from the optical bright modes, showing up as side peaks in the coupled-cavity spectrum in the previous section (Figs.~\ref{fig:cav-cav_QM_left} and \ref{fig:cav-cav_QM_right}). They are positioned close to $\pm\sqrt{v_1^2 + v_2^2}$, but have an additional (small) dependence on the atom-cavity couplings $g_1$ and $g_2$ \cite{White2019}. 

This relationship is also justified by the suppression of the bright-state contributions in the reflection spectrum when the system is driven from the right (similar to what can be observed for the side peaks in Figs.~\ref{fig:cav-cav_QM_left} and ~\ref{fig:cav-cav_QM_right}). As there is an asymmetry in the reflectances of the central mirrors in the setup the asymmetry in the fiber-cavity couplings show stronger bright-mode contributions when reflected from the left than from the right (Figs.~\ref{fig:cQED-cQED_QM_left} and \ref{fig:cQED-cQED_QM_right}). Overall, the QO and TM descriptions agree reasonably well for the parameters of the recent experiments, thus supporting the use of the single-mode description in these works \cite{Kato2019,White2019}.

\begin{figure*}[tb!]
\includegraphics[scale=0.45]{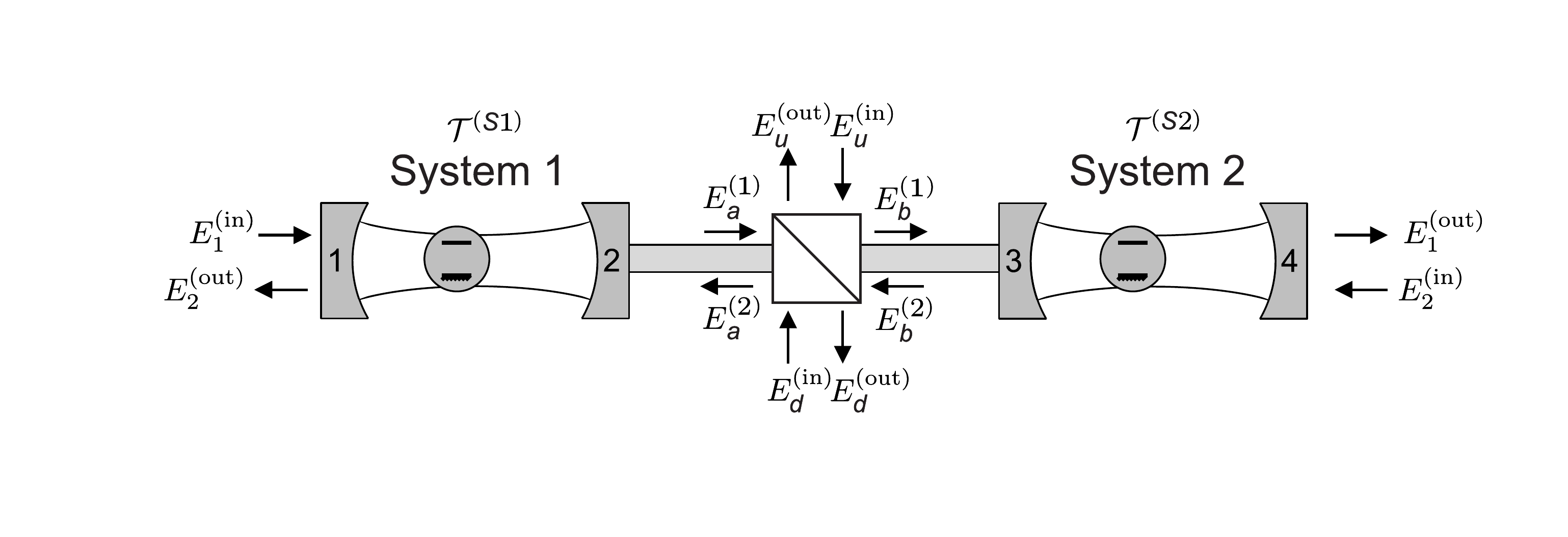}%
\vspace{-1.3cm}
\caption{Connected cavity QED systems with a 4-port beam splitter in the connecting fibre.
\label{fig:connected-cQED-fibre}}
\vspace{-.3cm}
\end{figure*}

\section{Including a beam splitter in the connecting fiber}

The spectra for the coupled cavity-QED system shown so far included only four peaks. There are, however, five normal modes, one of which does not have any cavity-field contribution, it is a cavity-dark mode. Nevertheless, it still involves the fibre field, thus by introducing an extra input-output channel which can be monitored, its presence has been detected by a recent experiment \cite{White2019}.

 Modelling this experiment requires a modified approach, as the beam splitter is characterized by a total of four input and four output ports (Fig.~\ref{fig:connected-cQED-fibre}). In order to facilitate the calculations, we collect the transfer matrices corresponding to cavity-QED setups 1 and 2 in the following way:

\begin{align}
\begin{pmatrix}
\E{in}_1\\
\E{out}_2
\end{pmatrix}
&=\transt{S1}
\begin{pmatrix}
\E{1}_a \\
\E{2}_a
\end{pmatrix} ,
\\
\begin{pmatrix}
\E{1}_b \\
\E{2}_b
\end{pmatrix}
&=\transt{S2}
\begin{pmatrix}
\E{out}_1 \\
\E{in}_2
\end{pmatrix} ,
\label{eq:Eab_with_TS12}
\end{align}
where
\begin{widetext}
\begin{align*}
\transt{S1} &= \transt{M1}\trans{l_1/2}\transt{A1}\trans{l_1/2}\transt{M2} \trans{l_f/2} 
\\
&= \frac{i}{\sqrt{T_1}}
\begin{pmatrix}
-1 & \sqrt{R_1} \\
-\sqrt{R_1} & T_1+R_1
\end{pmatrix}
\begin{pmatrix}
e^{-i\tilde{k}_1 l_1/2} &0 \\
0 & e^{i\tilde{k}_1 l_1/2}
\end{pmatrix}
\begin{pmatrix}
1-i\xi_1 & -i\xi_1 \\
i\xi_1 & 1+i\xi_1
\end{pmatrix}
\begin{pmatrix}
e^{-i\tilde{k}_1 l_1/2} &0 \\
0 & e^{i\tilde{k}_1 l_1/2}
\end{pmatrix}
\\
&\hspace{.5cm}\times
\frac{i}{\sqrt{T_2}}
\begin{pmatrix}
-1 & \sqrt{R_2} \\
-\sqrt{R_2} & T_2+R_2
\end{pmatrix}
\begin{pmatrix}
e^{-i\tilde{k}_f l_f/2} &0 \\
0 & e^{i\tilde{k}_f l_f/2}
\end{pmatrix} ,
\end{align*}
\begin{align*}
\transt{S2} &= \trans{l_f/2}\transt{M3}\trans{l_2/2}\transt{A2}\trans{l_2/2} \transt{M4} \\
&=\begin{pmatrix}
e^{-i\tilde{k}_f l_f/2} &0 \\
0 & e^{i\tilde{k}_f l_f/2}
\end{pmatrix}
\frac{i}{\sqrt{T_3}}
\begin{pmatrix}
-1 & \sqrt{R_3} \\
-\sqrt{R_3} & T_3+R_3
\end{pmatrix}
\begin{pmatrix}
e^{-i\tilde{k}_2 l_2/2} &0 \\
0 & e^{i\tilde{k}_2 l_2/2}
\end{pmatrix}
\begin{pmatrix}
1-i\xi_2 & -i\xi_2 \\
i\xi_2 & 1+i\xi_2
\end{pmatrix}
\\
&\hspace{.5cm}\times
\begin{pmatrix}
e^{-i\tilde{k}_2 l_2/2} &0 \\
0 & e^{i\tilde{k}_2 l_2/2}
\end{pmatrix}
\frac{i}{\sqrt{T_4}}
\begin{pmatrix}
-1 & \sqrt{R_4} \\
-\sqrt{R_4} & T_4+R_4
\end{pmatrix} .
\end{align*}
\end{widetext}
The different directions of the introduced fields $a$ and $b$ are named according to the original definition of field $1$ travelling from left to right and field $2$ from right to left.

The fields impinging on the fiber beam splitter can be described using a higher dimensional scattering matrix based on equation (\ref{eq:S_M}):
\begin{align*}
&\begin{pmatrix}
\E{2}_a\\
\E{out}_d\\
\E{1}_b\\
\E{out}_u
\end{pmatrix} 
=S^\BS
\begin{pmatrix}
\E{1}_a\\
\E{in}_d\\
\E{2}_b\\
\E{in}_u
\end{pmatrix} ,
\end{align*}
where
\begin{align*}
S^\BS
=\begin{pmatrix}
0 & i\sqrt{R^\BS} & \sqrt{T^\BS} & 0\\
i\sqrt{R^\BS} & 0 & 0 & \sqrt{T^\BS}\\
\sqrt{T^\BS} & 0 & 0 & i\sqrt{R^\BS}\\
0 & \sqrt{T^\BS} & i\sqrt{R^\BS} & 0
\end{pmatrix} .
\end{align*}

This relationship can also be expressed with two $2\times2$ matrices, which are the same for a symmetric beam splitter in the following way:
\begin{align}
\begin{pmatrix}
\E{2}_a\\
\E{out}_u
\end{pmatrix}&=S^{({\rm M/BS})}
\begin{pmatrix}
\E{in}_d\\
\E{2}_b
\end{pmatrix} ,
\label{eq:Eab_BS_1}\\
\begin{pmatrix}
\E{out}_d\\
\E{1}_b
\end{pmatrix}&=S^{({\rm M/BS})}
\begin{pmatrix}
\E{1}_a\\
\E{in}_u
\end{pmatrix} ,
\label{eq:Eab_BS_2}
\end{align}
where the corresponding scattering matrix is similar to the one in (\ref{eq:S_M}):
\begin{align}
S^{({\rm M/BS})} = 
\begin{pmatrix}
i\sqrt{R^\BS} & \sqrt{T^\BS}\\
\sqrt{T^\BS} & i\sqrt{R^\BS}
\end{pmatrix} .
\label{eq:S_MBS}
\end{align}
In the following, we look at various driving conditions to characterize this setup.

\begin{figure*}[tb!]
\includegraphics[width=0.9\textwidth]{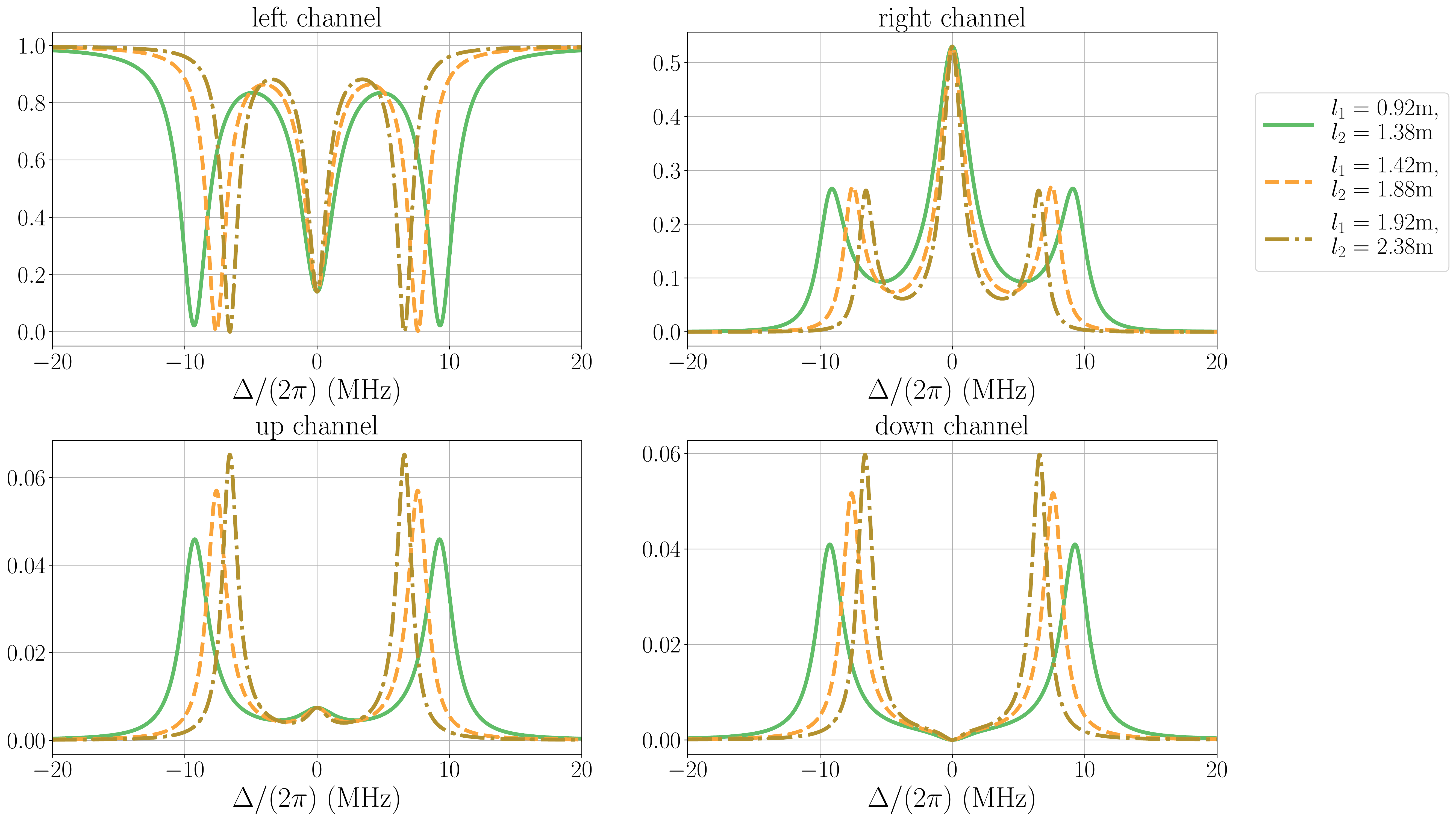}%
\vspace{-2mm}
\caption{Spectra of two coupled, empty Fabry-P\'erot cavities using the TM approach with weak driving from the left ($\E{in}_1$). The various curves correspond to the increasing lengths $l_1$ and $l_2$ of the cavities. The rest of the parameters are the same as in Table~\ref{tab:expTM_params} of Appendix \ref{sec:params} (empty-cavity case $g_1=g_2=0$).
\label{fig:CQCQ_ATM_empty}}
\vspace{3mm}
\includegraphics[width=0.9\textwidth]{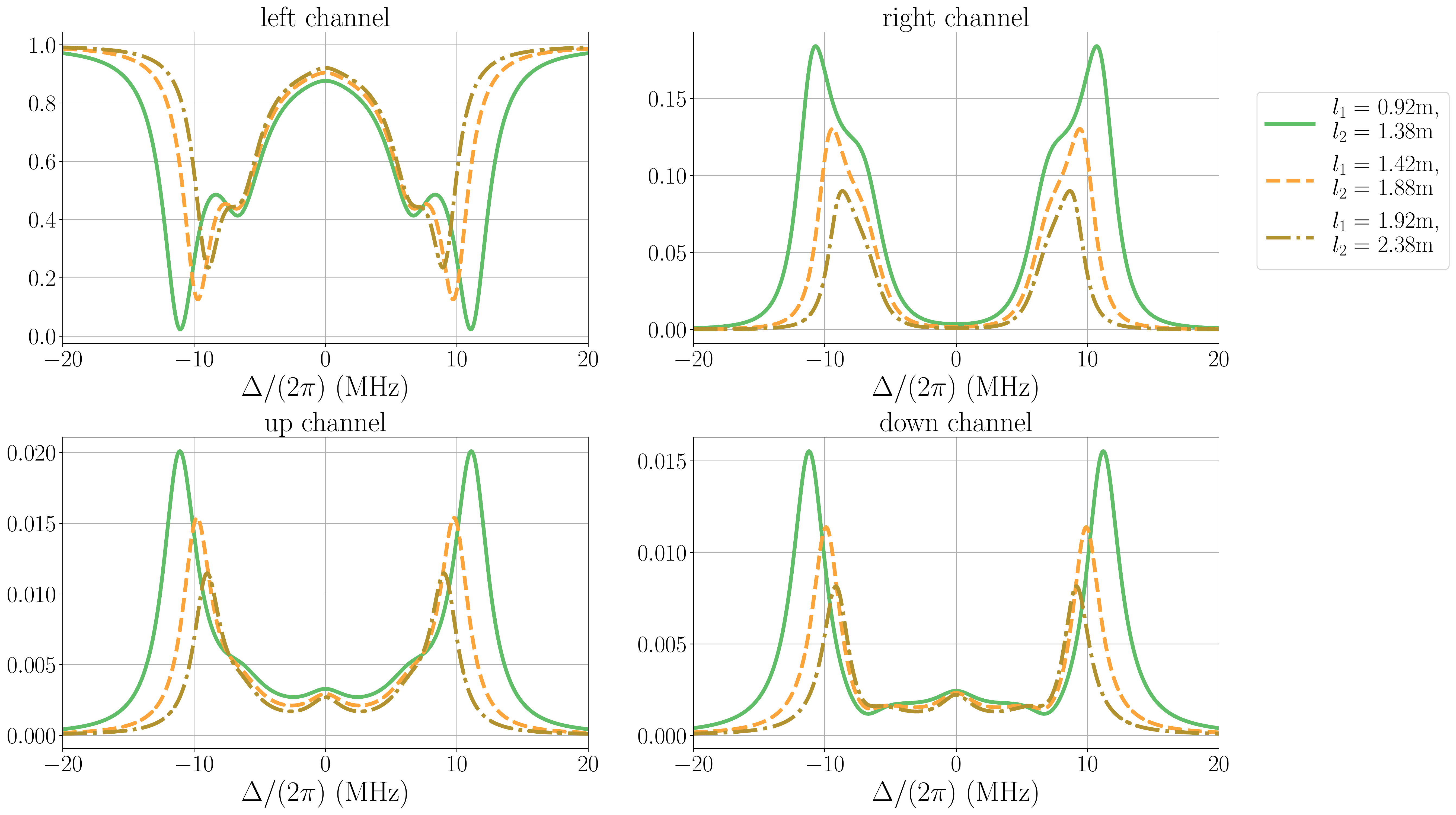}%
\vspace{-2mm}
\caption{Looking at the same spectra as in Fig.~\ref{fig:CQCQ_ATM_empty} but with atoms. The parameters for these spectra can be found in Table~\ref{tab:expTM_params} of Appendix \ref{sec:params}.
\label{fig:CQCQ_ATM_full}}
\vspace{-3mm}
\end{figure*}

\subsection{Driving from the left-hand side}

In the case when the system is only driven from the left,
\begin{align}
\E{in}_2 = \E{in}_u = \E{in}_d = 0 ,
\end{align}
the above equations simplify as
\begin{align}
\begin{pmatrix}
\E{in}_1\\
\E{out}_2
\end{pmatrix}&=
\transt{S1}\transt{\alpha}\transt{S2}
\begin{pmatrix}
\E{out}_1\\
0
\end{pmatrix}\nn
\\
&=\transt{2cQED,\alpha}
\begin{pmatrix}
\E{out}_1\\
0
\end{pmatrix} ,
\end{align}
which, using equation (\ref{eq:TtoS}) for the output fields on the left and right, translates into
\begin{align}
\label{eq:out_2}
\frac{\E{out}_2}{\E{in}_1} &= \frac{\transt{2cQED,\alpha}_{21}}{\transt{2cQED,\alpha}_{11}} ,\\
\frac{\E{out}_1}{\E{in}_1} &= \frac{1}{\transt{2cQED,\alpha}_{11}} .
\end{align}

The beam splitter output fields described by equations (\ref{eq:Eab_with_TS12}-\ref{eq:S_MBS}) also simplify in the following way:
\begin{align}
\label{eq:E2_a-1}
\E{2}_a &= \sqrt{T^\BS}\E{2}_b ,\\
\label{eq:Eout_d-1}
\E{out}_d &= i\sqrt{R^\BS}\E{1}_a ,\\
\label{eq:E1_b-1}
\E{1}_b &= \sqrt{T^\BS}\E{1}_a ,\\
\label{eq:Eout_u-1}
\E{out}_u &=i\sqrt{R^\BS}\E{2}_b= i\sqrt{\frac{R^\BS}{T^\BS}}\E{2}_a.
\end{align}

The field $\E{1}_a$ can be expressed in terms of $\E{in}_1$ using $\transt{S1}$ and equation (\ref{eq:out_2}),
\begin{align}
\begin{pmatrix}
\E{1}_a\\
\E{2}_a
\end{pmatrix} &= 
\left(\transt{S1}\right)^{-1}
\begin{pmatrix}
\E{in}_1\\
\E{out}_2
\end{pmatrix}
\nn\\
&=
\left(\transt{S1}\right)^{-1}
\begin{pmatrix}
1\\
\frac{\transt{2cQED,\alpha}_{21}}{\transt{2cQED,\alpha}_{11}}
\end{pmatrix}
\E{in}_1 .
\end{align}

Thus, we obtain the following for the output fields from the fiber:
\begin{align}
\frac{\E{out}_d}{\E{in}_1} &= \frac{i\sqrt{R^\BS}}{\det{\transt{S1}}}\left(\transt{S1}_{22}-\transt{S1}_{12}\frac{\transt{2cQED}_{21}}{\transt{2cQED}_{11}}\right) ,
\end{align}
and
\begin{align}
\frac{\E{out}_u}{\E{in}_1} &= \frac{i\sqrt{R^\BS}}{\sqrt{T^\BS}\det{\transt{S1}}}\ \nn\\
&\hspace{.5cm}\times\left(-\transt{S1}_{21}+\transt{S1}_{11}\frac{\transt{2cQED}_{21}}{\transt{2cQED}_{11}}\right) .
\end{align}

Looking at the output channels through the fiber beam splitter in Fig.~\ref{fig:CQCQ_ATM_empty} gives us some insight about the behaviour of the fiber mode. The central peak corresponds to the antisymmetric superposition of the fields in the two cavities, therefore little or no contribution can be detected in the fiber. The down channel ($\E{out}_d$) is a direct continuation of the laser drive on the left, whereas a reflection from mirror 3 is necessary for the light to reach the up channel ($\E{out}_u$). Therefore, the output field through the up channel shows a small peak on resonance confirming the build-up of a central fiber-mode. The modes reaching the down channel travel in the opposite direction, which gives an extra phase that transforms the small bump on resonance into a dip. The small bump and dip are due to the finite transmission rates of the outer mirrors resulting in overlapping normal modes and, thus, interference between them \cite{Shillito2019}. Increasing the lengths of both cavities shifts the side peaks closer to resonance.
Looking at the same spectra but with atoms in Fig.~\ref{fig:CQCQ_ATM_full}, only the outermost peaks corresponding to the bright modes show the same shift. The fiber channels show the signatures of all five normal modes.

\subsection{Driving through the beam splitter}

In the case when the system is driven only via the beam splitter, directly through the fiber, the following applies:
\begin{align}
\E{in}_1 = \E{in}_2 = \E{in}_d = 0 .
\end{align}

In terms of the fields scattered by the beam splitter this means that
\begin{align}
\label{eq:E2_a-f}
\E{2}_a &= \sqrt{T^\BS}\E{2}_b ,\\
\label{eq:Eout_d-f}
\E{out}_u &= i\sqrt{R^\BS}\E{2}_b ,\\
\label{eq:E1_b-f}
\E{1}_b &= \sqrt{T^\BS}\E{1}_a +i\sqrt{R^\BS}\E{in}_u ,\\
\label{eq:Eout_u-f}
\E{out}_d &=\sqrt{T^\BS}\E{in}_u+i\sqrt{R^\BS}\E{1}_a.
\end{align}

Considering the fields impinging on the beam splitter from the two cavities, we have the following expressions from  (\ref{eq:Eab_with_TS12}) using the corresponding scattering matrices,
\begin{align}
\begin{pmatrix}
\E{1}_a\\
\E{out}_2
\end{pmatrix} &= 
\scatt{S1}
\begin{pmatrix}
0\\
\E{2}_a
\end{pmatrix} ,
\\
\begin{pmatrix}
\E{out}_1\\
\E{2}_b
\end{pmatrix}&=
\scatt{S2}
\begin{pmatrix}
\E{1}_b\\
0
\end{pmatrix} ,
\end{align}
which gives extra conditions for fields $a$ and $b$,
\begin{align}
\label{eq:E1_a-f}
\E{1}_a &= \scatt{S1}_{12}\E{2}_a , \\
\label{eq:E2_b-f}
\E{2}_b &= \scatt{S2}_{21}\E{1}_b .
\end{align}

\begin{figure*}[tb!]
\includegraphics[width=0.9\textwidth]{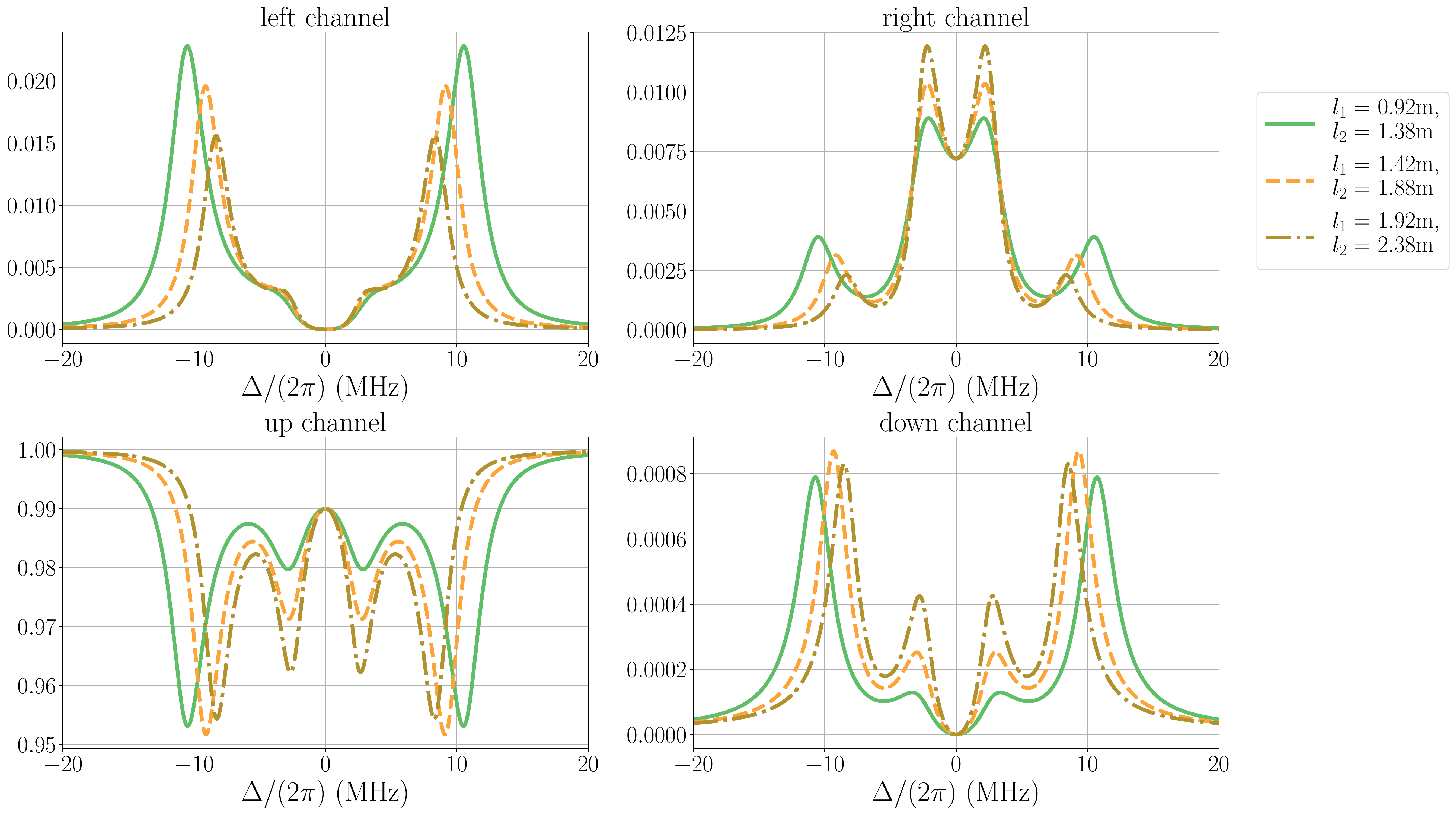}%
\vspace{-2mm}
\caption{Spectra of two coupled Fabry-P\'erot cavities using the TM approach with weak driving from the up channel ($E_u^{\rm (in)}$). Only cavity 1 contains an atom. The parameters are the same as in recent experiments (Table~\ref{tab:expTM_params}, Appendix \ref{sec:params}, $g_2=0$).
\label{fig:CQCQ_CTM_left}}
\vspace{3mm}
\includegraphics[width=0.9\textwidth]{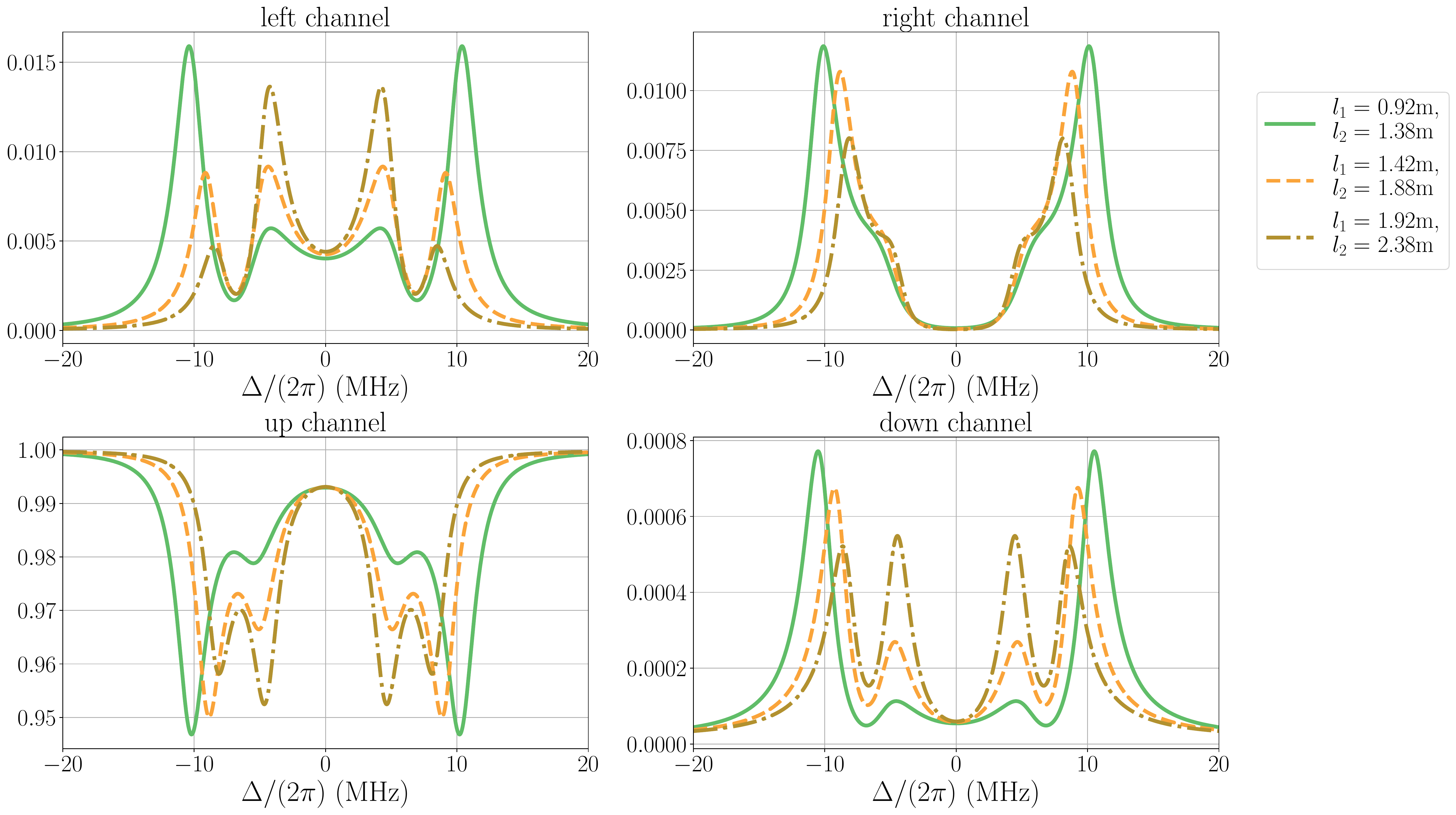}%
\vspace{-2mm}
\caption{Same as Fig.~\ref{fig:CQCQ_CTM_left}, but now only cavity 2 contains an atom. Parameters in Table~\ref{tab:expTM_params}, Appendix \ref{sec:params}, except for $g_1=0$.\label{fig:CQCQ_CTM_right}}
\vspace{-.3cm}
\end{figure*}

\begin{figure*}[tb!]
\includegraphics[width=0.9\textwidth]{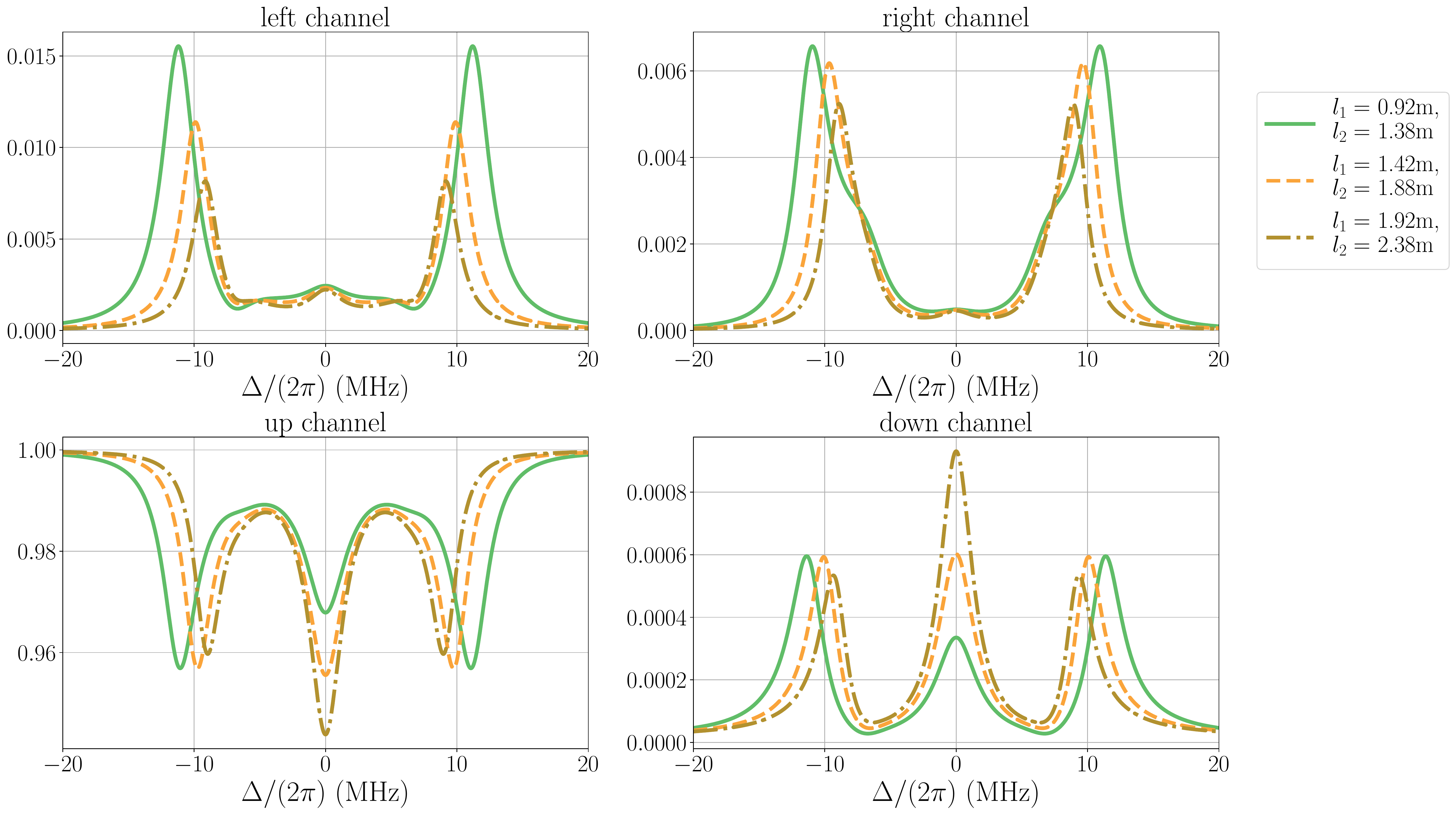}%
\vspace{-2mm}
\caption{Same as Fig.~\ref{fig:CQCQ_CTM_left} but both cavities contain atoms. Parameters in Table~\ref{tab:expTM_params}, Appendix \ref{sec:params}.
\label{fig:CQCQ_CTM_full}}
\vspace{-.3cm}
\end{figure*}

Finally, Eq.~(\ref{eq:E2_a-f}) provides the necessary link between fields $a$ and $b$. Putting Eqs.~(\ref{eq:E2_a-f}, \ref{eq:E1_b-f}) and (\ref{eq:E1_a-f}, \ref{eq:E2_b-f}) together thus gives
\begin{align}
\E{1}_b &= \sqrt{T^\BS}\scatt{S1}_{12}\sqrt{T^\BS}\scatt{S2}_{21}\E{1}_b + i\sqrt{R^\BS}\E{in}_u\nn
\\
        &= \frac{i\sqrt{R^\BS}}{1-T^\BS\scatt{S1}_{12}\scatt{S2}_{21}}\E{in}_u \nn
\\        
        &= i\sqrt{R^\BS}\xi\E{in}_u.
\end{align}

The output fields on the left and right have the expression of
\begin{align}
\E{out}_2 &= \scatt{S1}_{22}\E{2}_a \nn \\
&=\scatt{S1}_{22}\sqrt{T^\BS}\scatt{S2}_{21}i\sqrt{R^\BS}\xi\E{in}_u ,\\
\E{out}_1 &= \scatt{S2}_{11}\E{1}_b ,
\end{align}
which means
\begin{align}
\frac{\E{out}_2}{\E{in}_u} &=i\sqrt{R^\BS T^\BS} \frac{\det{\lsz\transt{S1}\rsz}}{\transt{S1}_{11}}\frac{\transt{S2}_{21}}{\transt{S2}_{11}}\xi , \\
\frac{\E{out}_1}{\E{in}_u} &= \frac{i\sqrt{R^\BS}\xi}{\transt{S2}_{11}} .
\end{align}
The output fields through the beam splitters are
\begin{align}
\frac{\E{out}_u}{\E{in}_u} &= -R^\BS\frac{\transt{S2}_{21}}{\transt{S2}_{11}}\xi , \\
\frac{\E{out}_d}{\E{in}_u} &= \sqrt{T^\BS}\lsz 1+R^\BS\frac{\transt{S1}_{12}}{\transt{S1}_{11}} \frac{\transt{S2}_{21}}{\transt{S2}_{11}}\xi\rsz .
\end{align}
In this case, as Figs.~\ref{fig:CQCQ_CTM_left}-\ref{fig:CQCQ_CTM_full} show, changing the lengths of the cavities not only shifts the outermost peaks but also affects the visibility of the central peaks. If atoms are only present in cavity 1 (Fig.~\ref{fig:CQCQ_CTM_left}), the inner peaks are much closer to resonance and much more pronounced in the fiber contributions than when both cavities are loaded. This is due to the broken spatial symmetry of the setup, which results in a different set of normal modes. In this setup, increasing the cavity lengths shifts the outer peaks closer to resonance, while the inner peaks are left in place but are enhanced.

In the case of cavity 2 being loaded instead of cavity 1 (Fig.~\ref{fig:CQCQ_CTM_right}), similar features can be observed as in Fig.~\ref{fig:CQCQ_CTM_left}. However, the atom is coupled slightly more strongly to its cavity field, which results in an increased splitting of the inner peaks in the spectrum. The spectra on the left ($\E{out}_1$) and right ($\E{out}_2$) are effectively exchanged. It is interesting to note that loading both cavities with atoms (Fig. \ref{fig:CQCQ_CTM_full}), the spectrum measured on the right ($\E{out}_2$) is more similar to the one corresponding to the case where only the right-hand-side cavity is loaded (Fig. \ref{fig:CQCQ_CTM_right}). 

Having both cavities loaded with atoms, the spectrum in Fig. \ref{fig:CQCQ_CTM_full} undergoes a qualitative change. Instead of four well-defined peaks, mostly two peaks corresponding to the bright modes are detected in the left ($\E{out}_1$) and right ($\E{out}_2$) channels. Signatures of the fiber-dark modes can be seen as small independent side peaks, as well. The fiber output channels, on the other hand, show an extra peak on resonance that corresponds to the cavity-dark mode.

As the lengths of the cavities are increased, the dark-mode contribution becomes more and more pronounced. Due to the asymmetry in the experimental setup, the fiber-dark modes are also driven via the fiber. They show up as small contributions with a phase that depends on the direction of propagation.

\subsection{Quantum-optical model}

\begin{figure*}[tb!]
\includegraphics[width=0.88\textwidth]{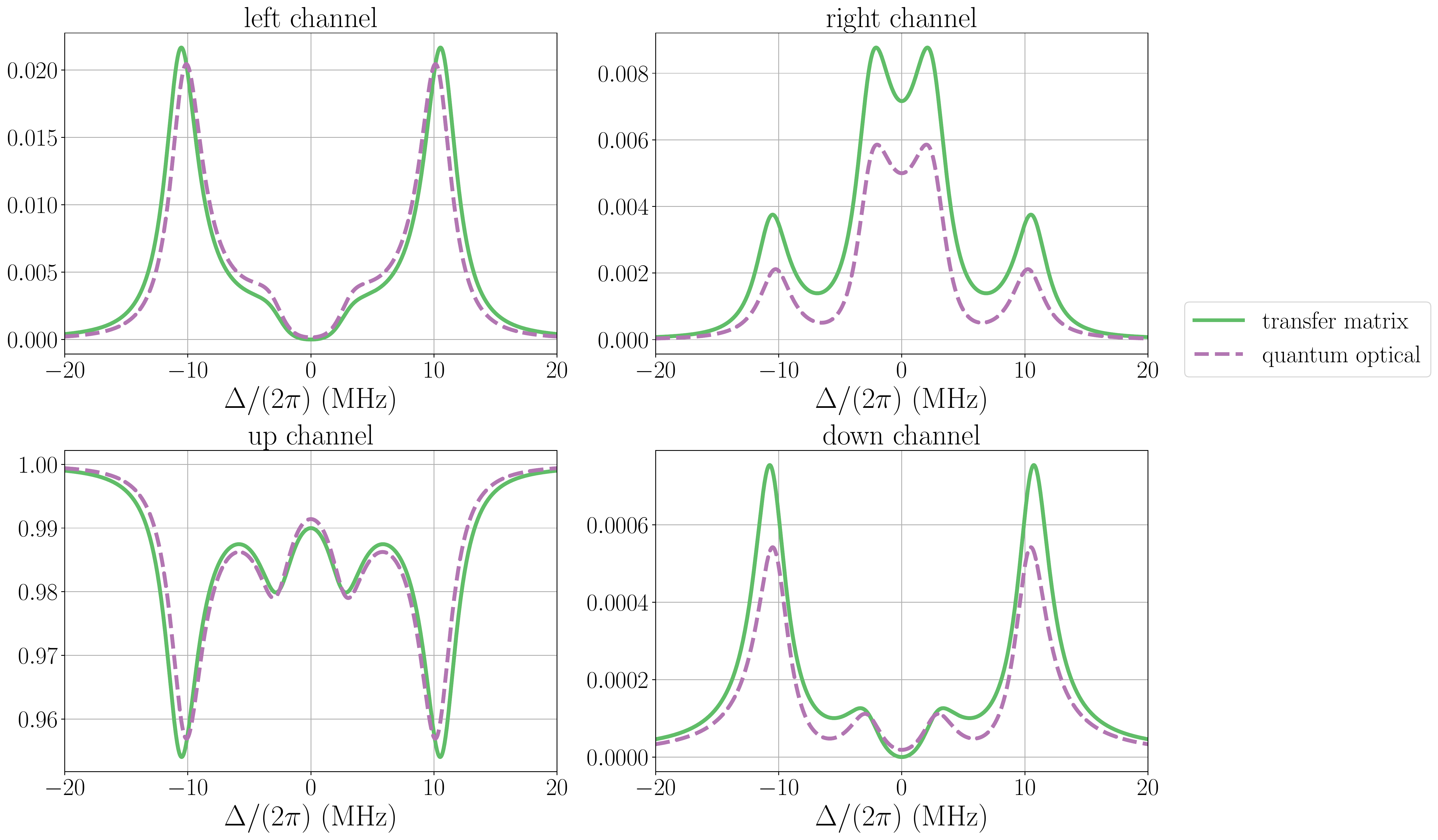}%
\vspace{-2mm}
\caption{Comparison of the spectra obtained from the TM and QO models when driving through the beam splitter input port $E_u^{\rm (in)}$ and when only cavity 1 contains an atom. The parameters for these spectra can be found in Tables~\ref{tab:expTM_params} and \ref{tab:expQO_params} of Appendix \ref{sec:params} (except for $g_2=0$).
\label{fig:CQCQ_CQM_left}}
\vspace{3mm}
\includegraphics[width=0.88\textwidth]{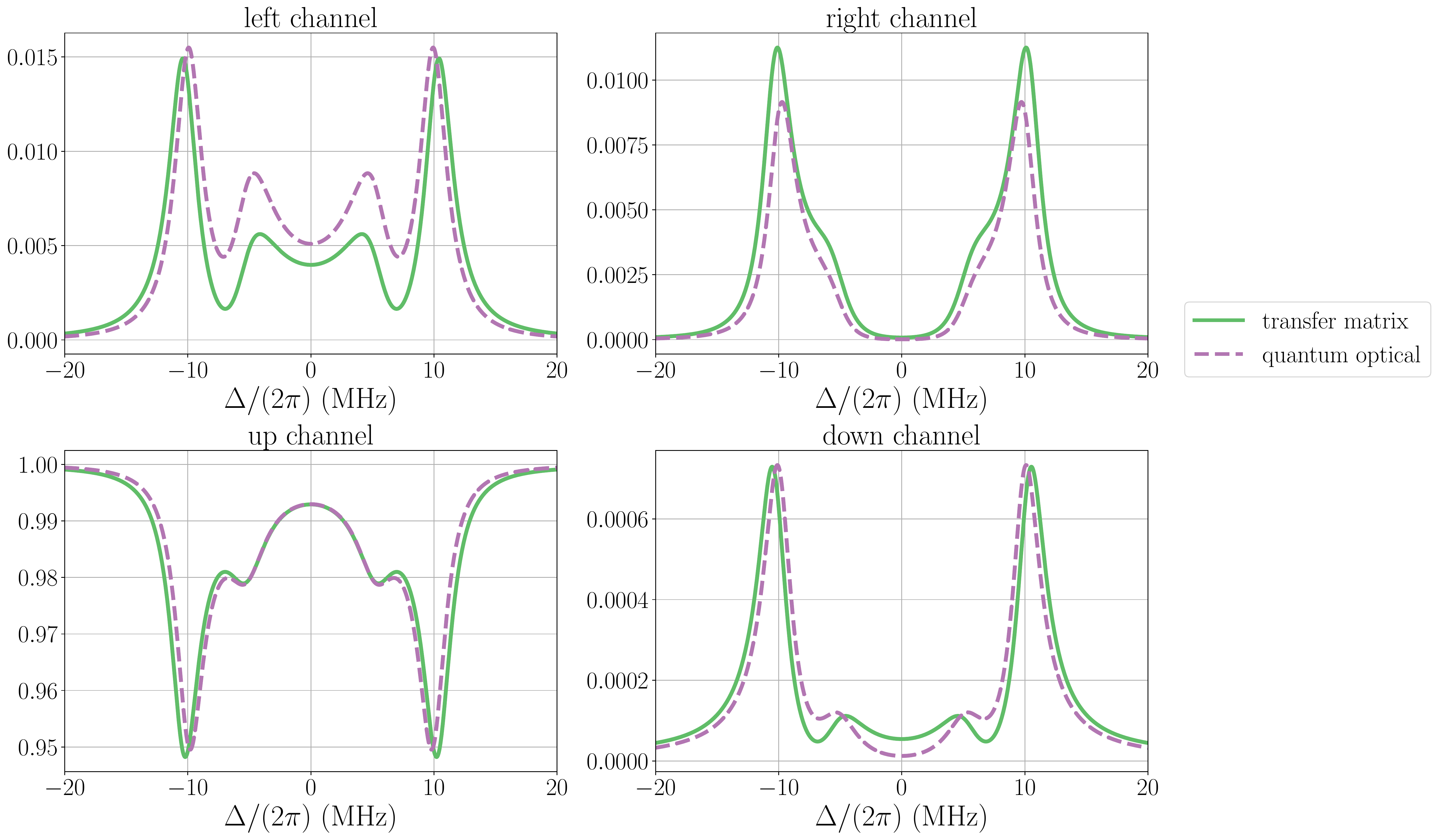}%
\vspace{-2mm}
\caption{Same as FIG.~\ref{fig:CQCQ_CQM_left}, but now only cavity 2 contains an atom. The parameters for these spectra can be found in Tables~\ref{tab:expTM_params} and \ref{tab:expQO_params} of Appendix \ref{sec:params} (except for $g_1=0$).
\label{fig:CQCQ_CQM_right}}
\vspace{-.3cm}
\end{figure*}
\begin{figure*}[tb!]
\includegraphics[width=0.895\textwidth]{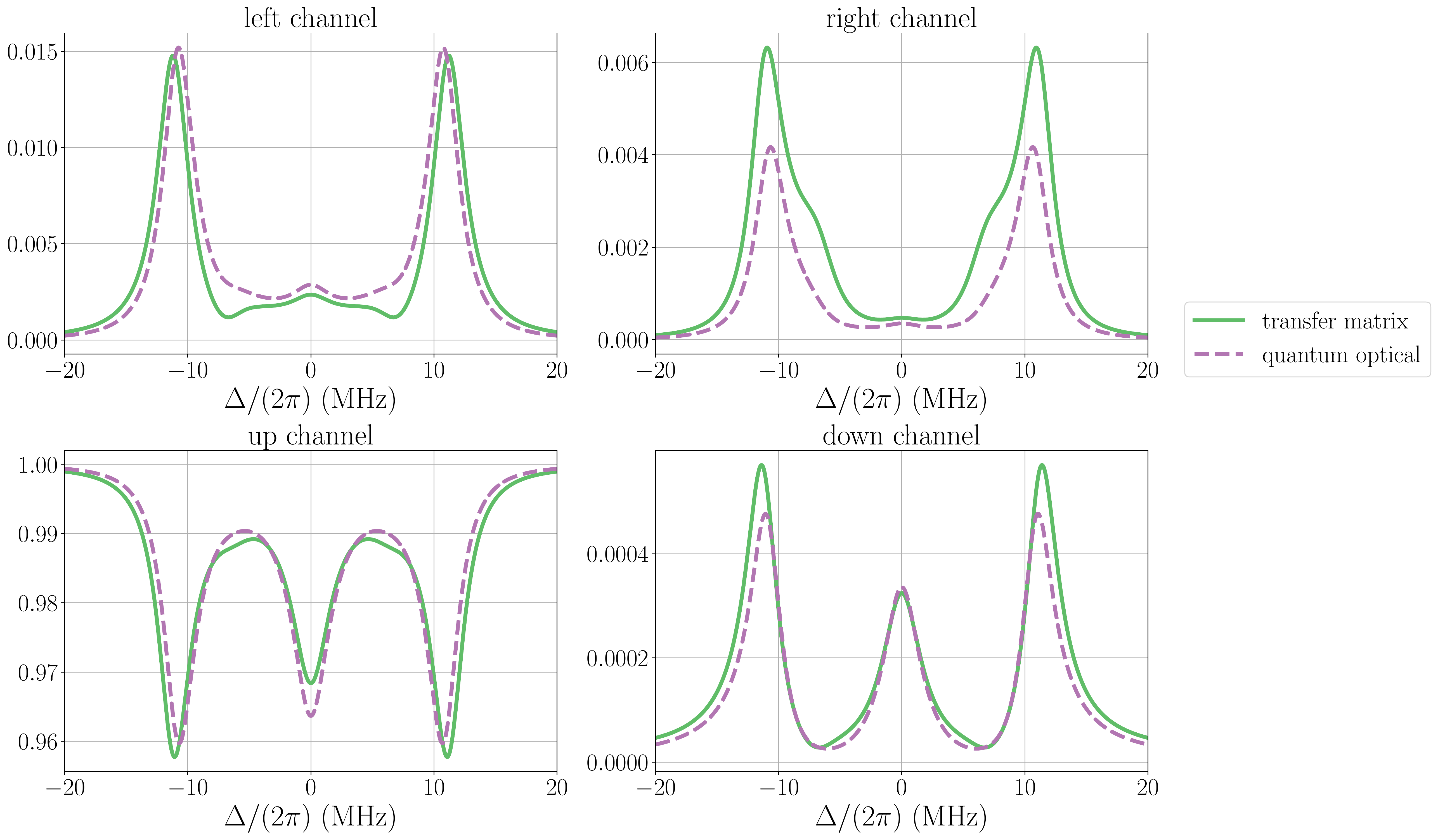}%
\vspace{-2mm}
\caption{Same as Fig.~\ref{fig:CQCQ_CQM_left} but both cavities are loaded with atoms. The parameters for these spectra can be found in Tables~\ref{tab:expTM_params} and \ref{tab:expQO_params} of Appendix \ref{sec:params}.
\label{fig:CQCQ_CQM_both}}
\vspace{.3cm}
\includegraphics[width=0.895\textwidth]{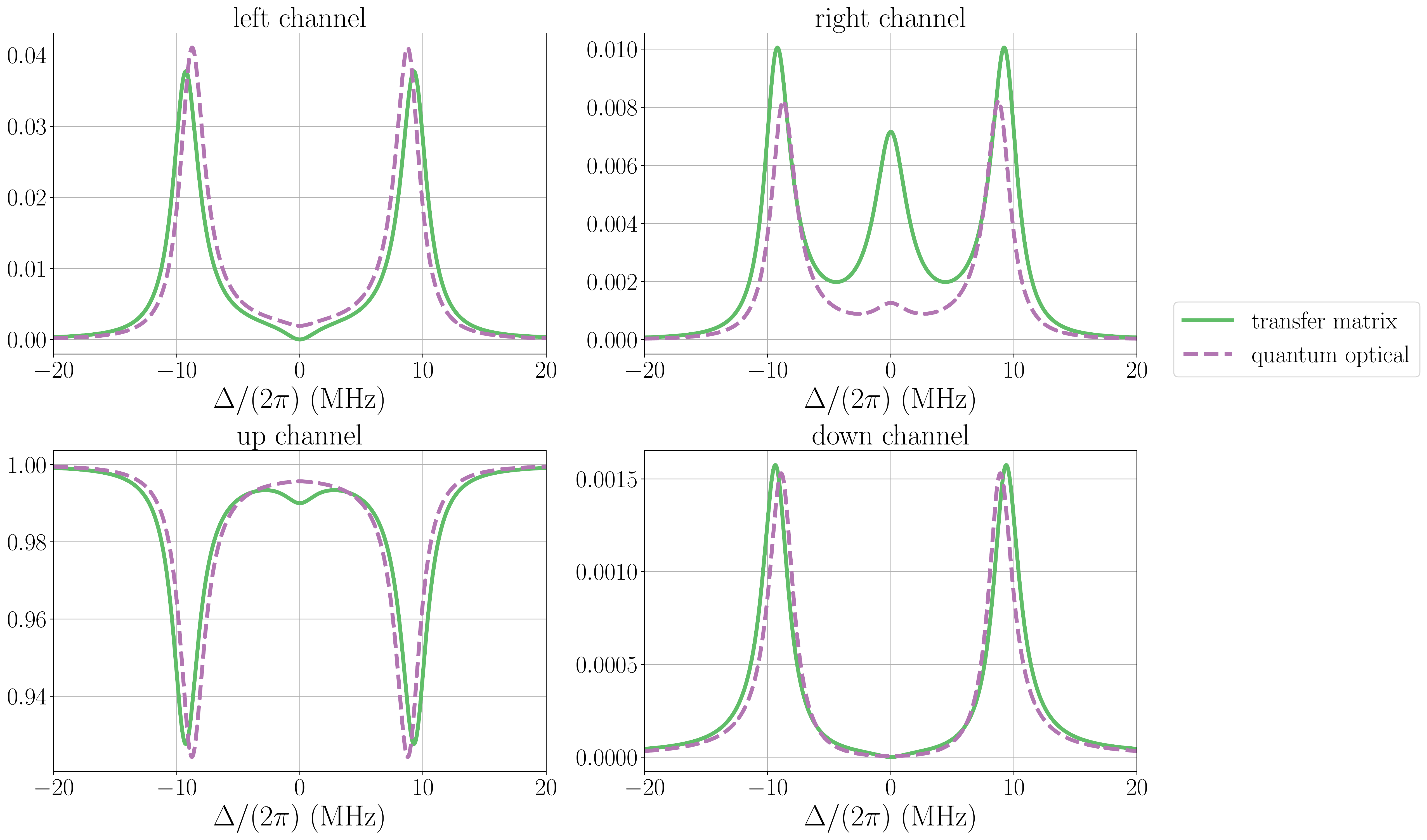}%
\vspace{-2mm}
\caption{Same as Fig.~\ref{fig:CQCQ_CQM_left} but considering empty cavities. The parameters for these spectra can be found in Tables~\ref{tab:expTM_params} and \ref{tab:expQO_params} of Appendix \ref{sec:params} (except for $g_1=g_2=0$).
\label{fig:CQCQ_CQM_empty}}
\vspace{-.3cm}
\end{figure*}

In this case we extend the master equation with a direct driving of the connecting fiber mode, which also introduces an extra loss channel with rate $\kappa_{BS}$. The master equation is thus
\begin{align}
&\frac{d}{dt}\rhop=-\frac{i}{\hbar}\lsz\Hop^{\rm (cQED)}_1+\Hop^{\rm (cQED)}_2+\Hop^{\rm (fiber-dr)},\rhop\rsz \nn
\\
&\hspace{.5cm}+\Lb{\rm (cQED)}{1}{\rhop}+\Lb{\rm (cQED)}{2}{\rhop}+\Lb{\rm (fiber-dr)}{}{\rhop} ,
\end{align}
where the terms describing the evolution of the fiber mode are adjusted to include the effective loss produced by the beam splitter and the direct laser driving, i.e.,
\begin{align}
\Lb{\rm(fiber-dr)}{}{\rhop} &= \lk\kappa_{\rm Cf,i}+\kappa_{\rm BS}\rk\D{\bop}\rhop = \beta^\prime\D{\bop}\rhop , \\
\Hop^{\rm (fiber-dr)}&=-\Delta_{\rm Cf}\bdop\bop+v_1\lk\bdop\aop_1+\adop_1\bop\rk\nn
\\
&\hspace{.3cm}+v_2\lk\bdop\aop_2+\adop_2\bop\rk + \Ed_{\rm u}\lk\bop+\bdop\rk ,
\end{align}
and $\Hop^{\rm (cQED)}_i$ and $\Lb{\rm (cQED)}{i}{\rhop}$ are defined as in the single cavity case with cavity operators $\aop_i$, and spin operators $\sigop^-_i$.

The master equation results in a similar set of dynamical equations as before. The only difference is in the equation for the fiber mode:
\begin{align}
\frac{d}{dt}\ev{\bop} &= (i\Delta_{\rm Cf}-\kappa_{\rm Cf})\ev{\bop}-iv_1\ev{\aop_1} \nn
\\
&\hspace{.3cm}-iv_2\ev{\aop_2}-i\Ed_{\rm u},
\end{align}
where $\kappa_{\rm Cf} = \kappa_{\rm Cf,i}+\kappa_{\rm BS}$.


\begin{figure*}[tb!]
\includegraphics[width=0.895\textwidth]{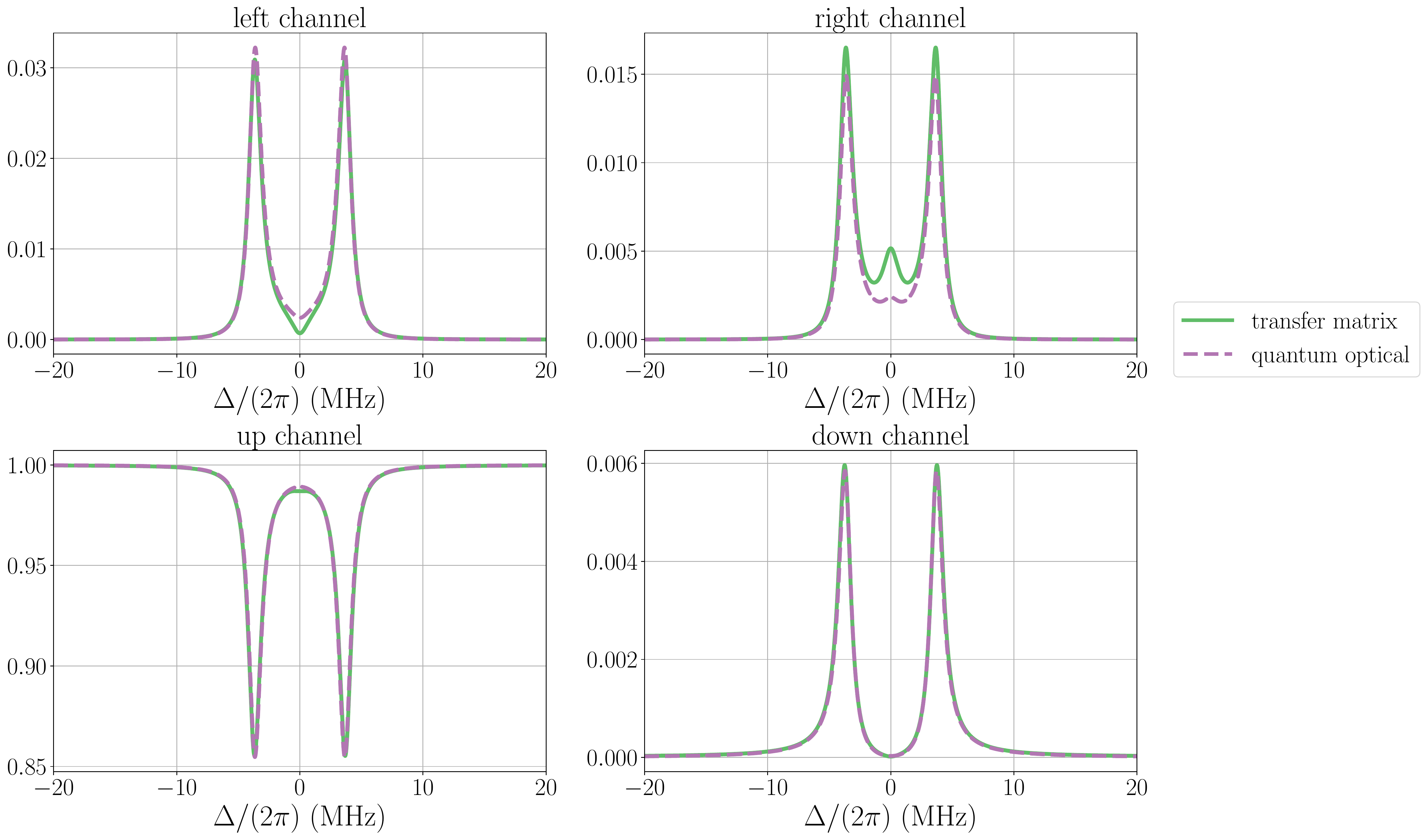}%
\vspace{-2mm}
\caption{Increasing the reflectivities of all the mirrors for the setup considered in Fig.~\ref{fig:CQCQ_CQM_empty}. The parameters for these spectra can be found in Tables~\ref{tab:expTM_params} and \ref{tab:expQO_params} of Appendix \ref{sec:params} except for $g_1=g_2=0$ and TM: $R_1 =0.95$, $R_2=0.95$, $R_3=0.95$, $R_4=0.95$, QO: $\kappa_1/(2\pi) = 0.446$ MHz, $\kappa_4/(2\pi) = 0.298$ MHz, $v_1/(2\pi) = 2.856$ MHz, $v_2/(2\pi) = 2.332$ MHz.
\label{fig:CQCQ_CQM_allhighR}}
\vspace{-.5cm}
\end{figure*}

Similarly to the previous subsection, using the steady state expectation values of the cavity fields we can determine the reflection and transmission spectra on the left and the right, as well as through the beam splitter.

The QM description assumes well-defined, standing-wave field modes, between which any spectral overlap is negligible. If the mirror reflectances are low and the cavity lengths are high, such well-defined standing-wave modes are not so ``cleanly'' formed. In particular, this is highlighted here by the case of driving through the beam splitter input channel $E_u^{\rm (in)}$. A traveling wave incident through this channel first propagates either to the right or downwards, and the appreciable transmittance of mirrors 3 and 4 ($R_3=0.80$ and $R_4=0.85$) means that a non-negligible fraction of the incident field can be ``lost'' from the system without contributing to the buildup of the connecting fiber ``mode''. This is why Fig.~\ref{fig:CQCQ_CQM_left} presents significant differences between the QM and TM descriptions in the right and down output channels ($E_1^{\rm (out)}$ and $E_d^{\rm (out)}$). 
The spectra for the left and up output channels ($E_2^{\rm (out)}$ and $E_u^{\rm (out)}$), however, show better agreement as a result of longer propagation, plus reflection, enabling interference between counter-propagating fields and better establishment of a standing-wave mode. Note that this is also related to the presence of an atom in cavity 1, and its effect on the overall transmission of this portion of the system.

If, instead, there is an atom only in cavity 2, as in Fig.~\ref{fig:CQCQ_CQM_right}, then the particularly high transmission of mirror 2 ($R_2=0.65$) plays a more significant role and substantial differences between QM and TM models also arise in the spectrum from the left output channel ($E_2^{\rm (out)}$).

\begin{figure*}[tb!]
\includegraphics[width=0.9\textwidth]{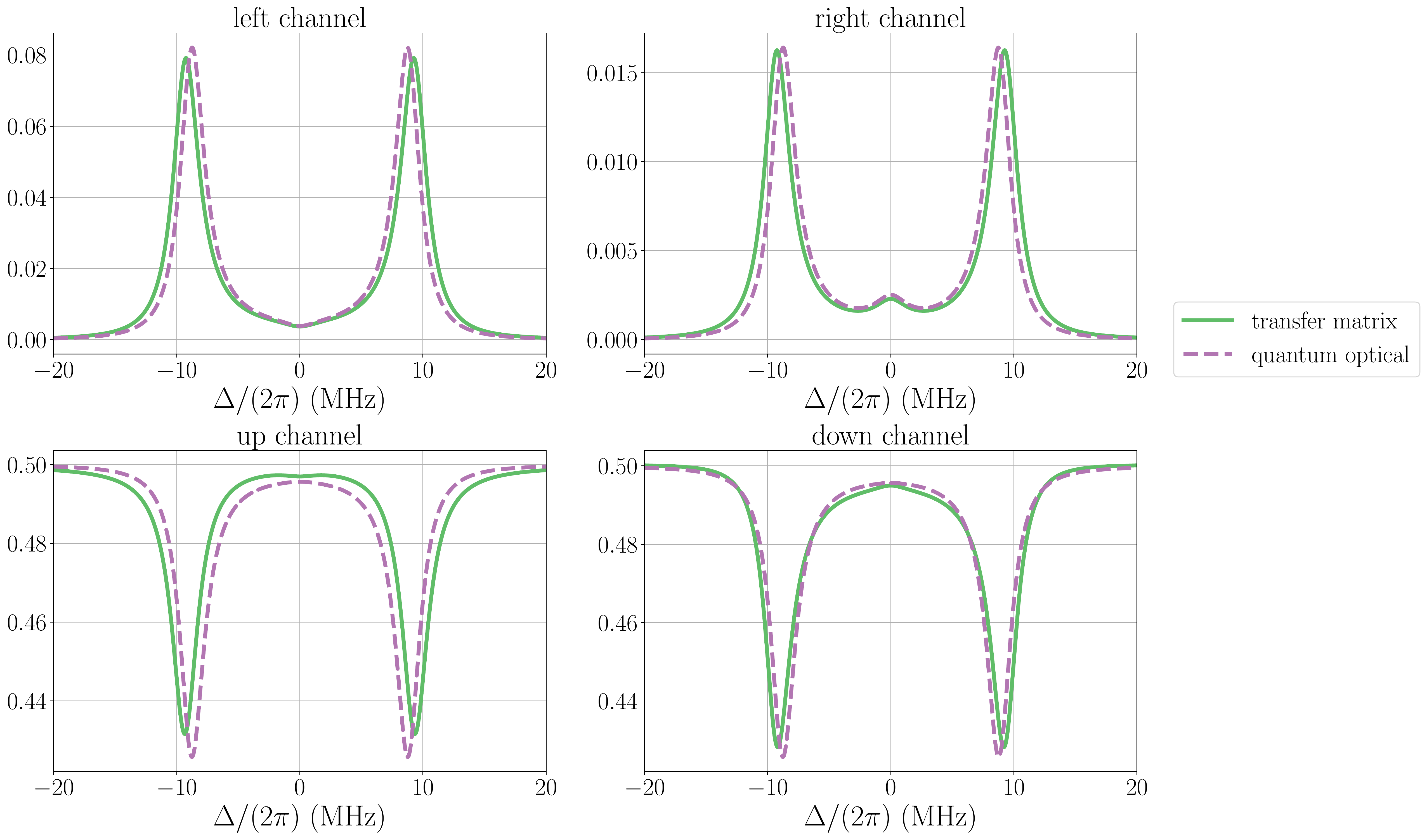}%
\vspace{-2mm}
\caption{Probing the system through both fiber beam splitter input ports (up $\E{in}_u$ and down $\E{in}_d$) simultaneously. The parameters for these spectra can be found in Tables~\ref{tab:expTM_params} and \ref{tab:expQO_params} of Appendix \ref{sec:params} (except for $g_1=g_2=0$).
\label{fig:CQCQ_CDQM_empty}}
\vspace{-.3cm}
\end{figure*}

The presence of strongly-coupled atoms in the cavities generally mitigates the differences between spectra obtained from the TM and QO models, and one observes at least qualitative agreement between the two. This can also be observed in Fig.~\ref{fig:CQCQ_CQM_both} where both cavities are loaded with atoms. However, without atoms at all, as illustrated in Fig.~\ref{fig:CQCQ_CQM_empty}, a striking quantitative {\em and qualitative} difference can be observed. In particular, where the QO model predicts only weak emission on resonance in the right output channel ($E_1^{\rm (out)}$), as one would also expect from the associated normal-mode analysis (resonant driving of the connecting fiber should not excite the fiber-dark mode), the TM model gives a pronounced peak and strong emission. Note that we have in fact also observed this resonance in the laboratory, confirming the TM prediction.

The reasoning based upon traveling- and standing-wave modes in the system also applies in this case. 
The high transmission rates of mirrors 2 and 3, as well as the symmetry-breaking directionality of the incident probe field, means that the simple picture of the probe field driving only a resonant, standing-wave fiber mode is no longer valid.
Viewed another way, a significant portion of the probe field incident through $E_u^{\rm (in)}$ can be transmitted on resonance directly to the right ($\E{out}_2$) in a single pass.

In order to achieve better agreement between the TM and single-mode QO models, we can consider higher reflectances for the mirrors, as in Fig.~\ref{fig:CQCQ_CQM_allhighR}, where $R_{1-4}=0.95$. This leads to a ``better-defined'' fiber mode and ``cleaner'' coupling of the probe to this mode, which clearly reduces the previous difference between the two approaches, but does not completely eliminate it.

\subsection{Driving through multiple ports}

In the previous subsection we argued that the discrepancies between the predictions of the QO and TM approaches are primarily due to the lower reflectances of the central mirrors, which reduce the effectiveness with which standing-wave modes are established or coupled to by an incident traveling-wave probe. To further justify this argument, we can also try to recover the standing-wave-like character of the field modes by driving the system in opposite directions simultaneously, i.e., through both the up ($\E{in}_u$) and down ($\E{in}_d$) input ports of the beam splitter.

Using equations (\ref{eq:Eab_with_TS12}-\ref{eq:S_MBS}) and the conversion between transfer and scattering matrices (\ref{eq:TtoS}), we can derive the following set of equations for the output spectra:

\begin{figure*}[tb!]
\includegraphics[width=0.895\textwidth]{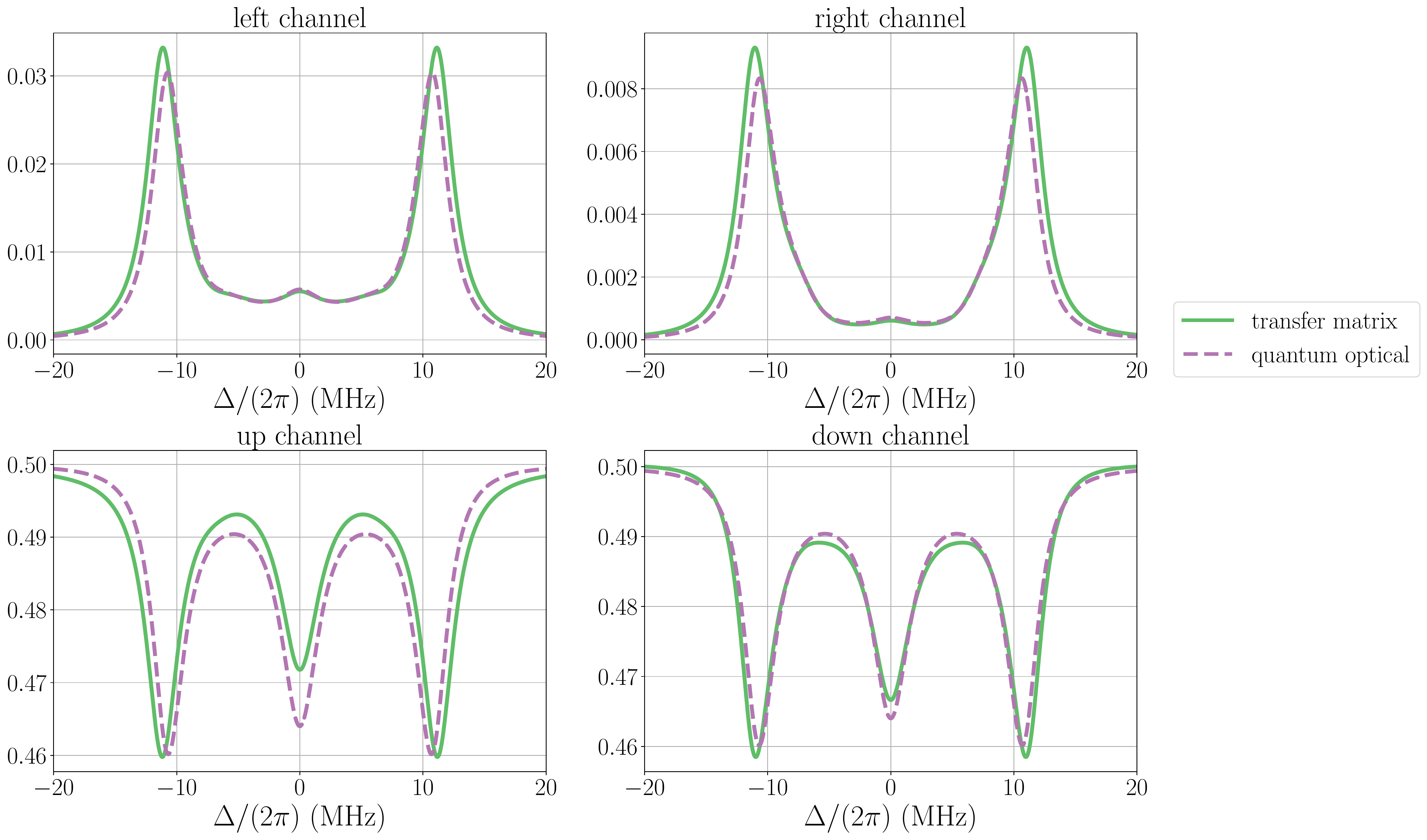}%
\vspace{-2mm}
\caption{Same as Fig.~\ref{fig:CQCQ_CDQM_empty} but both cavities are loaded with atoms. The parameters for these spectra can be found in Tables~\ref{tab:expTM_params} and \ref{tab:expQO_params} of Appendix \ref{sec:params}.
\label{fig:CQCQ_CDQM_both}}
\vspace{-.3cm}
\end{figure*}

\begin{widetext}
\begin{align}
\E{out}_1 &= \xi\lka \sqrt{T^\BS}\frac{1}{\transt{S1}_{11}}\frac{1}{\transt{S2}_{11}}\E{in}_1\
-\lsz T^\BS\frac{\transt{S1}_{12}}{\transt{S1}_{11}}\frac{\transt{S2}_{22}}{\transt{S2}_{11}}+\frac{\transt{S2}_{12}}{\transt{S2}_{11}}\rsz\E{in}_2 \right. \nn
\\
& \left. \hspace{.8cm} -i\sqrt{R^\BS T^\BS}\frac{\transt{S1}_{12}}{\transt{S1}_{11}}\frac{1}{\transt{S2}_{11}}\E{in}_d
+i\sqrt{R^\BS}\frac{1}{\transt{S2}_{11}}\E{in}_u \rka ,
\end{align}

\begin{align}
\E{out}_2 &= \xi\lka\lsz T^\BS\frac{\transt{S1}_{22}}{\transt{S1}_{11}}\frac{\transt{S2}_{21}}{\transt{S2}_{11}}+\frac{\transt{S1}_{21}}{\transt{S1}_{11}}\rsz\E{in}_1
+\sqrt{T^\BS}\frac{\det{\transt{S1}}}{\transt{S1}_{11}}\frac{\det{\transt{S2}}}{\transt{S2}_{11}}\E{in}_2 \right.\nn
\\
&\left.\hspace{.8cm}+i\sqrt{R^\BS}\frac{\det{\transt{S1}}}{\transt{S1}_{11}}\E{in}_d
+i\sqrt{R^\BS T^\BS}\frac{\transt{S2}_{21}}{\transt{S2}_{11}}\frac{\det{\transt{S1}}}{\transt{S1}_{11}}\E{in}_u\rka ,
\end{align}

\begin{align}
\E{out}_d &= \xi\lka \frac{i\sqrt{R^\BS}}{\transt{S1}_{11}}\E{in}_1
-i\sqrt{R^\BS T^\BS} \frac{\transt{S1}_{12}}{\transt{S1}_{11}}\frac{\det{\transt{S2}}}{\transt{S2}_{11}}\E{in}_2\right.\nn
\\
&\left.\hspace{.8cm}+R^\BS\frac{\transt{S1}_{12}}{\transt{S1}_{11}}\E{in}_d
+\sqrt{T^\BS}\lsz1+\frac{\transt{S1}_{12}}{\transt{S1}_{11}}\frac{\transt{S2}_{21}}{\transt{S2}_{11}}\rsz\E{in}_u\rka ,
\end{align}

\begin{align}
\E{out}_u &= \xi\lka i\sqrt{R^\BS T^\BS}\frac{1}{\transt{S1}_{11}} \frac{\transt{S2}_{21}}{\transt{S2}_{11}}\E{in}_1
+i\sqrt{R^\BS}\frac{\det{\transt{S2}}}{\transt{S2}_{11}}\E{in}_2 \right.\nn
\\
&\left.\hspace{.8cm}+\sqrt{T^\BS}\lsz1+\frac{\transt{S1}_{12}}{\transt{S1}_{11}}\frac{\transt{S2}_{21}}{\transt{S2}_{11}}\rsz\E{in}_d-R^\BS\frac{\transt{S2}_{21}}{\transt{S2}_{11}}\E{in}_u\rka .
\end{align}

\end{widetext}

Meanwhile, the QO model simply accounts for the effects of a second probe field with an additional driving term in the fiber Hamiltonian $\Hop^{\rm (fiber-dr)}$,
\begin{align}
    \Ed_{\rm d}\lk\bop+\bdop\rk,
\end{align}
where $\Ed_d$ is the driving strength corresponding to the other port.

Driving through both input ports of the beam splitter, we obtain very good agreement between the TM and QO approaches, even for the spectrum detected from the right output channel (Fig.~\ref{fig:CQCQ_CDQM_empty}). This is due to the counter-propagating contributions from the probe fields enhancing their effective ``overlap'' with the standing-wave modes upon which the QO approach is based. This is further supported by the good agreement in the case where both cavities are loaded with atoms as well (Fig.~\ref{fig:CQCQ_CDQM_both}).

\section{Conclusion}

In this work, we have presented a transfer matrix formalism that offers a straightforward and intuitive means of calculating the transmission and reflection spectra of weak probe fields incident upon a network of fibers, fiber cavities, beam splitters, and nanofiber-coupled atoms. It is intrinsically multimode in nature, accounting properly for propagation distance and the case of low mirror reflectance, and thus gives a more accurate description of such networks in its (linear response) regime of validity than quantum-optical models based upon single mode fields. 

For examples, we have focused on recent, topical configurations involving coupled, nanofiber-cavity-QED systems \cite{Kato2019,White2019}, for which single-mode quantum-optical models still provide good descriptions of experiments, but where differences with the transfer matrix approach, though typically quite small, are clearly noticeable. In cases of more substantial differences, we are able to trace the differences to the relatively low reflectances of mirrors, appreciable propagation distances (i.e., small free spectral range), and to the precise way in which the probe laser is incident upon the system, i.e., to an explicit directionality of excitation in the system, which is not accounted for in the quantum-optical model.

The simplicity and flexibility of the transfer matrix model make it well suited to examine quite general fiber-based networks, as well as other, topical atom-nanofiber systems, such as (possibly large) ring-cavity QED setups \cite{Ruddell2017,Johnson2019} with potential chiral coupling between the atoms and the waveguide. It might also prove useful in examining the behavior of fiber-based, coherent feedback setups with finite time delays, e.g., a version of the emitter-in-front-of-a-mirror problem (see, e.g., \cite{Dorner2002,Tufarelli2014}), where atoms couple to a portion of nanofiber, which is in turn linked to a long length of regular fiber terminated at the end by a highly reflecting FBG mirror.



\appendix
\section{Notations}
Each mirror in the schematics is numbered $j$ in increasing order from left to right. Similarly, each cavity and corresponding atom in the schematics is numbered $j$ in increasing order from left to right.

\begin{table}[h!]
\centering
\caption{Mirror and atomic parameters}
\label{tab:mirat_param}
\begin{tabular}{ c|c } 
 \hline\hline
 \multicolumn{2}{c}{Transfer Matrix} \\
 \hline\hline
 Reflectivities & $R_j$ \\ 
 Transmission & $T_j$ \\ 
 Atom's distance from the left mirror & $d_j$\\
 \hline\hline
 \multicolumn{2}{c}{Quantum Optical} \\
 \hline  \hline
Atomic resonance frequency & $\omega_{\rm A}$ \\ 
 Atomic detuning & $\Delta_{\rm A}=\omega-\omega_{\rm A}$ \\
 Atomic spontaneous emission & $\gamma$ \\
 Mirror transmission rates & $\kappa_{j}$ \\ 
 \end{tabular}
\end{table}

\begin{table}[h!]
\centering
\caption{Cavity parameters}
\label{tab:cav_param}
\begin{tabular}{ c|c } 
 \hline\hline
 \multicolumn{2}{c}{Transfer Matrix} \\
 \hline\hline
 Cavity length & $l_{j}$ \\ 
 Effective transmission & $\eta_j$\\
 Cavity loss ratio & $\alpha_j=1-\eta_j$\\
 Free spectral range & $\omega_{{\rm FSR}j}$\\
 \hline\hline
 \multicolumn{2}{c}{Quantum Optical} \\
 \hline\hline
 Resonance frequencies & $\omega_{{\rm C}j}$ \\ 
 Detunings & $\Delta_{{\rm C} j}=\omega-\omega_{{\rm C }j}$ \\ 
 Overall decay rates & $\kappa_{{\rm C }j}$ \\ 
 Intrinsic fibre loss & $\kappa_{{\rm C} j, {\rm i}}$\\
 Driving rate & $\Ed_j$\\
Atom-cavity coupling strength & $g_j$\\
Cavity-fiber coupling strength & $v_j$
 \end{tabular}
\end{table}

\begin{table}[h!]
\centering
\caption{Parameters of the connecting fiber}
\label{tab:fib_param}
\begin{tabular}{ c|c } 
 \hline\hline
 \multicolumn{2}{c}{Transfer Matrix} \\
 \hline\hline
 Cavity length & $l_{\rm f}$ \\ 
 Effective transmission & $\eta_{\rm f}$\\
 Cavity loss ratio & $\alpha_{\rm f}=1-\eta_{\rm f}$\\
 Free spectral range & $\omega_{\rm FSRf}$\\
 \hline\hline
 \multicolumn{2}{c}{Quantum Optical} \\
 \hline\hline
 Resonance frequency & $\omega_{\rm Cf}$ \\ 
 Detuning & $\Delta_{\rm Cf}=\omega-\omega_{\rm Cf}$ \\ 
 Overall decay rates & $\kappa_{\rm Cf}$ \\ 
 Intrinsic fiber loss & $\kappa_{\rm Cf,i}$\\
 Outcoupling rate & $\kappa_{\rm BS}$
 \end{tabular}
\end{table}

\newpage
\section{Parameters for the coupled cavity-QED experiments}\label{sec:params}
In order to demonstrate that the difference between the transfer matrix results and the quantum optical predictions is measurable, we have used the parameters of recent experiments.

\begin{table}[h!]
\centering
\caption{CCQED parameters in the transfer matrix (TM) formalism:}
\label{tab:expTM_params}
\begin{tabular}{l|l}
 \hline\hline
 \multicolumn{2}{c}{Transfer Matrix} \\
 \hline\hline
Detuning between the optical & $\Delta=\omega_{{\rm C}j}-\omega$\\ \hspace{.3cm} (atomic) and the driving field& $\hspace{.35cm}=\omega_{{\rm A}j}-\omega$\\
Mirror reflectances &$R_1=0.8$\\
					&$R_2=0.65$\\
					&$R_3=0.8$\\
					&$R_4=0.85$\\
Lengths of the cavities and &$l_1=0.92$~m\\
\hspace{.3cm}the connecting fiber &$l_2=1.38$~m\\
								&$l_{\rm f}=1.8$~m\\
Free spectral range of &$\omega_{{\rm FSR}1}=112.25$~MHz\\		\hspace{.3cm}the cavities and &$\omega_{{\rm FSR}2}=74.833$~MHz\\
\hspace{.3cm}the connecting fiber &$\omega_{{\rm FSRf}}=57.372$~MHz\\
Intrinsic transmission of the cavities & $\eta_1=\eta_2$\\
\hspace{.3cm}and the connecting fiber&$\hspace{.4cm}=\eta_{\rm f}=0.97$\\
Reflectance of the outcoupling BS & $R^\BS=0.01$\\
Energy decay rate into free space &$\Gamma^\prime=5.2\times 2\pi$ MHz
\end{tabular}
\end{table}

\begin{table}[h!]
\centering
\caption{CCQED parameters in the quantum-optical single-mode (QO) approach:}
\label{tab:expQO_params}
\begin{tabular}{l|l}
 \hline\hline
 \multicolumn{2}{c}{Quantum Optical} \\
 \hline\hline
Cavity and fiber & $\kappa_1/2\pi=1.787$~MHz\\
	\hspace{.3cm}outcoupling rate&$\kappa_4/2\pi=0.893$~MHz\\
$\hspace{.3cm}\kappa_j=\frac{1}{2}\frac{\omega_{{\rm FSR}}}{2\pi}(1-R_j)$&$\kappa_{\rm BS}/2\pi=0.046$~MHz\\
Intrinsic loss rates in the & $\kappa_{{\rm C}1,i}/2\pi=0.544$~MHz\\
\hspace{.3cm}cavities and the fiber	& $\kappa_{{\rm C}2,i}/2\pi=0.363$~MHz\\
$\hspace{.3cm}\kappa_{Cj,i}=-\frac{1}{2}\frac{c}{l_j}\ln{\eta_j}$& $\kappa_{{\rm Cf},i}/2\pi=0.278$~MHz\\
Cavity-fiber coupling strength & $v_{1} =7.556$~MHz\\
$\hspace{.3cm}v_j=\sqrt{\frac{\kappa_{j+1}}\pi\omega_{{\rm FSRf}}}$& $v_{2} =4.664$~MHz\\
Atom-cavity  & $g_1= 6\times2\pi$ MHz\\
\hspace{.3cm}coupling strengths&$g_2= 7\times2\pi$ MHz
\end{tabular}
\end{table}

\newpage
\bibstyle{apsrev4-1}
\bibliography{Citations}


\end{document}